\documentstyle[aps,preprint,epsfig]{revtex}

\def\rem{$R_{\mbox{\scriptsize EM}}$\ }       
\def\rsm{$R_{\mbox{\scriptsize SM}}$\ }       

\def\remeq{R_{\mbox{\scriptsize EM}} }       
\def\rsmeq{R_{\mbox{\scriptsize SM}} }       

\begin{document}

\draft

\preprint{nucl-th/0010025}


\title{Dynamical Study of the $\Delta$ Excitation in
$N(e,e^\prime \pi)$ Reactions}


\author{
T. Sato,$^{a,b}$%
and T.-S. H. Lee\,$^b$%
}


\address{
$^a$
Department of Physics, Osaka University, Toyonaka, Osaka 560-0043,
 Japan\\ 
$^b$
Physics Division, Argonne National Laboratory, Argonne,
Illinois 60439}

\maketitle


\begin{abstract}
The dynamical model developed in [Phys. Rev. C {\bf 54}, 2660 (1996)] 
has been applied to investigate the pion electroproduction reactions on 
the nucleon. It is found that the model can describe to a very large extent 
the recent data of $p(e,e^\prime \pi^0)$ reaction from 
Jefferson Laboratory and MIT-Bates.
The extracted magnetic dipole(M1), electric dipole(E2), and Coulomb(C2)
strengths of the $\gamma N \rightarrow \Delta$ transition are presented.
It is found that the C2/M1 ratio drops significantly with Q$^2$ and
reaches about -13$\%$ at Q$^2$=4 (GeV/c)$^2$, while the E2/M1 ratio
remains close to the value $\sim -3 \%$ at the $Q^2=0$ photon point.
The determined M1 transition form factor drops
faster than the usual dipole form factor of the proton.
We also find that the non-resonant
interactions can dress the $\gamma N \rightarrow \Delta$ vertex
to enhance strongly its strength at low $Q^2$, but
much less at high $Q^2$. 
Predictions are presented for future experimental tests. 
Possible developments of the model are discussed.
\end{abstract}

\pacs{PACS number(s): 13.75.Gx,21.45.+v,24.10-i,25.20.Lj}

\section{Introduction}

One of the purposes of investigating the nucleon resonances($N^*$) is
to understand the non-perturbative dynamics of 
Quantum Chromodynamics(QCD). One possible approach to realize this
is to compare the predictions of QCD-inspired models with 
the resonance parameters which can be extracted from
the data of pion photoproduction and electroproduction 
reactions. In recent years, precise data including polarization
observables have been obtained in the $\Delta$ region
for pion photoproduction  at LEGS\cite{LEGS} and Mainz\cite{MAINZ},
and for pion electroproduction at Thomas Jefferson National Accelerator
Facility(JLab)\cite{JLAB,Bur},
MIT-Bates\cite{MIT} and NIKHEF\cite{NIKHEF}. These data now
allow us to investigate more precisely the electromagnetic
excitation of the $\Delta$ resonance.

In Ref.\cite{sl} we have developed a dynamical model(henceforth 
called the SL model) to extract
the magnetic dipole(M1) and electric quadrapole(E2) strengths of
the $\gamma N \rightarrow \Delta$ transition
from the pion photoproduction data.
The precise polarization data from LEGS and Mainz were essential in
our analysis.
In this paper, we report on the progress we have made in extending
the SL model to investigate the pion electroproduction reactions 
in the $\Delta$ excitation region. We will make use of the recent data
from JLab and MIT-Bates to explore the Q$^2$-dependence of
the $\gamma N \rightarrow\Delta$ transition 
and make predictions for future experimental tests.

The dynamical content of the SL model has been given in detail 
in Ref.\cite{sl}. The essential feature of the model 
is to have a consistent
description of both the $\pi N$ scattering and the electromagnetic
production of pions. This is achieved by using a 
unitary transformation
method to derive an effective Hamiltonian
defined in the subspace $\pi N\oplus\gamma N\oplus\Delta$ from the
interaction Lagrangians for $N,\Delta,\pi,\rho,\omega$ and photon
fields. The resulting model 
has given a fairly successful description of the
very extensive data for pion photoproduction.
The extension of the SL model to investigate pion 
electroproduction is straightforward.
The formulae needed for calculating the current matrix elements
of pion electroproductions are identical to  that given in Ref.\cite{sl}
except that a form factor must be included at each photon vertex. 
Therefore no detailed presentation of our model will be repeated here.
Similarly, we will not give detailed formulae for calculating the
electroproduction cross sections since they are well 
documented\cite{donn,ras,nl,nl1,dre}.
 
The SL model is one of the dynamical models developed
\cite{sl,tanabe,yang,nbl,pearcelee,afn,surya,habe,nakaya,ky} 
in recent years. 
Compared with other approaches based on the tree-diagrams of
effective Lagrangians\cite{muko90,muko99,maid} or 
dispersion-relations\cite{disp,hdt98,a98}, the main objective of a dynamical
approach is to separate the reaction mechanisms from the excitation
of the internal structure of the hadrons involved. 
Within the SL model, this has been achieved by applying 
the well-established reaction theory within the Hamiltonian
formulation(see, for example, Ref.\cite{feshbach}). 
In particular, the off-shell non-resonant contributions to
the $\gamma N \rightarrow \Delta$ form factors can be 
calculated explicitly in a dynamical approach.
Only when such non-resonant contributions are separated,
the determined "bare" $\gamma N \rightarrow \Delta$ form factors
can be compared with the predictions from hadron models.
Within the SL model, this was explored in detail and 
provided a dynamical interpretation of
the long-standing discrepancy between 
the empirically determined magnetic M1 strength of the
$\gamma N \rightarrow \Delta$ transition and the predictions from 
constituent quark models. In this work, we further explore
this problem utilizing the $Q^2$-dependence accessible to
electroproduction reactions. Furthermore, the Coulomb(Scalar) 
component $C2$($S2$) of  
the $\gamma N \rightarrow \Delta$ form factor will be determined. 

In section II, we briefly review the essential ingredients of the
SL model and define various form factors which are needed to
describe  pion electroproduction reactions. 
With the Mainz data\cite{MAINZ},
we have slightly refined our model at the $Q^2=0$ photon point. This will be
reported in section III.
The electroproduction  results are presented and compared
with the data in section IV.
In section V, we give a summary and discuss possible future
developments.

\section{The SL model}

Within the SL model,
the pion photoproduction and electroproduction reactions  are described 
in terms of photon and hadron degrees of freedom.
The starting Hamiltonian is
\begin{eqnarray}
H = H_0 + H_I, \label{hamilt}
\end{eqnarray}
with 
\begin{eqnarray}
H_I = \sum_{M,B,B^\prime} \Gamma_{MB\leftrightarrow B^\prime}, \label{hamili}
\end{eqnarray}
where $H_0$ is the free Hamiltonian and
 $ \Gamma_{MB \leftrightarrow B^\prime}$ describes
the absorption and emission of a meson($M$) by a baryon($B$).
In the SL model, such a Hamiltonian is obtained 
from  phenomenological Lagrangians for $N, \Delta, \pi, \rho, \omega$ 
and photon fields. In a more microscopic approach, this Hamiltonian can be
defined in terms of a hadron model, as attempted, for example, 
in Ref. \cite{yoshi}.  

It is a non-trivial many body problem to 
calculate $\pi N$ scattering and $\gamma N \rightarrow \pi N$ 
reaction amplitudes from the Hamiltonian Eq. (\ref{hamilt}). 
To obtain a manageable reaction model,
a unitary transformation method\cite{sl,ksh} is used  
up to second order in $H_I$ to derive an effective Hamiltonian.
 The essential idea of the employed
unitary transformation method is to eliminate the unphysical vertex
interactions $MB \rightarrow B^\prime$ with $m_M +m_B < m_{B^\prime}$ 
from the Hamiltonian and absorb their effects into 
 $MB\rightarrow M^\prime B^\prime$ two-body interactions. In the SL model,
the resulting effective Hamiltonian is  
defined in a subspace spanned by the
 $\pi N$, $\gamma N$ and $\Delta$ states and has the following form
\begin{eqnarray}
H_{eff} & = & H_0 + v_{\pi N} + v_{\gamma\pi}
 + \Gamma_{\pi N \leftrightarrow \Delta} 
+ \Gamma_{\gamma N \leftrightarrow \Delta},   \label{hamile}
\end{eqnarray}
where
$v_{\pi N}$ is a non-resonant $\pi N$ potential, and $v_{\gamma\pi}$
describes the non-resonant $\gamma N \leftrightarrow \pi N$
transition. The $\Delta$ excitation
is described by the vertex interactions
 $\Gamma_{\gamma N \leftrightarrow \Delta}$
 for  the $\gamma N \leftrightarrow\Delta$ transition
 and $\Gamma_{\pi N \leftrightarrow \Delta}$ 
for the  $\pi N \leftrightarrow\Delta$ transition. 
The vertex interaction  
$\Gamma_{\gamma N\leftrightarrow \Delta}$ is illustrated in Fig. 1.
The non-resonant $v_{\gamma\pi}$ consists of the usual pseudo-vector
Born terms, $\rho$ and $\omega$ exchanges, and the crossed $\Delta$
term, as illustrated in Fig. 2(the non-resonant 
term due to an intermediate anti-$\Delta$ 
state was found to be very weak and can be neglected).
Most of the dynamical models have the above form of the Hamiltonian.
However, the SL model has an important feature that
the deduced effective Hamiltonian $H_{eff}$
is energy independent and hermitian.
Hence, the unitarity of the resulting reaction amplitudes
is trivially satisfied.  Furthermore, the 
non-resonant interactions $v_{\gamma\pi}$ and $v_{\pi N}$
are derived from the same $H_I$ of Eq. (\ref{hamilt}), and hence
the $\pi N$ and $\gamma N$ reactions 
can be described consistently. 
Such a consistency is lost if $v_{\pi N}$ is either
constructed purely phenomenologically as done in 
Refs.\cite{tanabe,yang,nbl} or taken from a different theoretical
construction as done in Refs.\cite{pearcelee,ky}. 
This consistency is essential in
interpreting the extracted 
$\gamma N \rightarrow \Delta$ form factors since 
the non-resonant interactions $v_{\pi N}$
and $v_{\gamma\pi}$ can dress the $\gamma N \rightarrow \Delta$
vertex. As discussed in Refs.\cite{sl,lee}, only the dressed 
$\gamma N \rightarrow \Delta$ transition can be identified with the data.
The importance of the non-resonant effects
on the $\gamma N\rightarrow \Delta$ transition is also stressed 
recently in Ref.\cite{ky}.

From the effective Hamiltonian Eq. (\ref{hamile}), it is straightforward 
to derive
a set of coupled equations for $\pi N$ and $\gamma N$ reactions.
The resulting pion photoproduction amplitude can be written as
\begin{eqnarray}
T_{\gamma\pi}= <\pi N \mid \epsilon\cdot J \mid \gamma N>, \label{tj}
\end{eqnarray}
where $J$ is the current operator and
$\epsilon$ is the photon polarization
vector. It can be decomposed into two parts
\begin{eqnarray}
T_{\gamma\pi}(E)  =  t_{\gamma\pi}(E) + 
\frac{
\bar{\Gamma}_{\Delta \rightarrow \pi N}
\bar{\Gamma}_{\gamma N \rightarrow \Delta}
}
{E - m_\Delta - \Sigma_\Delta(E)} . \label{tmatt}
\end{eqnarray}
The non-resonant amplitude $t_{\gamma\pi}$ is 
calculated from $v_{\gamma\pi}$
by
\begin{eqnarray}
t_{\gamma\pi}(E)=   v_{\gamma\pi} 
+  t_{\pi N}(E)G_{\pi N}(E)v_{\gamma\pi},  \label{tmatg}
\end{eqnarray}
where the $\pi N$ free propagator is defined by
\begin{eqnarray}
G_{\pi N}(E)=\frac{1}{E-E_N(k)-E_\pi(k)+i\epsilon} . \label{green}
\end{eqnarray} 
The  amplitude $t_{\pi N}$ in Eq. (\ref{tmatg}) is calculated from 
the non-resonant $\pi N$
interaction $v_{\pi N}$ by solving the following equation 
\begin{eqnarray}
t_{\pi N}(E) = v_{\pi N} + v_{\pi N} G_{\pi N}(E) t_{\pi N}(E) . \label{tmatp}
\end{eqnarray}

The dressed vertices in Eq. (\ref{tmatt}) are defined by 
\begin{eqnarray}
\bar{\Gamma}_{\gamma N \rightarrow \Delta}  &=&  
   \Gamma_{\gamma N \rightarrow \Delta} + 
\bar{\Gamma}_{\pi N \rightarrow \Delta} G_{\pi N}(E) 
v_{\gamma \pi},  \label{vertg} \\
\bar{\Gamma}_{\Delta \rightarrow \pi N}
 &=& [1+t_{\pi N}(E)G_{\pi N}(E)]\Gamma_{\Delta\rightarrow\pi N}. \label{vertp}
\end{eqnarray}
In Eq.(\ref{vertg}), we also have defined
\begin{eqnarray}
\bar{\Gamma}_{\pi N \rightarrow \Delta}
 &=&\Gamma_{\pi N \rightarrow\Delta}[1 + G_{\pi N}(E)t_{\pi N}(E) ].
\nonumber
\end{eqnarray}
The $\Delta$ self-energy in Eq. (\ref{tmatt}) is then calculated from
\begin{eqnarray}
\Sigma_\Delta(E) = 
 \Gamma_{\pi N\rightarrow \Delta}
G_{\pi N}(E)\bar{\Gamma}_{\Delta \rightarrow \pi N}.  \label{self}   
\end{eqnarray}

As seen in the above equations,
an important consequence of the dynamical model is that 
the influence of the non-resonant mechanisms on 
the resonance properties can be identified and calculated explicitly.
The resonance position of the amplitude defined by
Eq. (\ref{tmatt}) is shifted from 
the bare mass $m_\Delta$  by the self-energy $\Sigma_\Delta(E)$.
The bare vertex $\Gamma_{\gamma N\rightarrow \Delta}$ is modified 
by the non-resonant interaction $v_{\gamma\pi}$ to give the dressed vertex 
$\bar{\Gamma}_{\gamma N\rightarrow\Delta}$, as defined by Eq. (\ref{vertg}). 
In the SL model, it was found that the extracted M1 strength of the bare
vertex $\Gamma_{\gamma N \rightarrow \Delta}$ 
is very close to the values predicted by the 
constituent quark models\cite{capst,bil,bhf},
while the empirical values given by the  Particle 
Data Group(PDG)\cite{pdg} can only
be identified with the dressed vertex 
$\bar{\Gamma}_{\gamma N \rightarrow \Delta}$.

The above equations can be solved for arbitrary photon four-momentum
$Q^2=-q^2 >0$.  For investigating the electroproduction
reactions, we only need to
define a form factor at each photon vertex in Figs. 1-2. 
For the non-resonant interactions(Fig. 2), we follow the previous
work\cite{nl,nl1}. 
The usual electromagnetic nucleon form factors(given explicitly in the
Appendix A of Ref.\cite{nl}) are used
in evaluating  the direct and crossed nucleon terms. 
To make sure that the non-resonant term 
$v_{\gamma\pi}$ is gauge invariant, we set
\begin{eqnarray}
F_A(q^2)=F_{\gamma \pi\pi}(q^2)=F_1^V(q^2), \label{formv}
\end{eqnarray}
where $F_A(q^2)$ is the form factor for the contact
term, $F_{\gamma\pi\pi}(q^2)$ is the pion form factor for the
pion-exchange term, and $F_1^V(q^2)$ is the nucleon isovector
form factor(also given explicitly in the appendix A of
Ref.\cite{nl}). The form factors for the 
vector meson-exchange terms are chosen to be
\begin{eqnarray}
 g_{V\pi\gamma}(q^2)=g_{V\pi\gamma}/(1-q^2/m_V), \label{coupv}
\end{eqnarray}
 where $m_V$ is the vector meson mass and the coupling constants
$g_{V\pi\gamma}$ for $V=\rho,\omega$ are deduced from
 the $V\rightarrow \gamma \pi$ decay widths and are given in Ref.\cite{sl}.
The prescriptions Eqs. (\ref{formv})-(\ref{coupv}) 
have been commonly used in most of
the previous investigations such as that in Refs.\cite{nl,nl1}. Undoubtly,
this is an unsatisfactory aspect of this work. On the other hand, 
"dynamically" sound progress in solving the gauge invariance problem
 cannot be made unless a microscopic theory
of hadron structure is implemented consistently into our model.
This is beyond the scope of this work.

For the $\gamma N \rightarrow \Delta$ form factors, we extend Eq. (4.15) of
Ref.\cite{sl} to write in the rest frame of the $\Delta$
\begin{eqnarray}
<\Delta\mid \Gamma_{\gamma N\rightarrow \Delta} \mid q>
  &=& -\frac{e}{(2\pi)^{3/2}}\sqrt{\frac{E_N(\vec{q})+m_N
}{2E_N(\vec{q})}}
       \frac{1}{\sqrt{2\omega}} 
       \frac{3 (m_\Delta+m_N) }{ 4m_N( E_N(\vec{q})+m_N)}T_3 
\nonumber \\
&\times& 
{}  [i G_M(q^2)\vec{S}\times\vec{q}\cdot \vec{\epsilon}
      +G_E(q^2)
       (\vec{S}\cdot\vec{\epsilon} \vec{\sigma}\cdot\vec{q}
   +\vec{S}\cdot\vec{q}\vec{\sigma}\cdot\vec{\epsilon}) \nonumber \\
 & & +
        \frac{G_C(q^2)}{m_\Delta}
      \vec{S}\cdot\vec{q} \vec{\sigma}\cdot\vec{q} \epsilon_0], \label{vertgd}
\end{eqnarray}
where $e=\sqrt{4\pi/137}$, $q=(\omega,\vec{q})$ is the photon four-momentum,
and $\epsilon=(\epsilon_0,\vec{\epsilon})$ is the photon polarization
vector.
The transition operators $\vec{S}$ and $\vec{T}$ are defined
by the reduced matrix element $<\Delta ||\vec{S}|| N> =
<\Delta || \vec{T} || N > = 2$ in Edmonds' convention\cite{edm}. 
The parameterizations of the form factors $G_M(q^2)$, $G_E(q^2)$ and
$G_C(q^2)$ will be specified in section IV. 

With the form factors defined in Eqs.(12)-(14), 
both the non-resonant term $v_{\gamma\pi}$ and
the bare vertex $\Gamma_{\gamma N \rightarrow \Delta}$ are gauge invariant.
However the full amplitude defined by Eq. (\ref{tmatt}) involves
off-shell $\pi N$
scattering, as defined by Eqs. (\ref{tmatg})-(\ref{self}), is not gauge 
invariant. There exists
a simple prescription to eliminate this problem phenomenologically. 
This amounts to defining the conserved 
currents for Eq.(\ref{tj}) as
\begin{eqnarray}
J^\mu = J^\mu(SL) - \frac{q\cdot J(SL)}{n\cdot q} n^\mu , \label{gauj}
\end{eqnarray}
where $J(SL)$ is calculated from our model defined above,
and $n$ is an arbitrary four vector. It is obvious that the
currents $J^\mu$  defined by Eq. (\ref{gauj}) 
satisfies the
gauge invariance condition $q\cdot J =0$. If we use the  standard choice
of the photon momentum $q=(\omega,0,0,\mid\vec{q}\mid)$ and 
choose $n=(0,0,0,1)$, we then have
\begin{eqnarray}
J_0 &=& J_0(SL), \nonumber \\
J_x &=& J_x(SL), \nonumber \\
J_y &=& J_y(SL),           \label{gau0}
\end{eqnarray}
and
\begin{eqnarray}
J_z &=& J_z(SL) - \frac{\omega J_0(SL)-\mid\vec{q}\mid J_z(SL) }
{-\mid\vec{q}\mid}\cdot 1 \nonumber \\
&=& \frac{\omega}{\mid\vec{q}\mid}J_0(SL). \label{gau00}
\end{eqnarray}
The above equations mean that within our approach any 
$N(e,e'\pi)$ observable depending on the z-component of the 
current is determined by 
Eq. (\ref{gau00}) using the time component of the
SL model, not by $J_z(SL)$. This is very similar to
the prescription used in many nuclear calculations. 
We find that the data we have considered in this work 
can be described by the conserved currents $J^\mu$ defined by
Eqs. (\ref{gau0}) and (\ref{gau00}).
We have briefly investigated the model dependence due to the 
freedom in choosing $n$. We have found that the choice
$n=(1,0,0,0)$, which leads to $\vec{J}=\vec{J}(SL)$ and
 $J_0=\frac{\mid \vec{q}\mid}{\omega} J_z(SL)$ 
gives very similar results
at high $Q^2$. The differences at low $Q^2$ also appear to be not
so large. All results presented in this paper are from using
the choice Eqs. (\ref{gau0})-(\ref{gau00}).
We emphasize that this choice
is a phenomenological part of our model, simply
because we have not implemented any substructure dynamics
 of hadrons into our formulation. 
In fact this is also the case for all existing approaches based on
the prescription similar to the form of Eq. (\ref{gauj}). 
For example $n=q$ is chosen
 in the recent work by Kamalov and Yang\cite{ky} 
using the dynamical model
developed in Ref.\cite{yang}. 
There exist other prescriptions to fix the gauge invariance problem,
 such as those suggested in Refs.\cite{gross,ohta,habe1}. 
We have not explored those possibilities, since they are 
also not related microscopically to the 
substructure of the hadrons involved, and are 
as phenomenological as the prescription
defined by Eqs.(16)-(17).
We will return to this
problem when our model is further developed, as discussed in
section V.

For our later discussions on the $\gamma N \rightarrow \Delta$
 transition, we define some quantities in terms of more
commonly used  conventions.
As discussed in  detail in Ref.\cite{sl}, 
if we replace the $\pi N$ propagator $G_{\pi N}$ in 
Eqs. (\ref{tmatt})-(\ref{self})
by  $G^K_{\pi N}(E)=\frac{P}{E-E_N(k) -E_\pi(k)}$ 
with $P$ denoting the principal-value integration, 
the resulting dressed vertex 
$\bar{\Gamma}^K_{\gamma N \rightarrow \Delta}$ is real and
can be directly compared with the bare vertex 
$\Gamma_{\gamma N\rightarrow \Delta}$. 
The usual E2/M1 ratio \rem and C2/M1 ratio
\rsm for the dressed $\gamma N \rightarrow \Delta$ 
vertex are then defined by 
\begin{eqnarray}
\remeq&=&\frac{[\bar{\Gamma}^{K}_{\gamma N \rightarrow \Delta}]_{E2}}{
{}[\bar{\Gamma}^{K}_{\gamma N \rightarrow \Delta}]_{M1}}, \label{remg} \\
\rsmeq&=&\frac{[\bar{\Gamma}^{K}_{\gamma N \rightarrow \Delta}]_{C2}}{
{}[\bar{\Gamma}^{K}_{\gamma N \rightarrow \Delta}]_{M1}}.  \label{rsmg}
\end{eqnarray}
One can also show\cite{sl} that at the resonant
energy where the invariant mass $W=$1236 MeV and the
$\pi N$ phase shift in the $P_{33}$ channel  goes through 90 degrees, 
the multipole components of the dressed
vertex $\bar{\Gamma}^K_{\gamma N \rightarrow \Delta}$ are related to
the imaginary($ImM$) parts of the $\gamma N \rightarrow \pi N$
 multipole amplitudes(as defined in Ref.\cite{donn}
in the $\pi N$ $P_{33}$ channel 
\begin{eqnarray}
A_{M}  &=&[\bar{\Gamma}^K_{\gamma N \rightarrow \Delta}]_{M1}=
 N \times \mbox{Im}(M_{1+}^{3/2}), \label{am}\\
A_{E}  &=&[\bar{\Gamma}^K_{\gamma N \rightarrow \Delta}]_{E2}=
 N \times \mbox{Im}(E_{1+}^{3/2}),  \label{ae}\\
A_{C}  &=&[\bar{\Gamma}^K_{\gamma N \rightarrow \Delta}]_{C2}=
 N \times \mbox{Im}(S_{1+}^{3/2}),   \label{ac} 
\end{eqnarray}
where $N$ is a kinematic factor.
From the above relations, we obtain a very useful relation that the
E2/M1 ratio $R_{EM}$ and C2/M1 ratio $R_{SM}$ of the dressed 
$\gamma N \rightarrow \Delta$ transition at
$W=1236$ MeV can be evaluated directly by
using the $\gamma N \rightarrow \pi N$ multipole amplitudes
\begin{eqnarray}
R_{EM} &=& \frac{\mbox{Im} (E_{1+}^{3/2})}{\mbox{Im} (M_{1+}^{3/2})},
         \label{rema} \\
R_{SM} &=& \frac{\mbox{Im} (S_{1+}^{3/2})}{\mbox{Im} (M_{1+}^{3/2})}.
         \label{rsma}
\end{eqnarray}

The formula for calculating the  multipole amplitudes and various
cross sections from the total amplitudes $T_{\gamma \pi}$ will not
be given here, since they are well
documented\cite{donn,ras,nl,nl1,dre}.

\section{The results at $q^2=0$ photon point}

To determine the $\gamma N \rightarrow \Delta$ form factors 
defined by Eq. (\ref{vertgd}),
it is necessary to first fix their values at $q^2=0$ by investigating
the pion photoproduction reactions.
This was done in Ref.\cite{sl} by applying the formulation outlined in
section II.  The first step was to investigate the $\pi N$ scattering
from threshold to the $\Delta$ excitation region.
 By fitting the $\pi N$ phase
shifts, the parameters characterizing
the strong interaction vertices except the $\omega NN$ vertex in
Fig.2 were determined. 
The pion photoproduction data were then
used to determine $G_M(0)$ and $G_E(0)$ of the $\gamma N \rightarrow
\Delta$ transition(Eq. (\ref{vertgd})) and the coupling 
constant $g_{\omega NN}$
of the $\omega NN$ vertex of Fig. 2. In Ref.\cite{sl} we considered
previous pion photoproduction data\cite{bonn0} 
and the LEGS data\cite{LEGS} 
of photon asymmetry  defined by
\begin{eqnarray}
\Sigma = \frac{d\sigma_\perp - d \sigma_\parallel}
              {d\sigma_\perp + d\sigma_\parallel}, \label{sigma}
\end{eqnarray}
where $d\sigma_\perp$($d\sigma_\parallel$) are the cross sections with
photons linearly polarized in the direction perpendicular(parallel) to
the reaction plane. To refine our model, we consider the recent  Mainz 
data\cite{MAINZ} here.
For the photon asymmetry $\Sigma$, the data from Mainz agree very well 
with that
from LEGS. The main improvement we have made is from using the Mainz 
differential cross section($d\sigma/d\Omega$) data which are much more
precise than the data\cite{bonn0} used in Ref.\cite{sl}. 
 
With the $\pi N$ amplitudes calculated from the Model-L of Ref.\cite{sl}, 
we find that the Mainz data can be best reproduced by setting
$G_M(0)=1.85$, $G_E(0)=0.025$, and $g_{\omega NN}=11.5$. These values of
$G_M(0)$ and $G_E(0)$ are identical to that determined in Ref.\cite{sl}.
The $\omega NN$ coupling constant is also only slightly larger than the
value $10.5$ determined there.
Such a small change in
the determined parameters is due to the fact that the 
previous photoproduction data\cite{bonn0} 
are close to the Mainz data except that their errors are larger.

Our results for $\gamma p \rightarrow \pi^0 p$ and
$\gamma p \rightarrow \pi^+ n$ reactions are compared with the data in
Figs. 3 and 4 respectively. Clearly the agreement is satisfactory
in general. For the $\pi^0$ production(Fig. 3),
we also show the dependence of
 the calculated asymmetry $\Sigma$ on
the E2/M1 ratio of the bare $\gamma N \rightarrow \Delta$ vertex.
The value $R_{EM}(bare)=-1.3\%$, which corresponds to $G_M(0)=1.85$ and
$G_E(0)=0.025$, seems to be favored by the data. 
On the other hand, such a dependence is much less for the $\pi^+$
production(Fig. 4)
since the non-resonant interactions play a more important role in this
channel.
There are still some discrepancies with the data.
In particular, the differential cross sections at high energies
$E_\gamma > 350 $ MeV are underestimated. This could be due to
the neglect of the coupling with higher mass nucleon resonances and
two-pion production channels.
For the $\pi^+$ production(Fig. 4), the cross
sections at low energies are also underestimated slightly. This
could be mainly due to the deficiency of our nonresonant
amplitude which plays a much more important role in $\pi^+$ production
than in $\pi^0$ production. Possible improvements of our model
will be discussed in section IV. 

The results presented in the
rest of this paper are from calculations with 
$G_M(0)=1.85$ and $G_E(0)=0.025$.
In Fig. 5, we show that the predicted $M_{1+}^{3/2}$ and $E_{1+}^{3/2}$ 
amplitudes of the $\gamma N \rightarrow \pi N$ reactions
agree very well with the data from the Mainz98\cite{hdt98} 
and SM95\cite{vpi} analyses.
By using Eqs.(\ref{rema})-(\ref{rsma}) and reading
the results at resonant energy $E_\gamma = 340 $ MeV displayed in Fig. 5,
we find that the E2/M1 ratio for the dressed 
$\gamma N \rightarrow \Delta$ transition is $R_{EM}=-2.7\%$.
The dotted curves in Fig. 5 are obtained from setting the non-resonant
interaction  $v_{\gamma \pi}$ to zero in the calculations. 
Clearly the non-resonant mechanism has a 
crucial role in determining the electromagnetic excitation of the
$\Delta$.  
At the resonance energy  $E_\gamma = 340$ MeV the bare $\Delta$ amplitude 
is about 60\% of the full amplitude for the M1 transition
and almost a half for the E2 transition.
As discussed in Refs.\cite{sl,lee}, this large difference 
between the bare $\Delta$ and full amplitudes is the
source of the discrepancies between the quark model predictions of the
 $\gamma N \rightarrow \Delta$ transition and 
the values determined from the empirical amplitude analyses such as 
those listed by PDG\cite{pdg}.

In Table I, we list our results for the helicity amplitude $A_{3/2}$
and E2/M1 ratio of the $\gamma N \rightarrow \Delta$ transition and 
compare them with the results from other approaches. We see that the
 dressed E2/M1 values from different approaches are very close.
For our model and the model of Ref.\cite{ky},
the large differences between the dressed and bare values are evident,
indicating the importance of the non-resonant mechanisms in determining
the $\gamma N \rightarrow \Delta$ transition.
The bare values are clearly close to the quark model predictions.
With $G_M(0)$ and $G_E(0)$ determined, we can then investigate
the pion electroproduction reactions.

\section{Pion Electroproduction}

With the matrix element $T_{\gamma \pi}$ calculated by using
 the formula outlined in section II, it is straightforward to calculate
various observables for pion electroproduction reactions. The needed
formulation is well documented; see, for example, 
Refs.\cite{donn,ras,nl,nl1,dre}. 
We therefore will only give explicit formula which are needed for discussing our
results. 

We first consider the unpolarized differential cross sections
of the $\gamma^* N \rightarrow \pi N$ transition, where $\gamma^*$ denotes
the virtual photon. In the usual convention\cite{nl}, it is defined by
\begin{eqnarray}
\frac{d\sigma}{d\Omega_\pi}
 & = & \frac{d\sigma_T}{d\Omega_\pi} + \epsilon \frac{d\sigma_L}{d\Omega_\pi}
   + \sqrt{2\epsilon(1+\epsilon)} \frac{d\sigma_I}{d\Omega_\pi}\cos\phi_{\pi}
        + \epsilon \frac{d\sigma_P}{d\Omega_\pi} \cos 2\phi_{\pi},
   \label{crse}
\end{eqnarray}
where various differential cross sections depend on the pion scattering angle
$\theta_\pi$, photon momentum-square $q^2=-Q^2=\omega^2- \vec{q}^{\,2}$, 
and the invariant mass $W$ of the final $\pi N$ system.
$\phi_{\pi}$ is the off-plane scattering angle between the $\pi-N$ plane
and the $e-e'$ plane. The dependence of Eq.(26) on the angle $\theta_e$ 
between the outgoing and incoming electrons is in the parameter
 $\epsilon=[1+\frac{2\mid \vec{q}^2\mid}{Q^2}\tan^2\frac{1}{2}\theta_e]^{-1}$.
Recalling Eq.(2.14) of Ref.\cite{nl}, we note here 
that the transverse cross section
 $\sigma_T$ and polarization cross section $\sigma_P$ are only
determined by the transverse currents $J_x$ and $J_y$ and
the longitudinal cross section $\sigma_L$ 
only by the longitudinal current $J_z$. On the other hand,
the interference cross section $\sigma_I$ is determined by the real part of
the product $J_xJ_z^*$.
In general, the contributions from the 
longitudinal current are much weaker than
that from the transverse currents. Thus the longitudinal current 
can be more effectively studied by investigating the
 observables which are sensitive to $\sigma_I$. 
This has been achieved in 
the recent experiments at MIT-Bates and JLab by utilizing the 
$\phi_\pi$-dependence in Eq. (\ref{crse}).

To proceed, we need to define the strength of the charge form factor at
$q^2=0$. First we use  
the low momentum limit to set 
$G_C(0)=\frac{2 m_\Delta}{m_N-m_\Delta} G_E(0)$. With $G_E(0)=+ 0.025$
determined in  section III, we thus have 
$G_C(0)=-0.209$.
Next we consider the JLab $p(e,e'\pi^0)$ data\cite{JLAB} 
at $Q^2=2.8$, 4 (GeV/c)$^2$. 
Since the data are extensive enough, we are able to extract each 
$\phi_\pi-$dependent term in Eq. (\ref{crse}). We then adjust $G_M(Q^2), G_E(Q^2)$
and $G_C(Q^2)$ to fit the data of these extracted components.
Our best fits are the solid curves shown in Fig. 6. We also show that
the interference cross section $d\sigma_I/d\Omega$(which is
determined by $Re(J_xJ_z^*)$) is sensitive to the
charge form factor $G_C$ of the bare $\gamma N \rightarrow\Delta$ vertex. 
In Figs. 7 and 8, we show the comparison with the
original data at several off-plane-angle $\phi_\pi$. Clearly the agreement
is satisfactory. Similar good agreement with the data at different W are
also found. Some typical results are shown in Fig.9 
for W=1115, 1145, 1175, 1205 MeV.

We follow Refs.\cite{nl,ky}
to fit the determined form factor values at $Q^2=0$, 2.8 and 4 (GeV/c)$^2$ 
with the following simple parameterization
\begin{eqnarray}
G_\alpha(Q^2)= G_\alpha(0)G_D(Q^2)R_{\alpha}(Q^2),  \label{frmd}
\end{eqnarray}
with $\alpha= M, E, C$ and
\begin{eqnarray}
G_D(Q^2)_= \left( \frac{1}{1 + Q^2/0.71(GeV/c)^2} \right)^2 
    \label{frmdip}
\end{eqnarray}
is the usual proton form factor.
We find that our results can be fitted by choosing 
\begin{eqnarray}
R_\alpha(Q^2)=(1+aQ^2) \exp(-bQ^2), \label{frme}
\end{eqnarray}
with $a=0.154$ and $b=0.166$ (GeV/c)$^2$ for $\alpha=M$ and $C$.
The unpolarized cross section data are less sensitive to $G_E$.
Nevertheless, the allowed $G_E(Q^2)$ values at  $Q^2=0$, 2.8 and
4 (GeV/c)$^2$ seem to also follow Eq.(\ref{frme}). For simplicity,
we use Eq.(\ref{frme}) for $\alpha=M, E$ and $C$ in all of the 
calculations presented below.
This simple parameterization is similar to that used in Ref.\cite{ky}.
Eq. (\ref{frmd}) then allows us to make predictions for other 
values of $Q^2$.  Note that with $a=0.154$ and $b=0.166$ (GeV/c)$^2$, the
$Q^2$-dependence due to $R_\alpha(Q^2)$ is very small compared with
the dipole form factor $G_D(Q^2)$ in Eq.(\ref{frmdip}). 

To further explore the $\Delta$ excitation 
within our model, we show in Fig. 10 
the $Q^2$-dependence of the predicted $\gamma^* N \rightarrow \pi N$ 
multipole amplitudes $M^{3/2}_{1^+}$, $E^{3/2}_{1^+}$ and $S^{3/2}_{1^+}$ at
 $W=1236$ MeV where the $\pi N$ phase shift in $P_{33}$ channel 
reaches $90^0$. Hence, their real parts are negligibly small and are
omitted in Fig.10.
These amplitudes are proportional to the dressed 
$\gamma N \rightarrow \Delta$ 
transition strengths $A_{M}$, $A_{E}$ and $A_{C}$, as defined by 
Eqs. (\ref{am})-(\ref{ac}).
We also show the results(dotted curves)
from neglecting the non-resonant interaction $v_{\gamma\pi}$ in
 the calculations. 
It is interesting to note from Fig. 10 that the non-resonant
interaction $v_{\gamma\pi}$ enhances strongly 
these amplitudes at low $Q^2$,
but much less at high $Q^2$. 

From the results(solid curves) 
shown in Fig. 10 and Eqs. (\ref{rema}) and (\ref{rsma}), we obtain the
the $Q^2$-dependence of the E2/M1 ratio
\rem and C2/M1 ratio \rsm for the dressed $\gamma N \rightarrow \Delta$
transition. The results are shown in Fig. 11 and Table II.
We see that \rsm drops significantly with $Q^2$ and reaches $\sim 13 \%$
at $Q^2=4$ (GeV/c)$^2$, while \rem remains $\sim - 3\%$ in the
entire considered $Q^2$ region. 
This difference reflects an non-trivial consequence of
our dynamical treatment of the non-resonant interaction, as 
seen in Eq.(9). It will be interesting to test our predictions in the
entire $Q^2$ region. 
Clearly, the  Q$^2 < 4$ (GeV/c)$^2$ region is still far away from
the perturbative QCD region where 
\rem is expected to approach unity \cite{carl}.

The dressed $\gamma N \rightarrow \Delta$ vertex defined by
 Eq. (\ref{vertg})  can be
cast into the form of Eq. (\ref{vertgd}) at the resonance energy 
$W=1236$ MeV. This allows us to extract
the dressed M1 form factor $G^*_M(Q^2)$ from our results and compare it with
the bare form factor $G_M(Q^2)$ of Eq. (\ref{vertgd}). This is shown in Fig. 12
where we measure their $Q^2$-dependence against the proton dipole form
factor $G_D(Q^2)$ defined in Eq.(\ref{frmdip}).  We see that both 
bare and dressed M1 form factors
 drop faster than the  proton form factor. The difference between 
the solid and dotted curves is due to
the non-resonant term(see Eq. (\ref{vertg}))
 which can be interpreted as the effect due to the
pion cloud around the bare quark core. This meson cloud effect
accounts for about 40 $\%$ of the dressed form factor at $Q^2=0$, but
becomes much weaker at high $Q^2$. This implies that the future
data at higher Q$^2$ will be more effective in exploring the
structure of the bare quark core which can be identified with
the current hadron models, as discussed in Ref.\cite{yoshi}. 

With the $\gamma N \rightarrow \Delta$ form factors given by 
Eqs. (\ref{frmd})-(\ref{frme}),
we then can test our model by comparing our predictions 
with the data at other values of $Q^2$. We first
consider the MIT-Bates data\cite{MIT} at $Q^2=0.126$ (GeV/c)$^2$.
In Fig. 13, we show the results(solid curves) for $A_{LT}$ 
which is defined as(see Eq.(26))
\begin{eqnarray}
A_{LT}&=&\frac{\frac{d\sigma}{d\Omega_\pi}(\phi_\pi=180^0) 
-\frac{d\sigma}{d\Omega_\pi}(\phi_\pi=0^0)}
{\frac{d\sigma}{d\Omega_\pi}(\phi_\pi=180^0) 
+\frac{d\sigma}{d\Omega_\pi}(\phi_\pi=0^0)}\nonumber \\
&=&\frac{-\sqrt{2\epsilon(1+\epsilon)}\frac{d\sigma_I}{d\Omega_\pi}}
{\frac{d\sigma_T}{d\Omega_\pi}+\epsilon \frac{d\sigma_L}{d\Omega_\pi}
+\epsilon\frac{d\sigma_P}{d\Omega_\pi}}. \label{alt}
\end{eqnarray}
In the same figure, we also show the results(dotted curves) from setting
$G_E=G_C=0$ for the bare $\gamma N \rightarrow 
\Delta$ vertex(Eq. (\ref{vertgd})). 
The differences between the solid and dotted curves indicate
the accuracy needed to extract these two  quantities of
the bare $\gamma N \rightarrow \Delta$ transition within our model.

Clearly, our predictions are close to the data at $W=1.171$, 1.232 GeV.
The result at $W=$1.292 GeV appears to disagree 
with the data. However our model is expected to
be insufficient at this higher energy,
as already seen in the
results Figs. 3-4 for photoproduction. It is necessary to extend our
model to include additional reaction mechanisms such as the
$\gamma N \rightarrow \pi \Delta \rightarrow \pi N$  transition 
and higher mass nucleon resonances.

In Fig. 14, we compare our results(solid curve) with the data for  
\begin{eqnarray}
R_{TT}=\frac{q_c}{\mid\vec{k}_c\mid}
[\frac{d\sigma_T}{d\Omega}+\epsilon
\frac{d\sigma_L}{d\Omega}]_{\theta_\pi=180^0},  \label{rtt}
\end{eqnarray}
where $k_c$ and $q_c$ are the
momenta for the pion and photon in the $\pi N$ center of mass frame.
In the same figure,
we also show the results(dotted curve)
obtained from neglecting the nonresonant interaction
$v_{\gamma\pi}$ which renormalizes the
$\gamma N \rightarrow \Delta$ vertex and generate non-resonant 
amplitude $t_{\pi\gamma}$, as seen in Eqs. (\ref{tmatt}), (\ref{tmatg})
 and (\ref{vertg}). The importance of the non-resonant interaction
is evident.
Our predictions reproduce the position of the $\Delta$
peak which is shifted significantly by the non-resonant interaction, but
underestimate the magnitude at the peak by about 15 $\%$.

In Fig. 15, we present our results for the 
the induced
proton polarization $P_n$ for $\theta_\pi=180^0$ and the polarization
vector $\vec{n}$ perpendicular to the momentum of 
the recoiled proton.
The data at $W=$1232 MeV is also
from the measurement at MIT-Bates. Our model clearly only
agrees with the data in sign, but not in magnitude. More experimental
data for this observable are needed to test the energy-dependence
of our predictions.

The results shown in Figs. 13-15 are from the calculations using the
$\Delta$ form factors given by Eqs. (\ref{frmd})-(\ref{frme}) and fitted to
the data at $Q^2=$ 0, 2.8, 4.0 (GeV/c)$^2$. The chosen parameterization
is rather arbitrary and there is no strong reason to believe it should
give reliable predictions for the $\Delta$ form factors at the
considered $Q^2=0.126$ (GeV/c)$^2$.
We therefore have explored whether the discrepancies seen in
Figs. 13-15 can be 
removed by adjusting $G_M(Q^2), G_E(Q^2)$ and $G_C(Q^2)$ for
the bare $\gamma N \rightarrow \Delta$ transition. It turns out that 
we are not able to improve our results. It is necessary to also
modify the nonresonant amplitude $t_{\gamma\pi}$. 
We will discuss possible improvements in the next section.

In Fig. 16, we compare our predictions with some of the Bonn 
data\cite{bonn}
at $Q^2= 0.45$, 0.75 GeV/c$^2$. Our predictions are in
good agreement with these data, but these data have large errors. 
The new high-accuracy data\cite{Bur} from JLab in this
$Q^2$ region will give a more critical test of our predictions.
For the forthcoming inclusive $\vec{p}(\vec{e},e')$ data 
from NIKHEF, we also have made the predictions at
Q$^2$=0.11 (GeV/c)$^2$ for
two polarization observables $A_{TT}$ and $A_{TL}$ which are clearly
 defined by Eqs. (2.25b) and (2.25c) of 
Ref.\cite{nl1}. Our predictions(solid curves) are given in Fig. 17.  
The dotted curves are from setting the $\gamma^* N \rightarrow \pi N$
multipole amplitudes $E_{1^+}^{3/2}$ and $S_{1^+}^{3/2}$ to zero. 
This gives an estimate of the required 
experimental accuracy in using  the forthcoming data of
$A_{TT}$ and $A_{TL}$ to extract these
two amplitudes which contain information about $G_E$ and $G_C$ of
the $\gamma N \rightarrow \Delta$ transition.

\section{Summary and Future Developments}
In this work, we have extended the dynamical model developed in
Ref.\cite{sl} to investigate th pion electroproduction reactions.
The model is first refined at the $Q^2=0$ photon point by taking into account
the recent pion photoproduction data from Mainz\cite{MAINZ}.
It is found that the extracted 
M1 strength $G_M(0)=1.85$ and E2 strength $G_E(0)=0.025$ 
of the bare $\gamma N \rightarrow \Delta$ vertex are 
identical to that determined in Ref.\cite{sl}. By using the long wave
length limit, we then obtain $G_C(0)=-2.09$ for the charge
form factor of the bare $\gamma N \rightarrow \Delta$ transition.

For the investigation of pion electroproduction,
we follow the previous work to define the
form factor at each photon vertex in 
the non-resonant interaction $v_{\gamma\pi}$
illustrated in Fig. 2. At each value of $Q^2$, 
the $\gamma N \rightarrow \Delta$ transition strengths $G_M(Q^2)$,
$G_E(Q^2)$ and $G_C(Q^2)$ are the only free parameters 
in our calculations. 
We find that the recent $p(e,e'\pi^0)$
data at $Q^2=$ 2.8, 4 (GeV/c)$^2$ from JLab\cite{JLAB}, 
and at $Q^2=0.126$ (GeV/c)$^2$ from MIT-Bates\cite{MIT} 
can be described to a very large extent
if the bare $\gamma N \rightarrow \Delta$ 
form factors defined by Eqs. (\ref{frmd})-(\ref{frme}) are used in the
calculations. It is found that the remaining discrepancies can not
be resolved by only adjusting these $\Delta$ form factors. 

We focus on the investigation of the $Q^2$-dependence of
the $\Delta$ excitation mechanism. It is found that the non-resonant
interactions can
 dress the $\gamma N \rightarrow \Delta$ vertex
to enhance strongly its strength at low $Q^2$, but
much less at high $Q^2$(Fig. 10). 
The determined C2/M1 ratio (\rsm) drops significantly with $Q^2$ and
reaches $\sim 13 \%$ at $Q^2=4$ (GeV/c)$^2$, while the E2/M1 ratio
(\rem) remains at $\sim -3 \%$ of 
the value at the $Q^2=0$ photon point(Fig. 11 and table II).
The determined M1 form factor drops
faster than the usual dipole form factor of the proton(Fig. 12).
This is in agreement with the previous findings\cite{JLAB,ky}.

To end, we turn to discussing possible future developments 
within our formulation.
The model we developed in Ref.\cite{sl} and applied in this work is
defined by the effective Hamiltonian Eq. (\ref{hamile}). It is derived from
using a unitary transformation method up to  
second order in the vertex interaction
$H_I$(Eq. (\ref{hamili})). We further assume that the $\pi N$ and $\gamma N$ reactions 
can be described within a subspace $ \Delta \oplus\pi N \oplus \gamma N$.
From the point of view of the general Hamiltonian 
Eqs. (\ref{hamilt})-(\ref{hamili}), which
can be identified\cite{yoshi} with 
a hadron model, our model is clearly just a starting model. 
For this reason, no attempt is
made here to adjust the parameters of the non-resonant interaction
 $v_{\gamma\pi}$ to perform a
$\chi^2-$fit to the electroproduction data.
To resolve the remaining  discrepancies between our predictions and the data, 
it is necessary to improve our model in several directions. 

To improve our results in the higher energy region(Figs. 3-4,13), we need
to include the coupling with other reaction channels. 
An obvious step is to 
extend the effective Hamiltonian Eq. (\ref{hamile})
to include the transitions to $\eta N$, $\pi \Delta$
and $\rho N$ states and to include some higher mass $N^*$ states. The resulting
scattering equations will be more complex than what are given in
Ref.\cite{sl} and outlined in 
Eqs. (\ref{tmatt})-(\ref{self}). In particular, it will have the
$\pi\pi N$ cut structure due to the $\Delta\rightarrow \pi N$ decay in the
$\pi \Delta$ channel and the $\rho \rightarrow \pi\pi$ decay in the $\rho N$
channel. This $\pi\pi N$ cut must be treated exactly in any attempt to
explore the structure of $N^*$ resonances. This was well recognized
in the early investigations\cite{aaron} and must be pursued in a dynamical 
approach. 
 
The second necessary improvement is to use more realistic form factors
in defining the photon vertices of non-resonant interaction $v_{\gamma\pi}$.
The prescription Eq. (\ref{formv}) must be relaxed.
In particular we should use
a form factor $F_{\gamma\pi\pi}(Q^2)$ predicted from 
a calculation which accounts for the  off-mass-shell properties of the
exchange pion in Fig. 2.
Similarly, the vector meson form factor
 $g_{V\pi\gamma}$(Eq. (\ref{coupv})) must also
be improved since the exchanged vector meson is  also off its mass-shell.
The improvement in this direction could be possible in the near future,
since the calculations for such off-mass-shell form factors can now be 
performed within some QCD models of light mesons\cite{tandy}.

This work was supported by U.S. DOE Nuclear Physics Division, Contract 
No. W-31-109-ENG and by Japan Society for the Promotion of Science,
 Grant-in-Aid for Scientific Research (C) 12640273.

\newpage

\begin{table}[htb]
\caption[]{Helicity amplitude $A_{3/2}$ and E2/M1 ratio \rem for
the $\gamma N \rightarrow \Delta$ transition at $Q^2=0$ photon point.
 $A_{3/2}$ is in unit of $10^{-3} GeV^{-1/2}$ and \rem in \%. 
The references are:
a(this work), b(\cite{ky}), c(\cite{muko99}),
 d(\cite{hdt98}), e(\cite{capst}), f(\cite{bil}), g(\cite{bhf}).}
\begin{tabular}{lccccl}
\hline
              & \multicolumn{2}{c}{$A_{3/2}$}  
              & \multicolumn{2}{c}{$R_{EM}$}      & Refs.\\
                &  Dressed  & Bare & Dressed  & Bare & \\ \hline
Dynamical Model & -228      & -153 & -2.7     & -1.3 & a\\
                & -256      & -136 & -2.4     & 0.25 & b\\
K-Matrix        & -255      &  $-$    & -2.1     &  $-$    & c\\
Dispersion      & -252      &   $-$   & -2.5     &  $-$    & d\\
Quark Model     &  $-$      & -186 &   $-$       & $\sim 0$ & e \\
                &  $-$      & -157 &    $-$     & $\sim 0$ & f\\
                &  $-$      & -182 &   $-$       & -3.5   & g\\
                \hline
\end{tabular}
\end{table}

 \newpage
\begin{table}[htb]
\caption{ The $Q^2$-dependence of the E2/ M1 ratio \rem and C2/M1 ratio
\rsm for the dressed $\gamma N \rightarrow \Delta$ transition calculated
 from this work.}
\begin{tabular}{lccccccc}
\hline
  $Q^2 (GeV/c)^2$   & 0     & 0.1   & 1    &    2 & 3     & 4  \\ \hline
\rem     (\%)       & -2.7  & -3.2  & -2.2 & -1.9 & -2.0 & -2.3 \\
\rsm     (\%)       & $-$   & -4.0  & -6.7 & -8.9 & -11  & -13
 \\ \hline
\end{tabular}
\end{table}

\newpage

\begin{figure}[h]
 \centerline{\epsfig{file=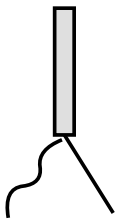,width=1.5cm}}
\vspace*{0.5cm}
 \caption{Graphical representation of the 
$\gamma N \leftrightarrow \Delta$
interaction.}
\end{figure}

\begin{figure}[h]
 \centerline{\epsfig{file=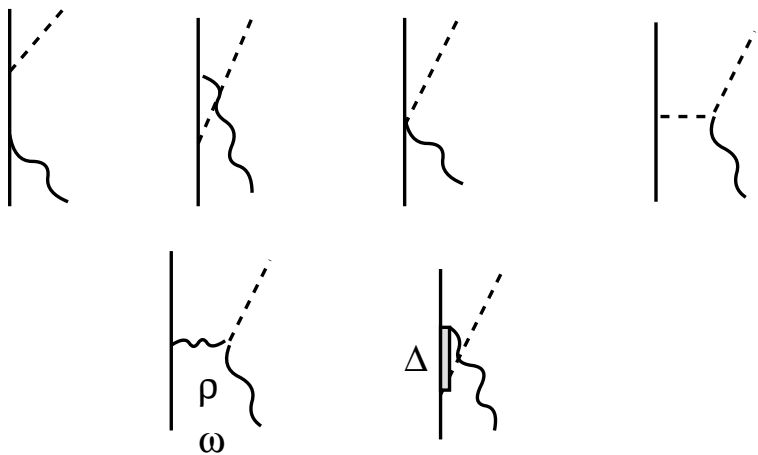,width=10cm}}
\vspace*{0.5cm}
 \caption[]{Graphical representation of the non-resonant 
interaction of $v_{\gamma \pi}$ of Eq. (\ref{hamile}). The 
waved, dashed, and solid lines denote the
photon, pion, and nucleon respectively.}
\end{figure}

\newpage

\begin{figure}[h]
 \centerline{\epsfig{file=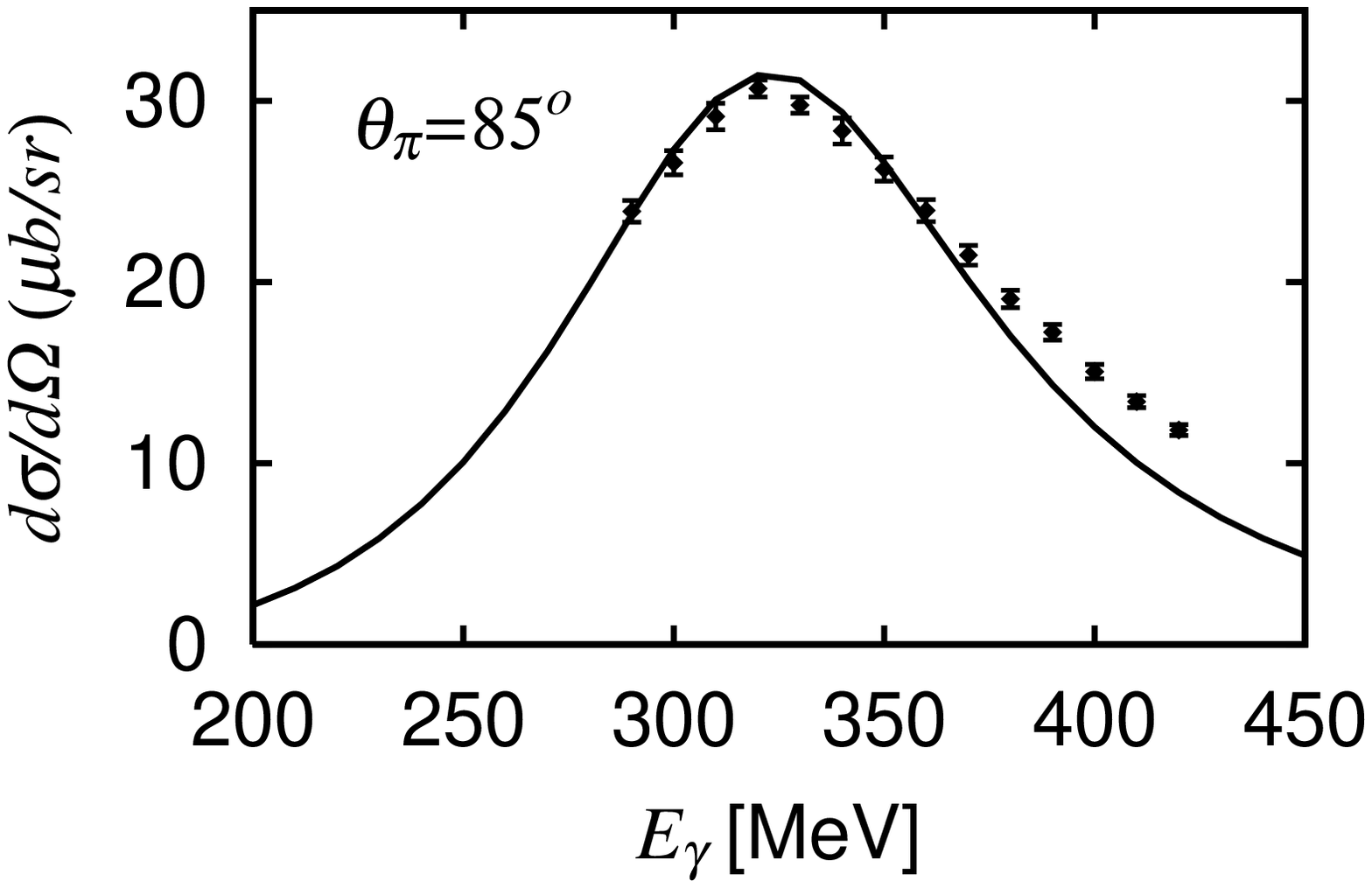,width=7.5cm}
             \epsfig{file=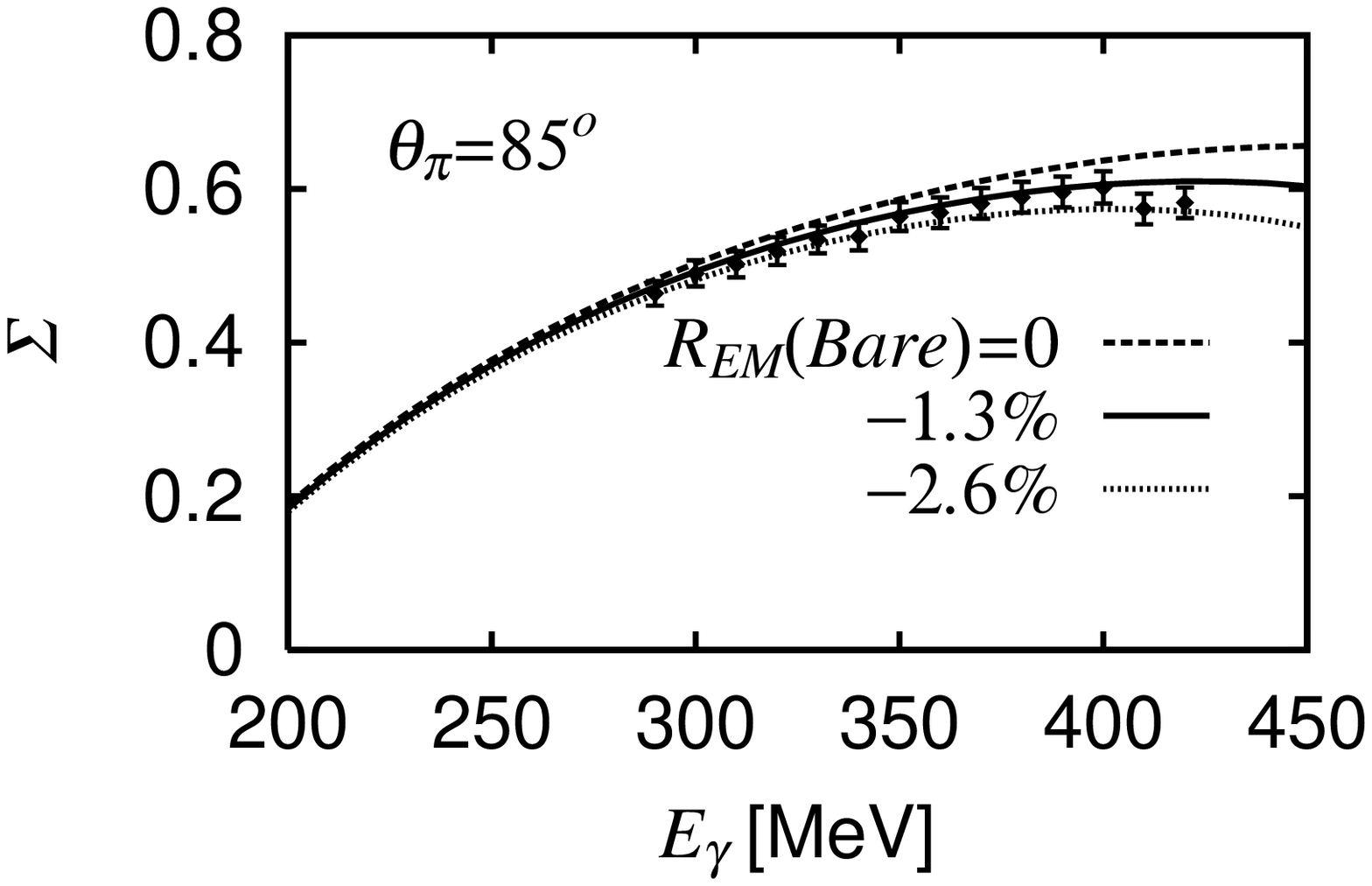,width=7.5cm}}
 \centerline{\epsfig{file=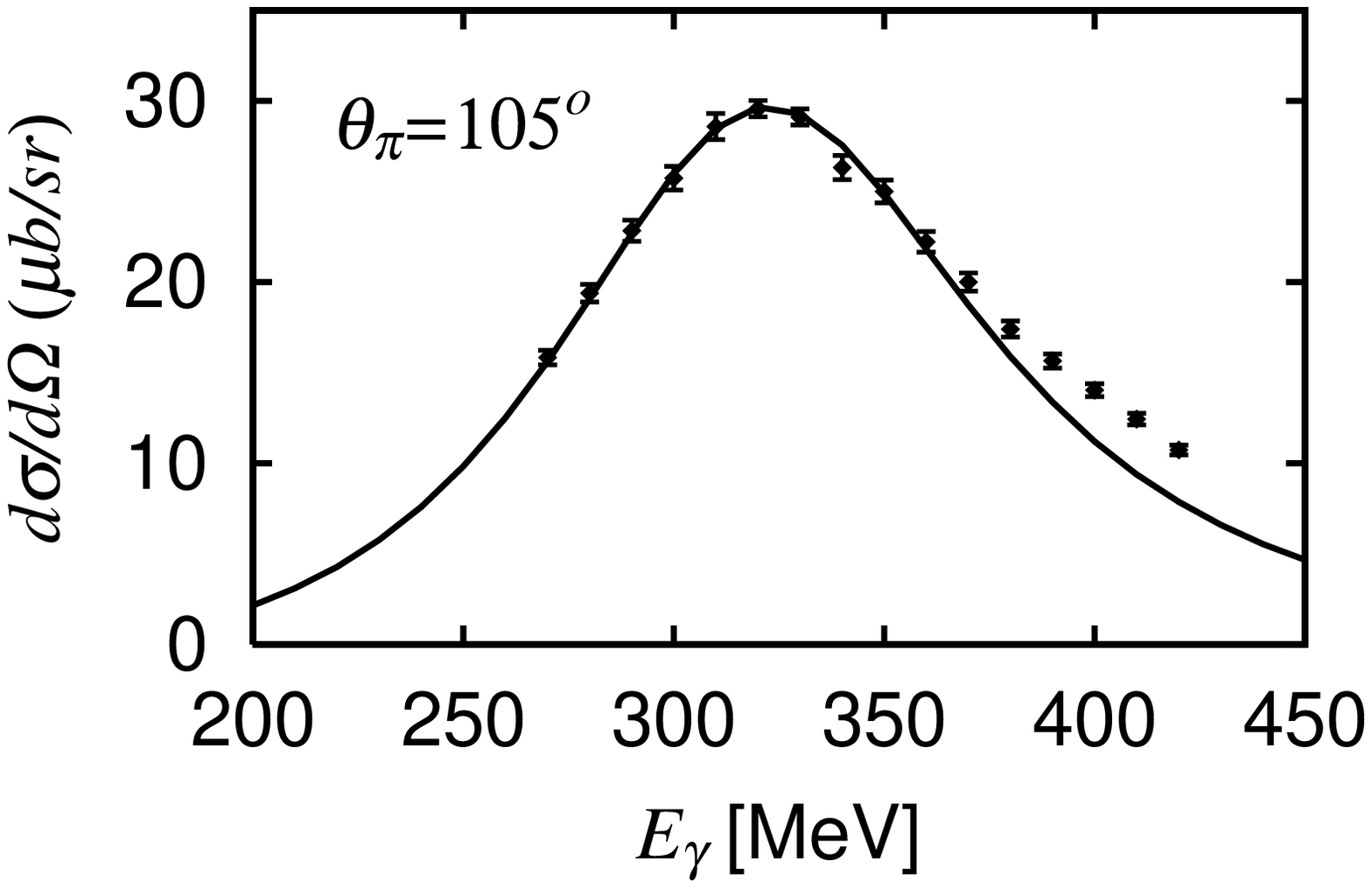,width=7.5cm}
             \epsfig{file=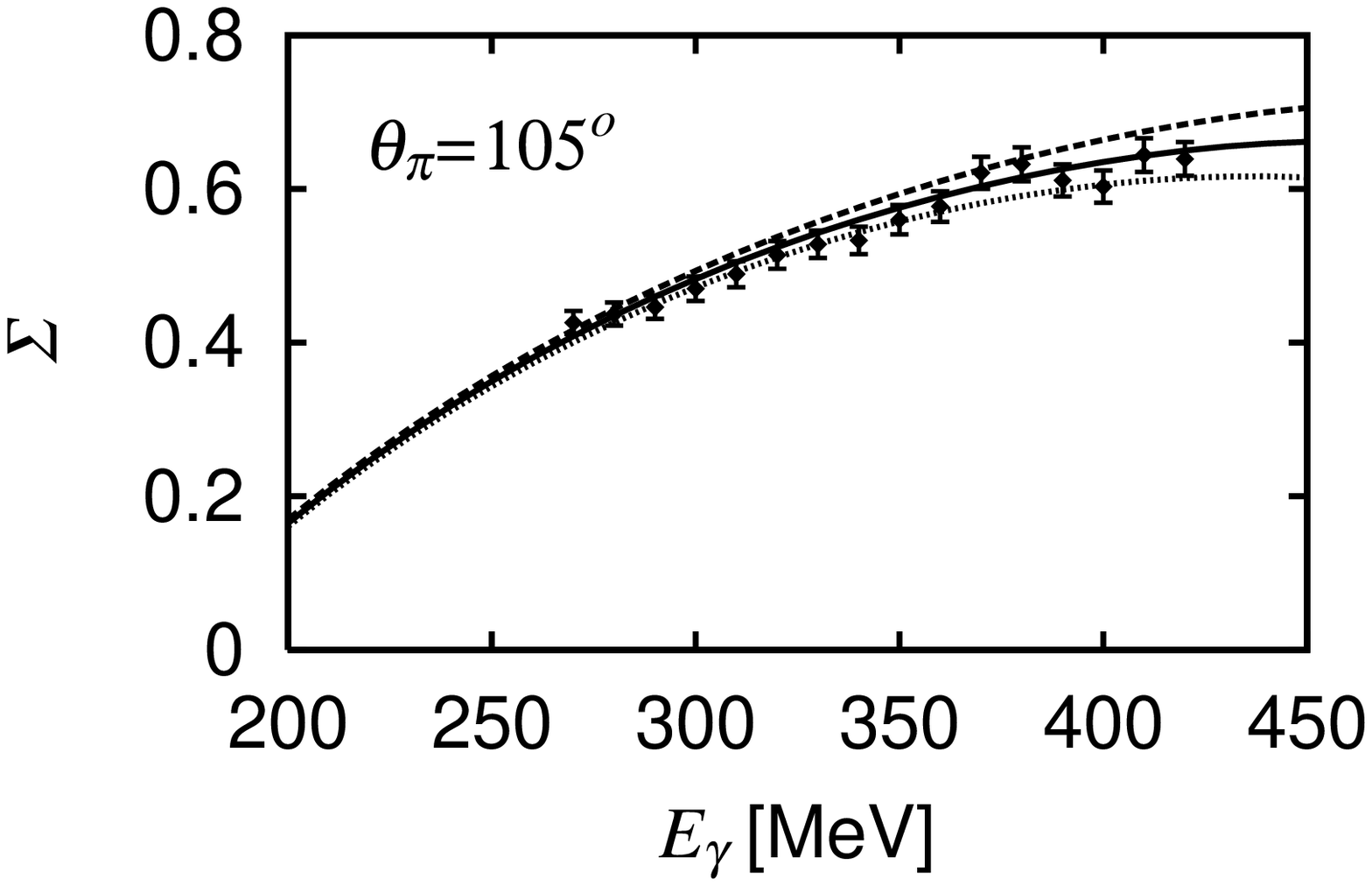,width=7.5cm}}
 \centerline{\epsfig{file=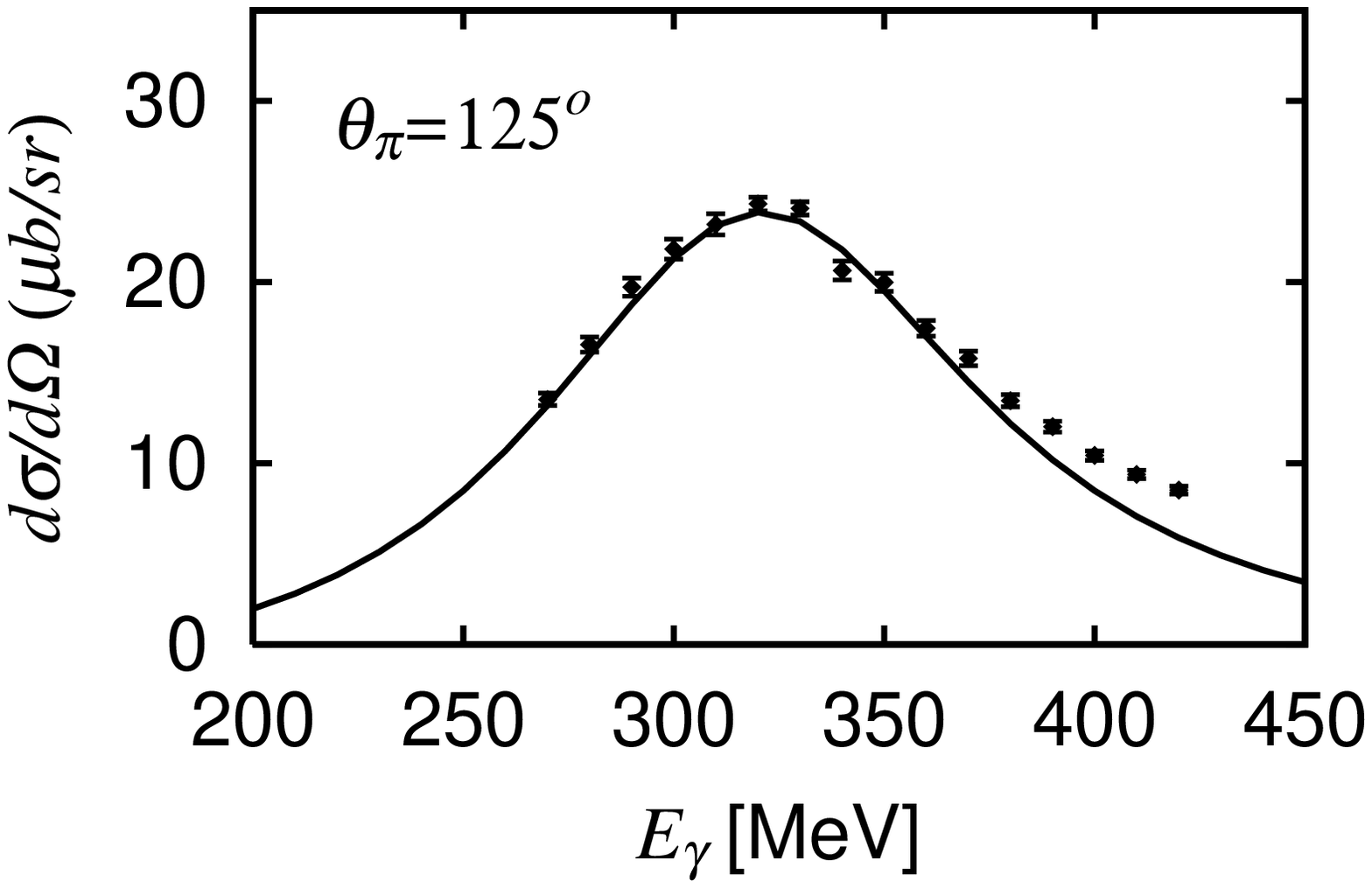,width=7.5cm}
             \epsfig{file=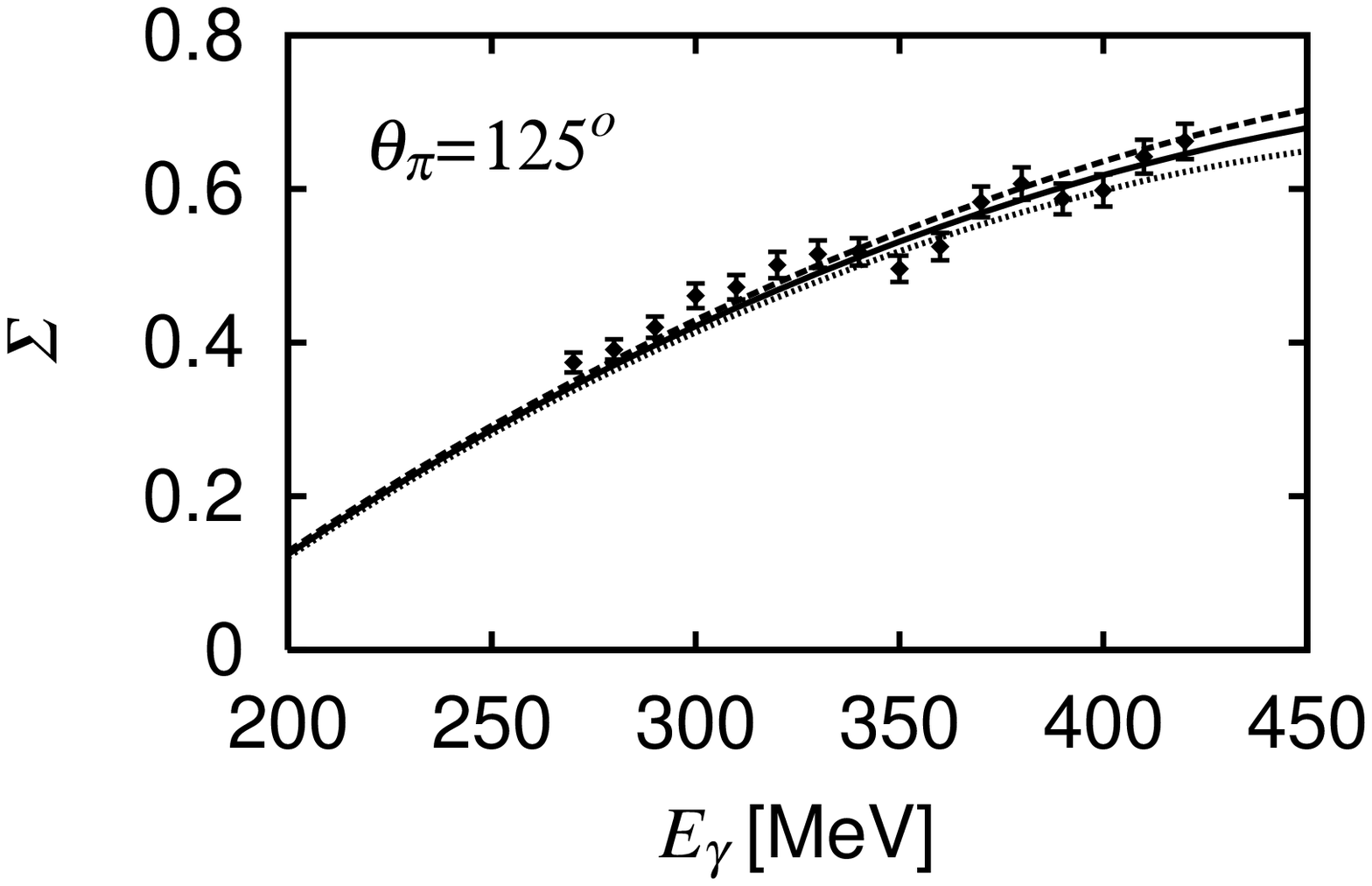,width=7.5cm}}
\vspace*{0.5cm}
 \caption[]{The calculated differential cross section($d\sigma/d\Omega$) and 
           photon asymmetry($\Sigma$) of the $p(\gamma,\pi^0)p$ reaction 
          are compared with the  Mainz data \cite{MAINZ}. The results
          from setting $R_{EM}(bare)=\frac{G_E(0)}{G_M(0)}$ =0, -1.3, and -2.6
          $\%$ with $G_M(0)=1.85$ 
          for the bare $\gamma N \rightarrow \Delta$ vertex are
          indicated in the figure. The three results for
          the differential cross section are not distinguishable.}
\end{figure}

\newpage
\begin{figure}[h]
 \centerline{\epsfig{file=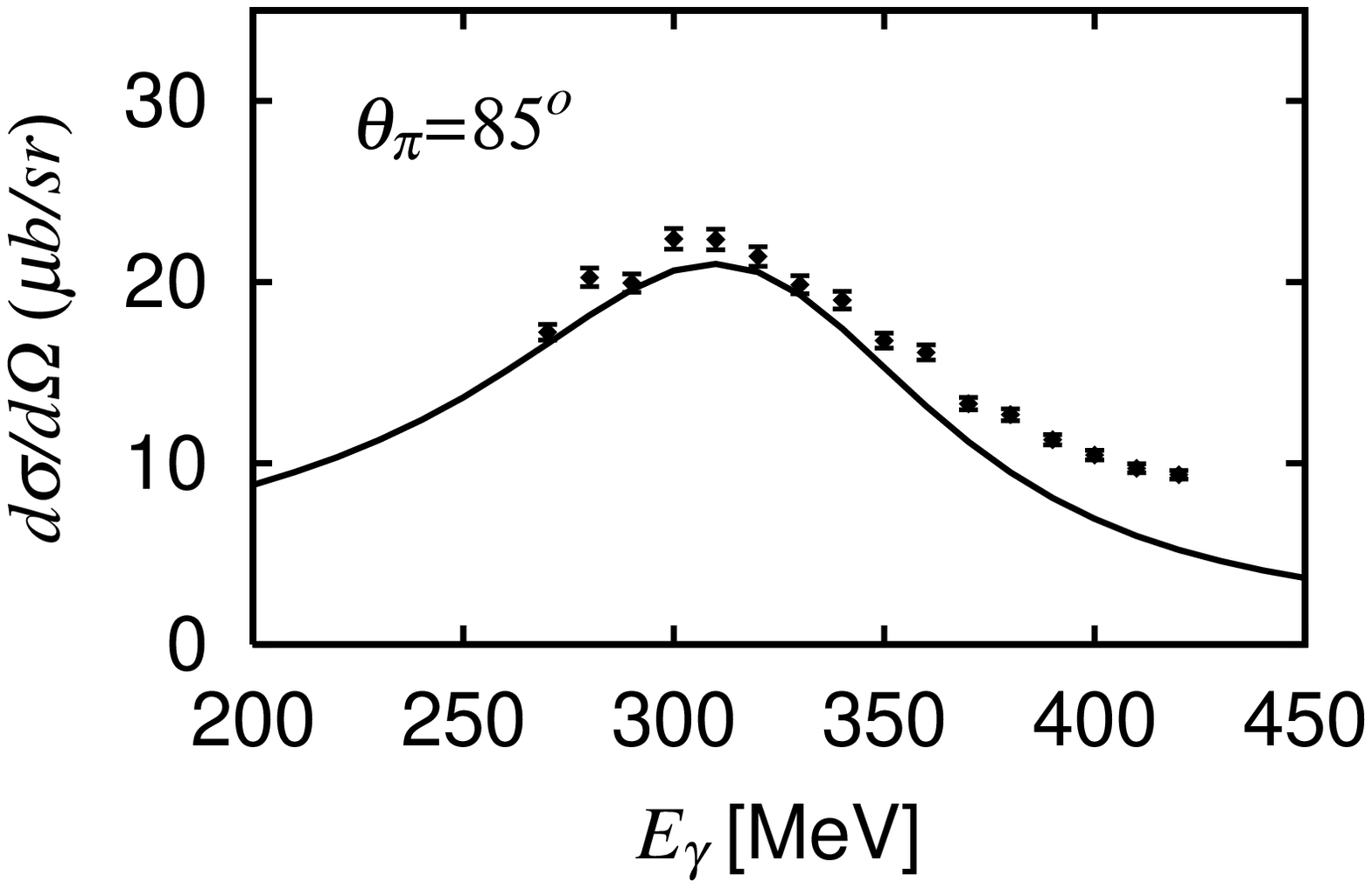,width=7.5cm}
             \epsfig{file=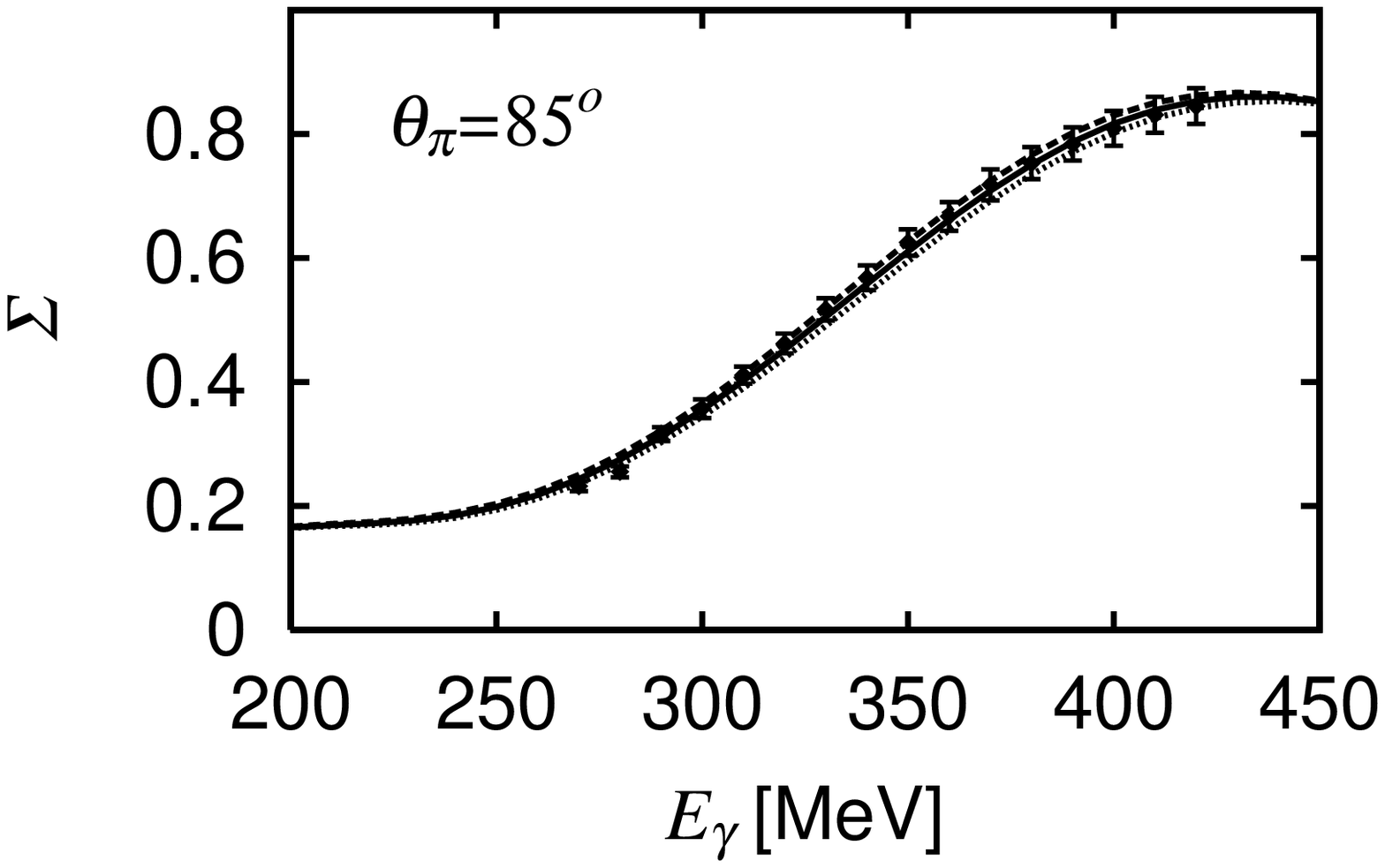,width=7.5cm}}
 \centerline{\epsfig{file=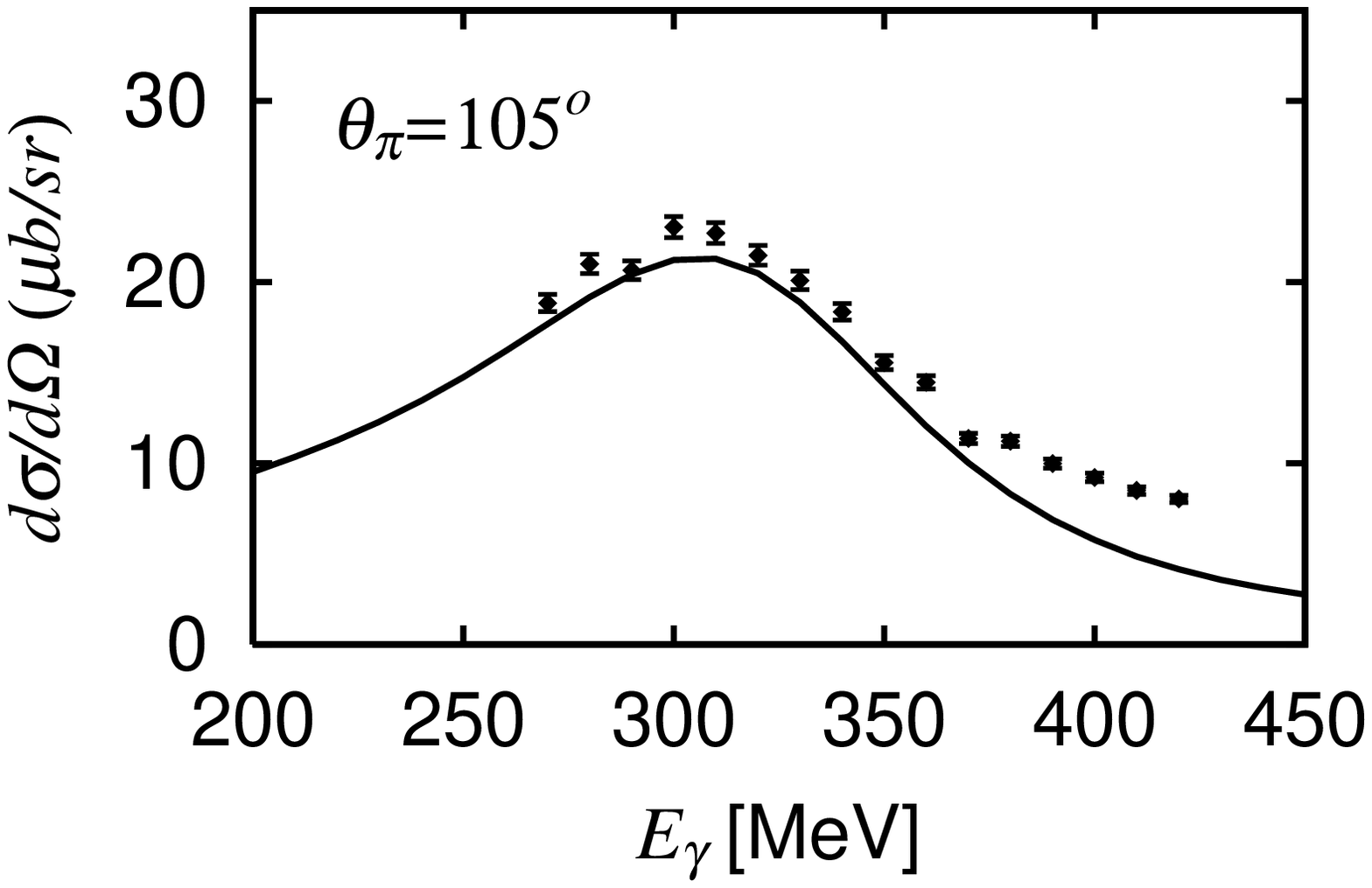,width=7.5cm}
             \epsfig{file=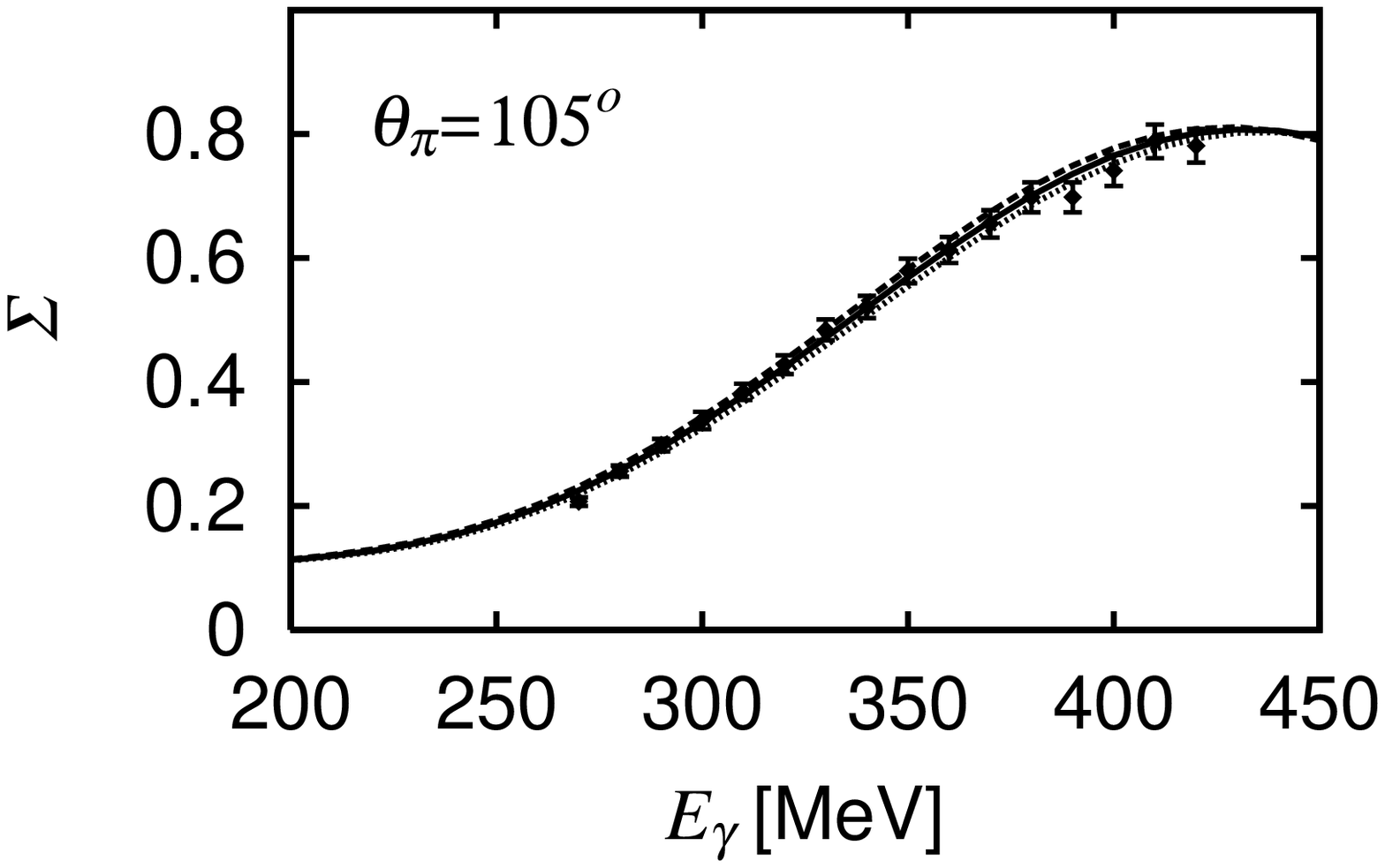,width=7.5cm}}
 \centerline{\epsfig{file=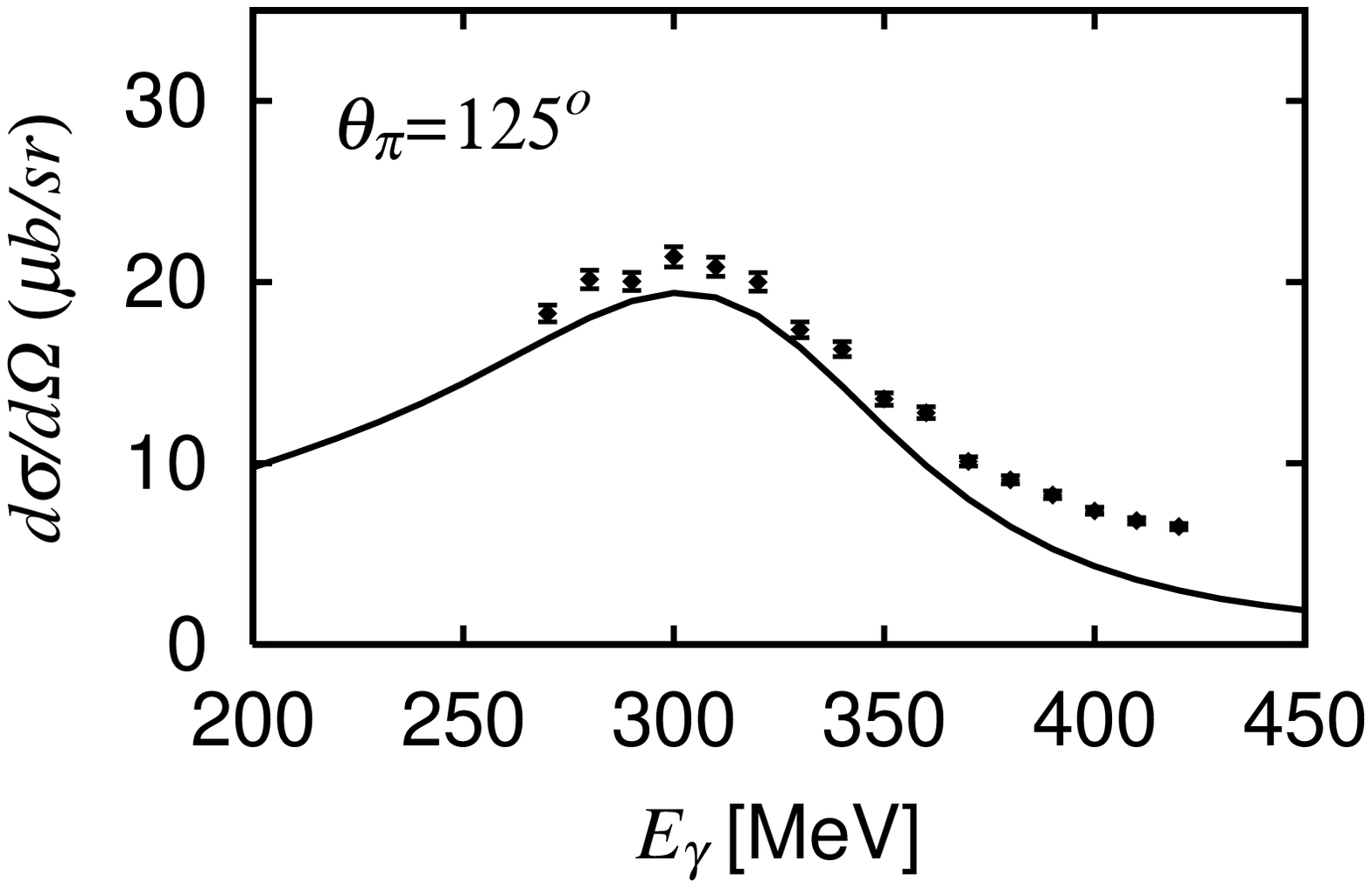,width=7.5cm}
             \epsfig{file=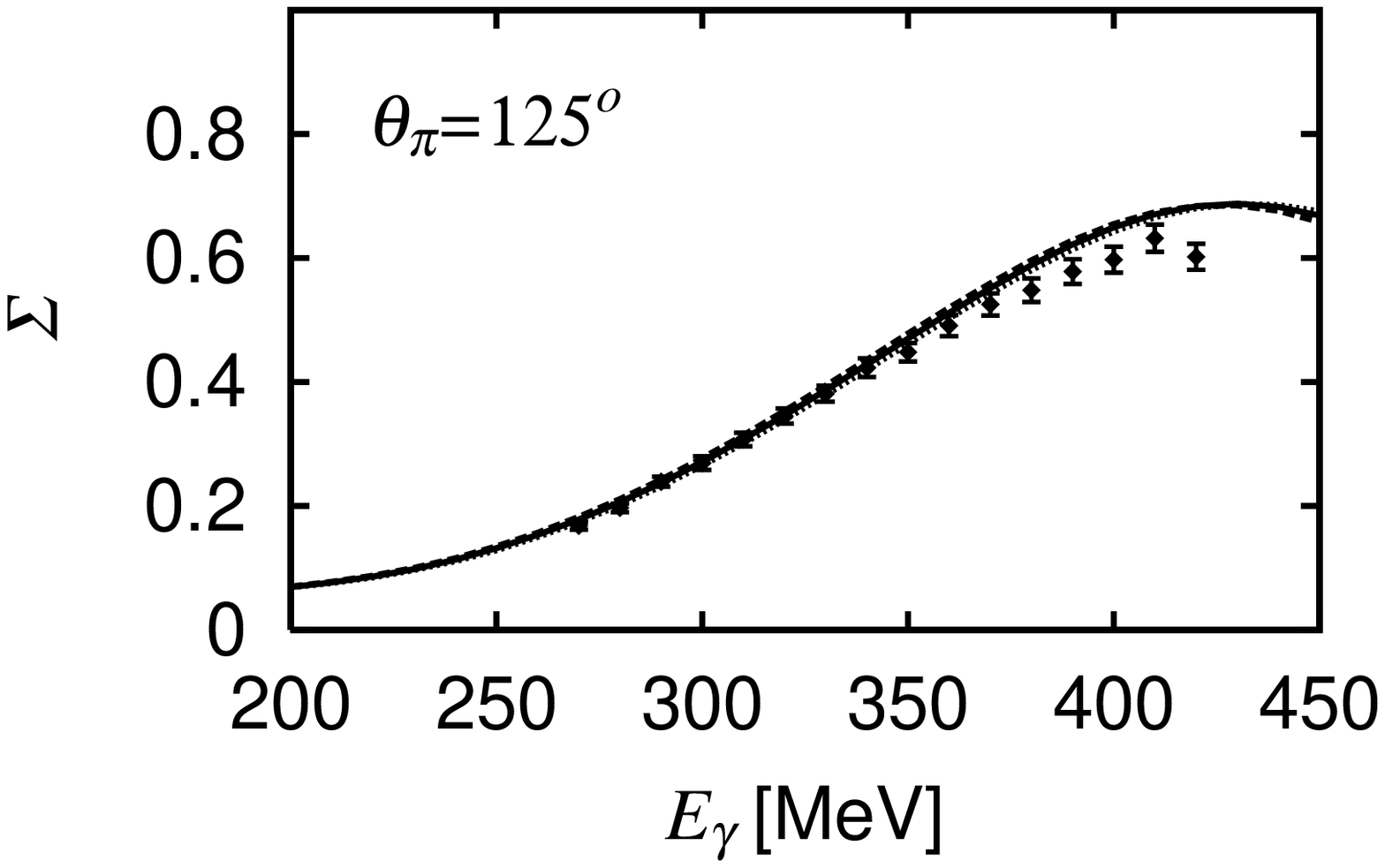,width=7.5cm}}
\vspace*{0.5cm}
 \caption[]{Same as Fig. 3 except for 
          the $p(\gamma,\pi^+)n$ reaction. Three results are not 
          distinguishable here.} 
\end{figure}

\newpage
\begin{figure}[h]
\centerline{\epsfig{file=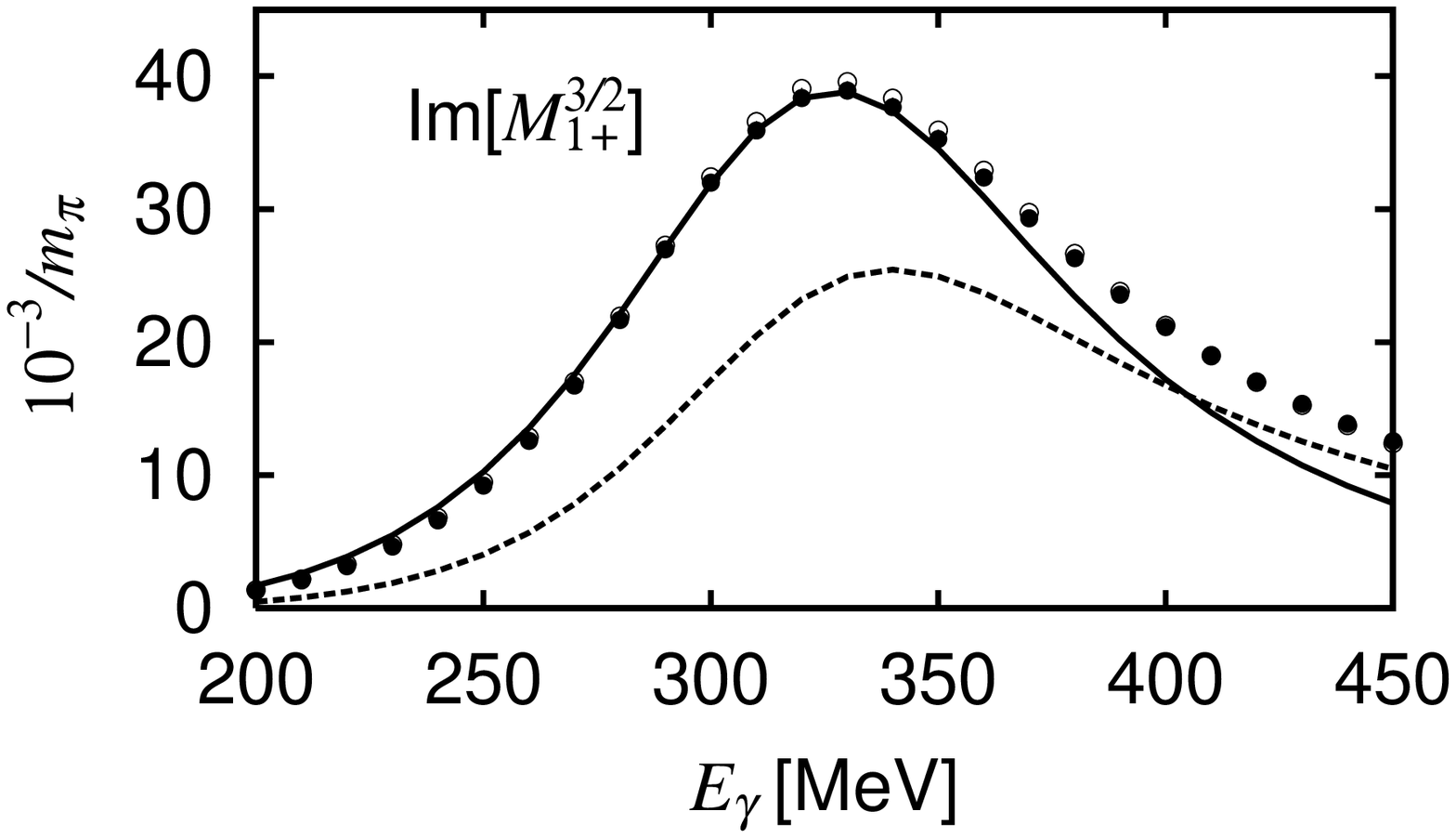,width=7.5cm}
            \epsfig{file=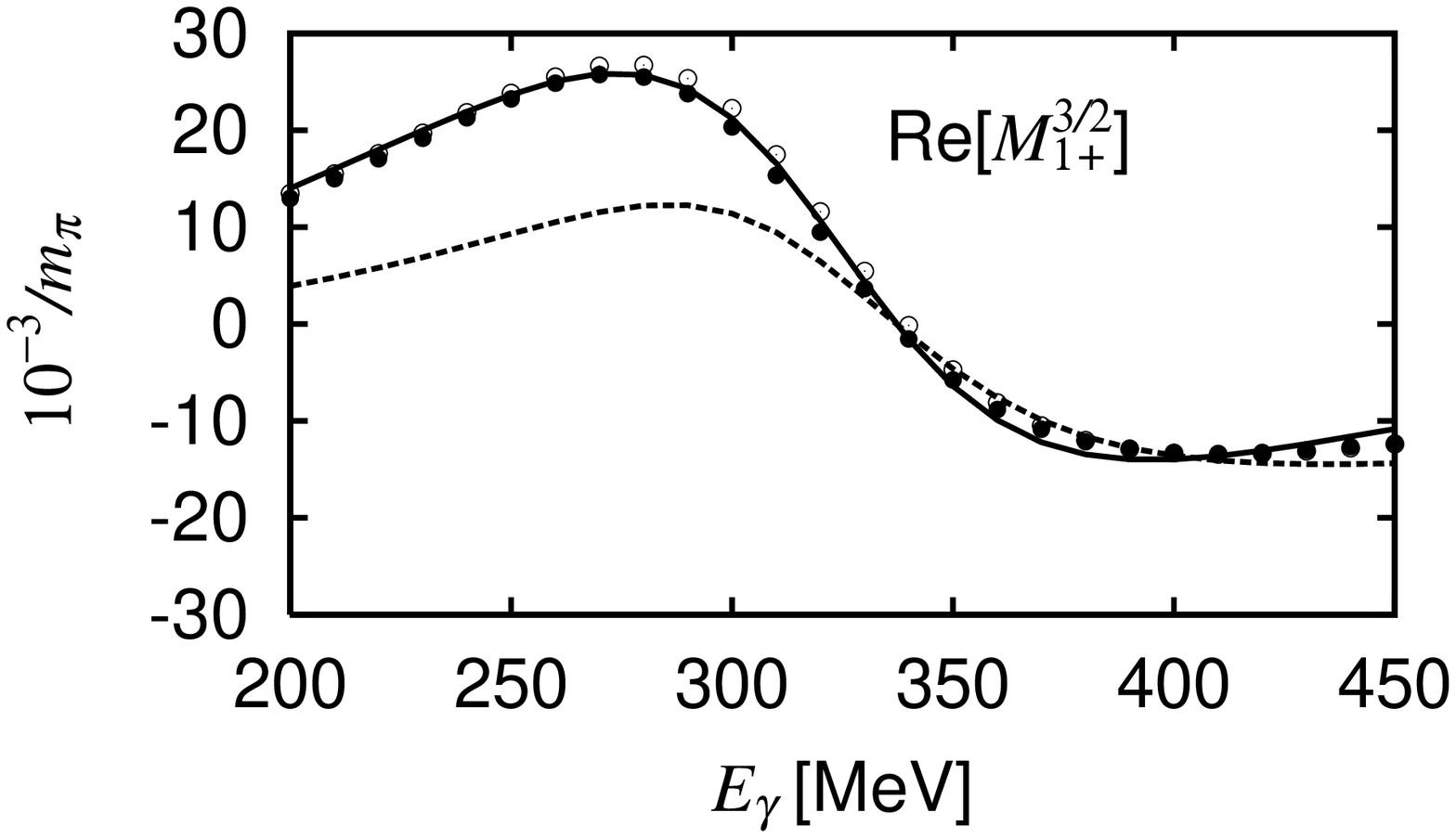,width=7.5cm}}

\centerline{\epsfig{file=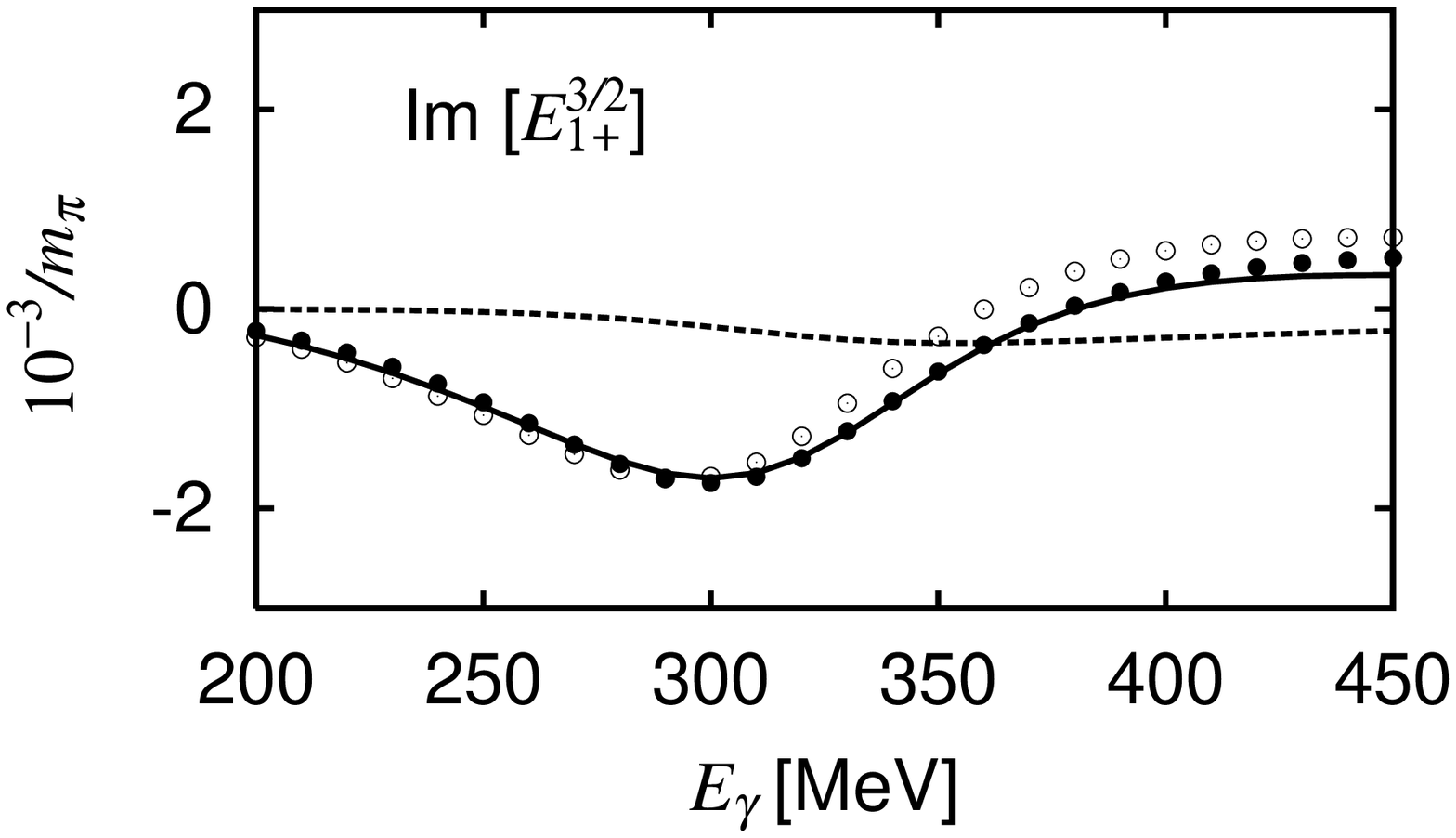,width=7.5cm}
            \epsfig{file=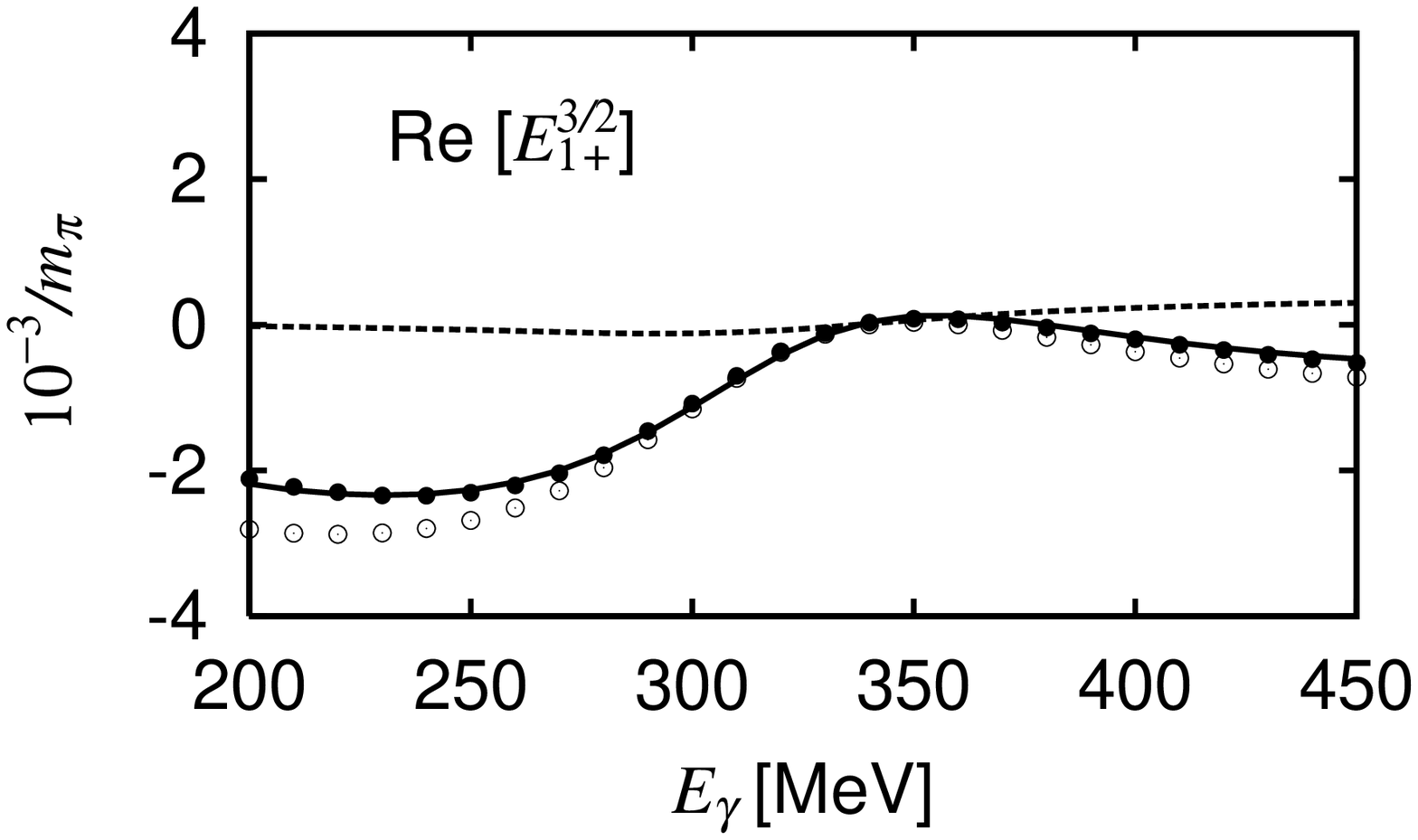,width=7.5cm}}

\vspace*{0.5cm}
\caption[]{The predicted $M_{1+}^{3/2}$ and $E_{1+}^{3/2}$ amplitudes
         for the $\gamma N \rightarrow \pi N$ reaction are compared
         with the results from the empirical amplitude analyses.
         The dotted curves are from the calculations neglecting the
         non-resonant interaction $v_{\gamma\pi}$. See text for more
         detailed description. The open circle data are from SM95 \cite{vpi}
         and solid circle data are from Mainz98 \cite{hdt98}.} 
\end{figure}

\newpage
\begin{figure}[h]
\centerline{
\epsfig{file=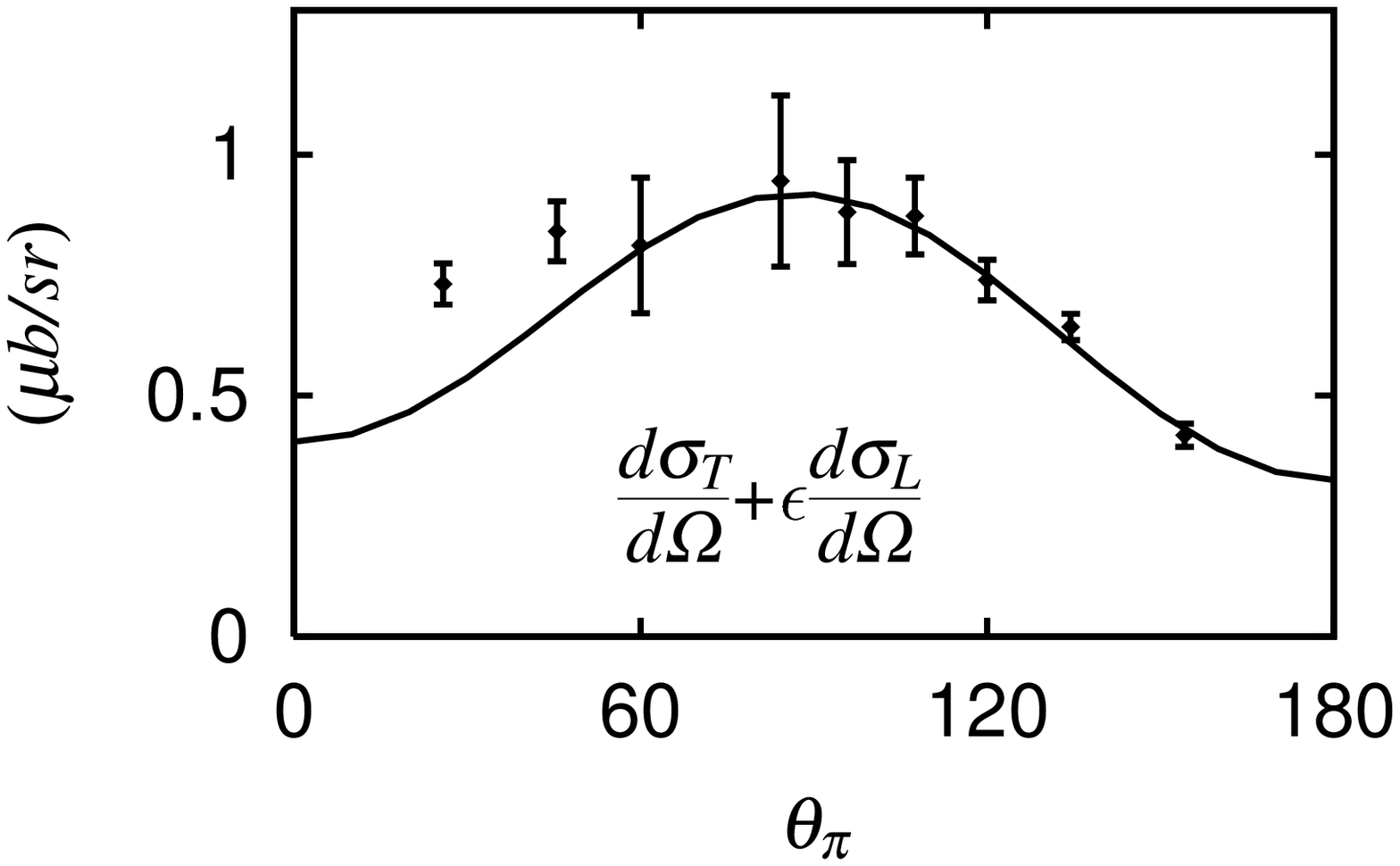,width=5.1cm}
\epsfig{file=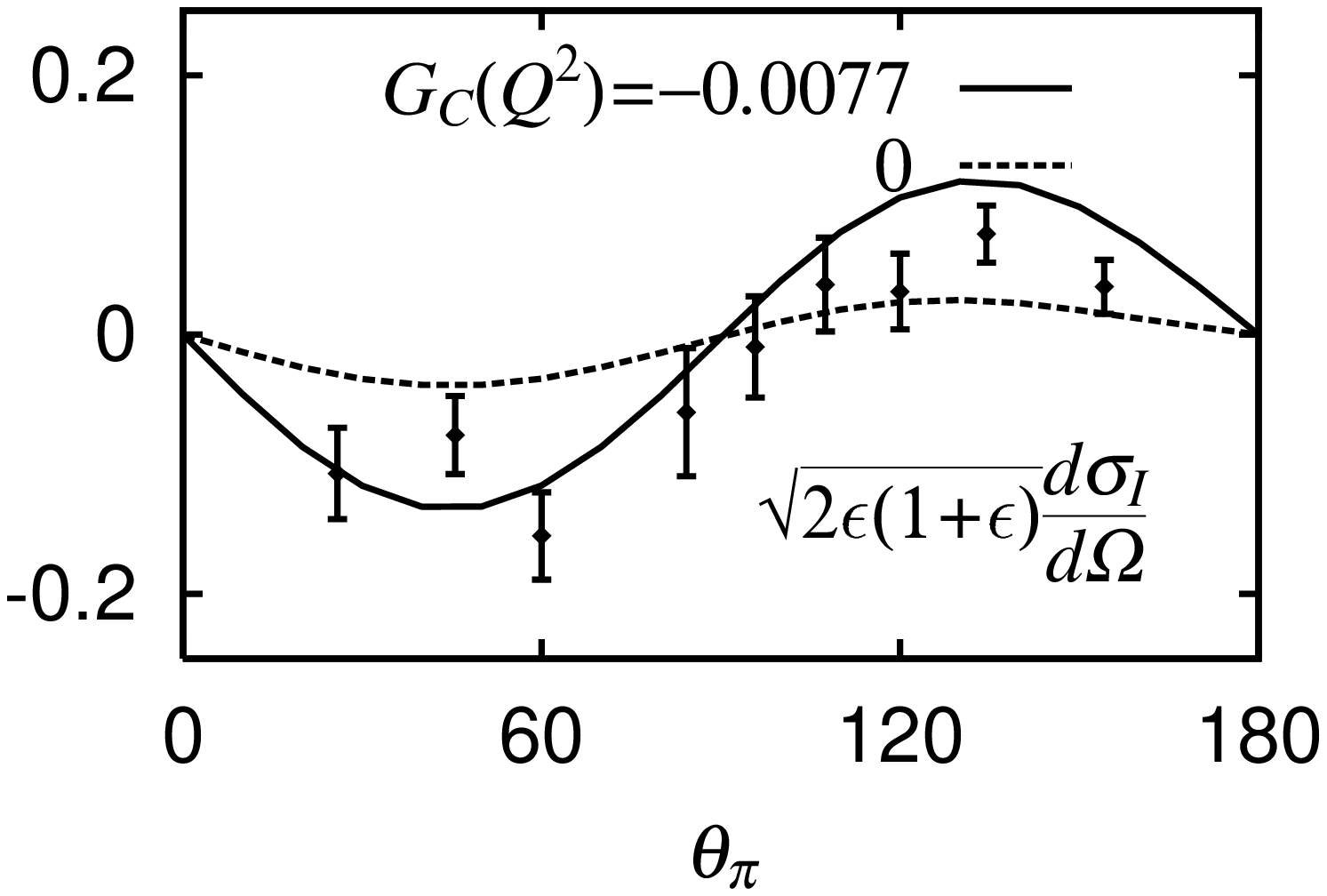,width=5.1cm}
\epsfig{file=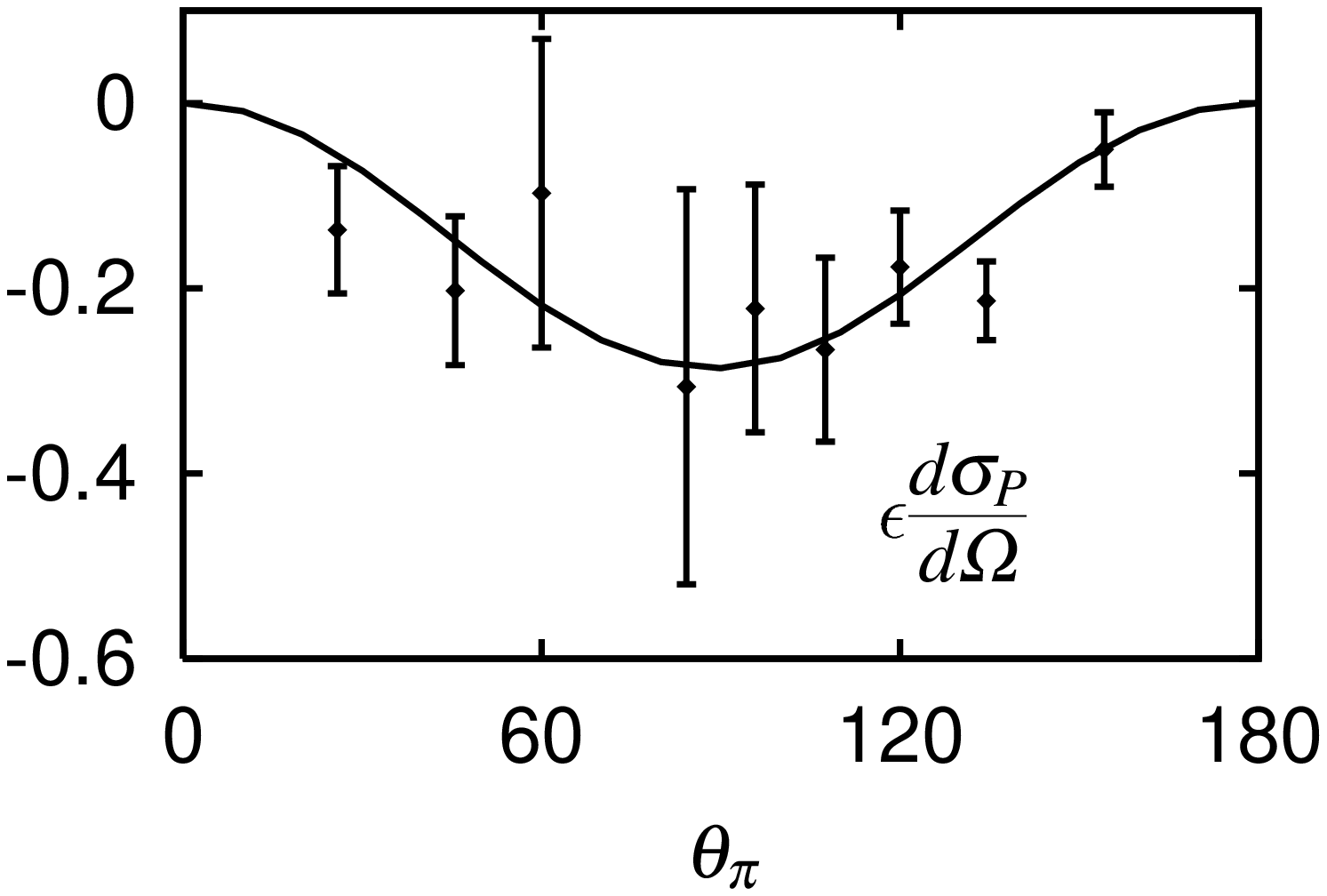,width=5.1cm}
}
\centerline{
\epsfig{file=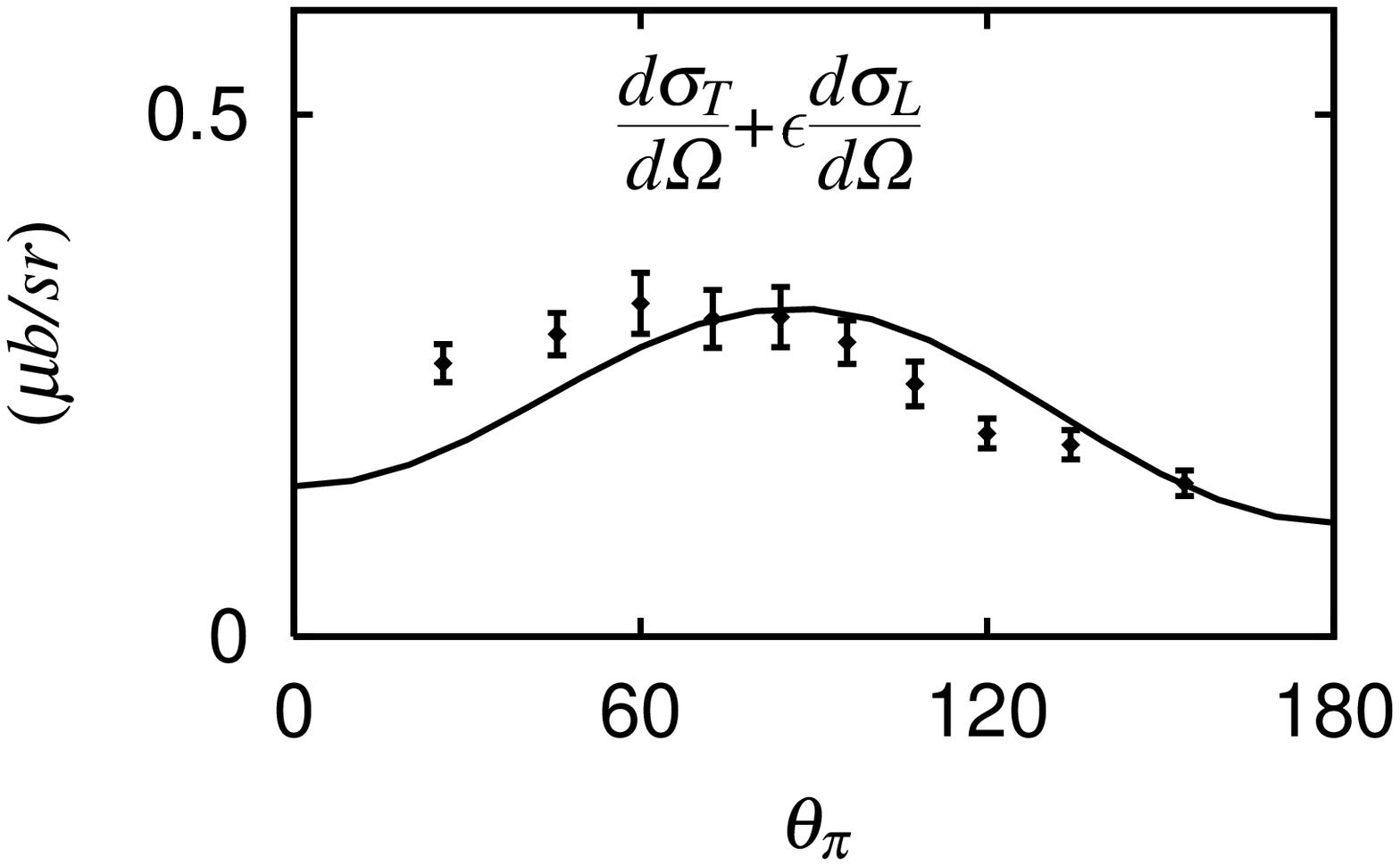,width=5.1cm}
\epsfig{file=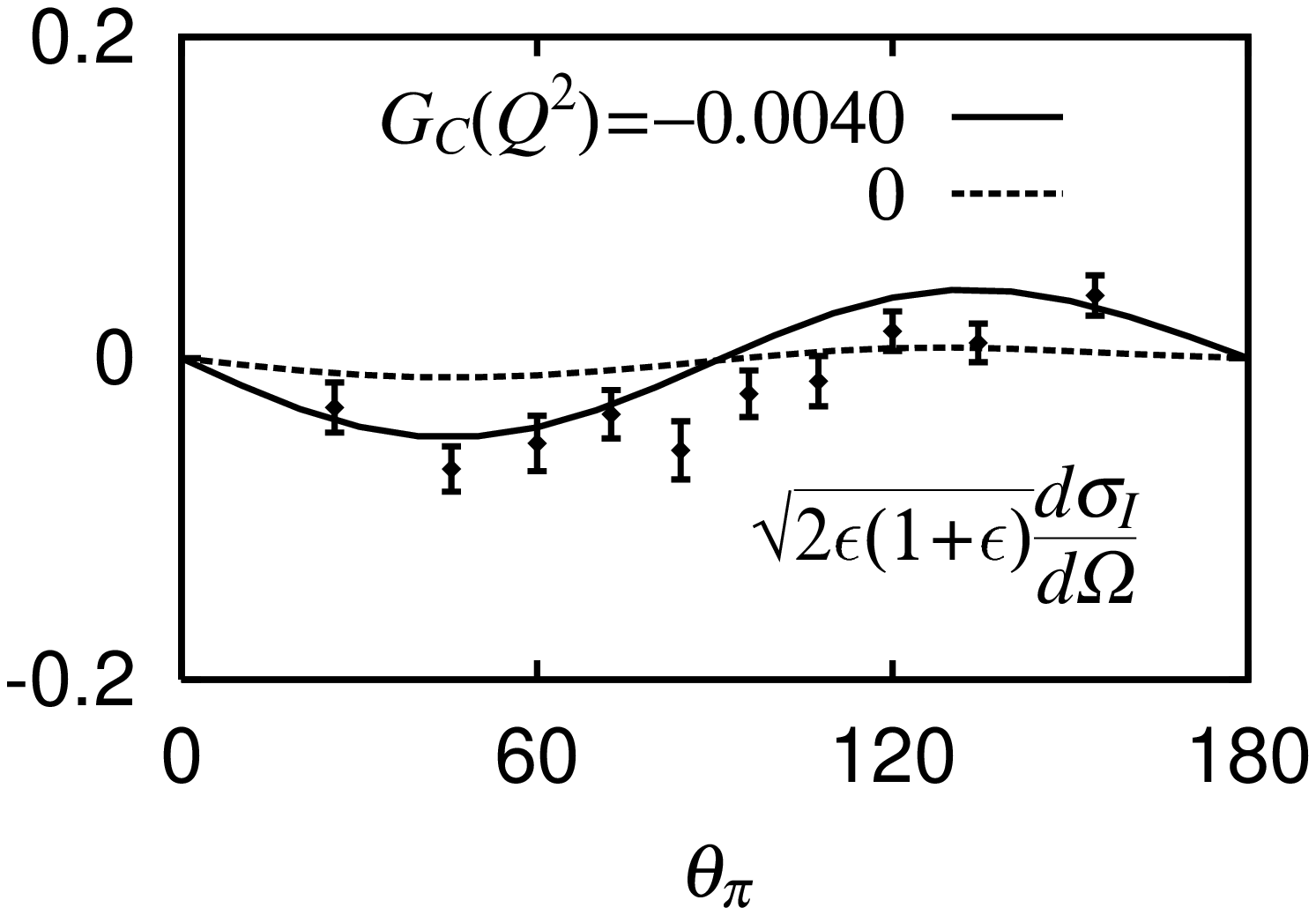,width=5.1cm}
\epsfig{file=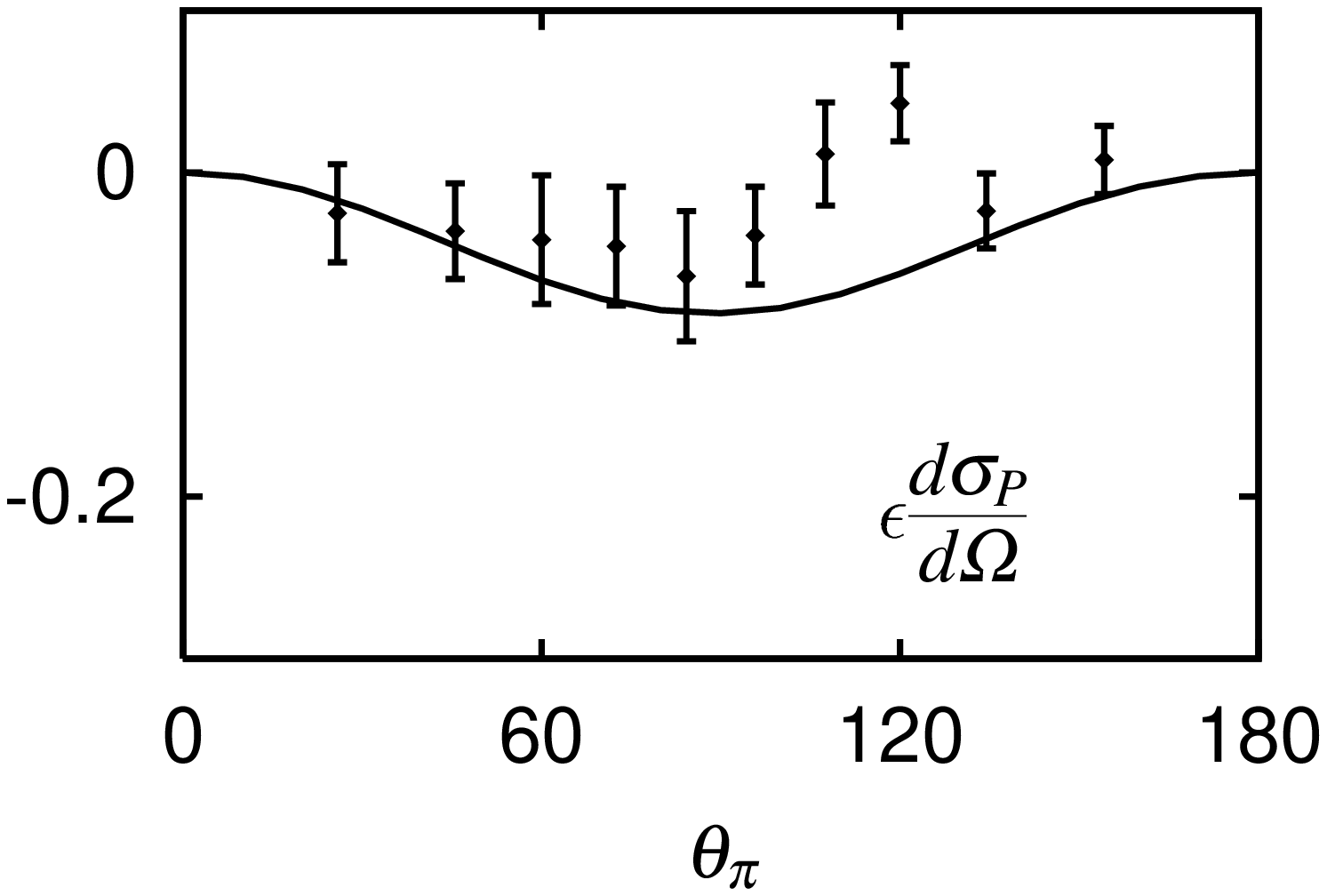,width=5.1cm}
}

\vspace*{0.5cm}
\caption[]{Three components of the 
         calculated $p(e,e'\pi^0)$ differential cross section(Eq. (\ref{crse}))
         at $Q^2=2.8$(upper row), $4.0$(lower row ) (GeV/c)$^2$
         and $W=1235$ MeV are compared with the data which are extracted from
         the JLab data \cite{JLAB}(some of
         these original data are shown in Figs. 7-9 ).}
\end{figure}

\newpage
\begin{figure}
\centerline{
\epsfig{file=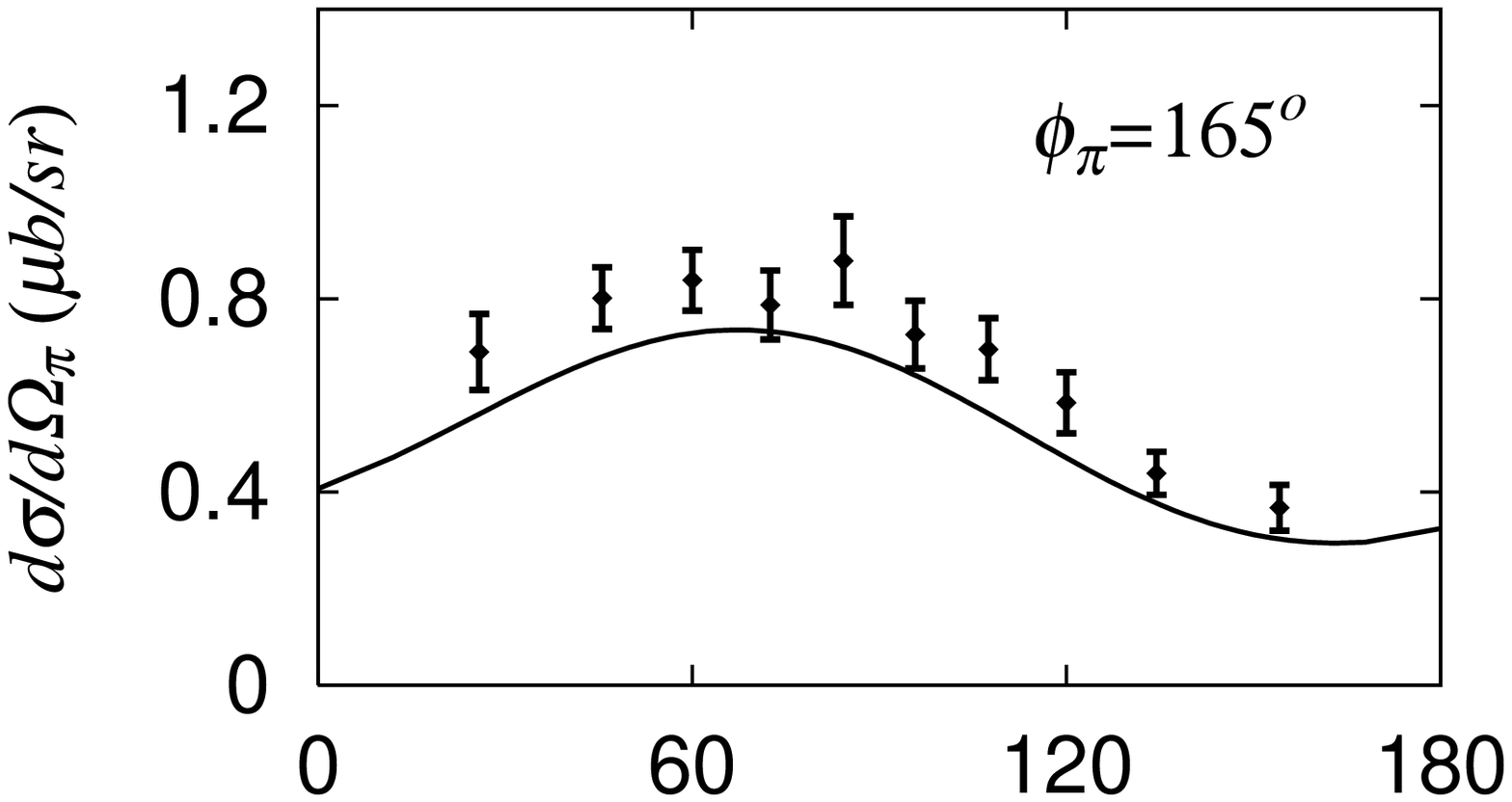,width=7cm}
\epsfig{file=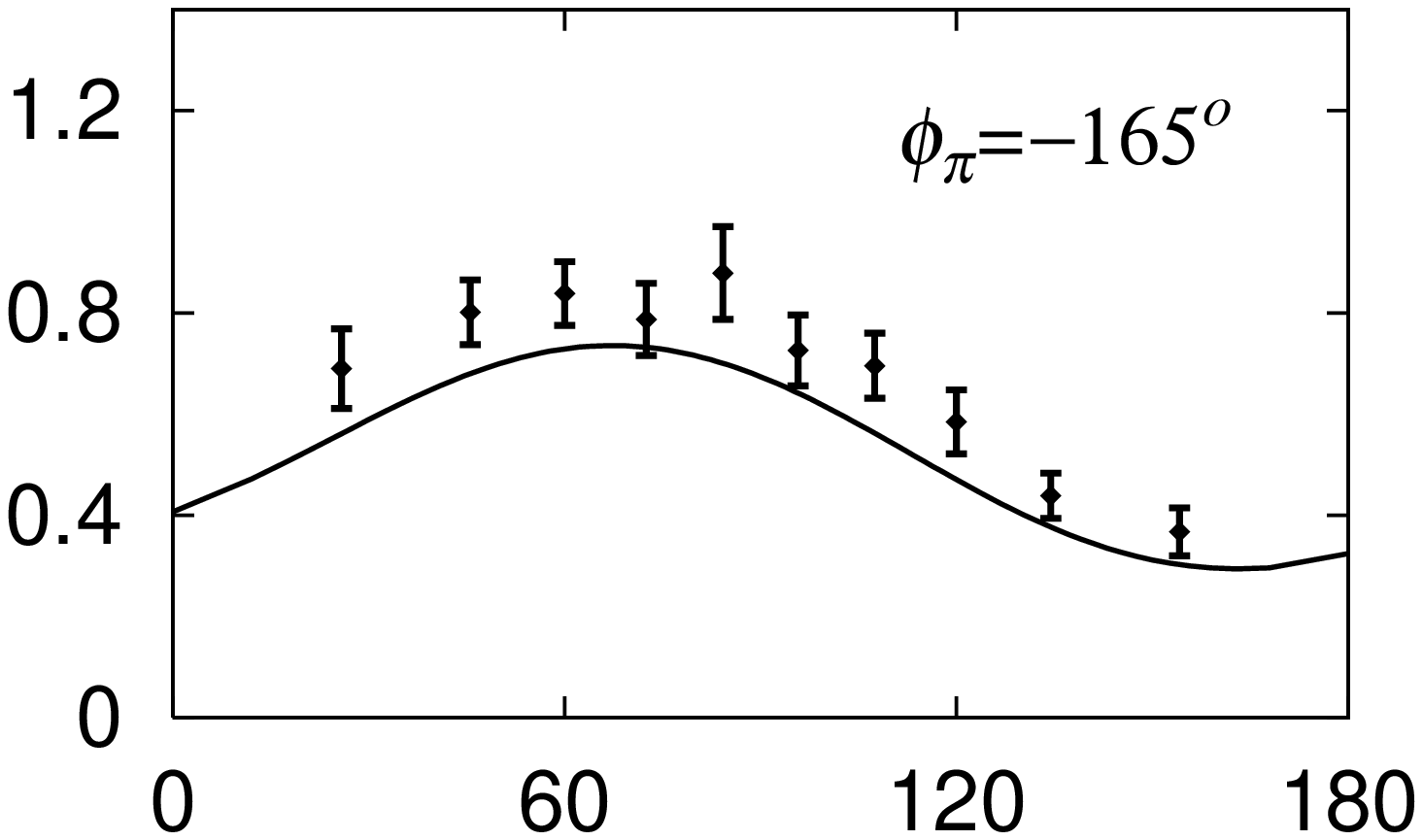,width=7cm}
           }
\vspace*{-0.5cm}
\centerline{
\epsfig{file=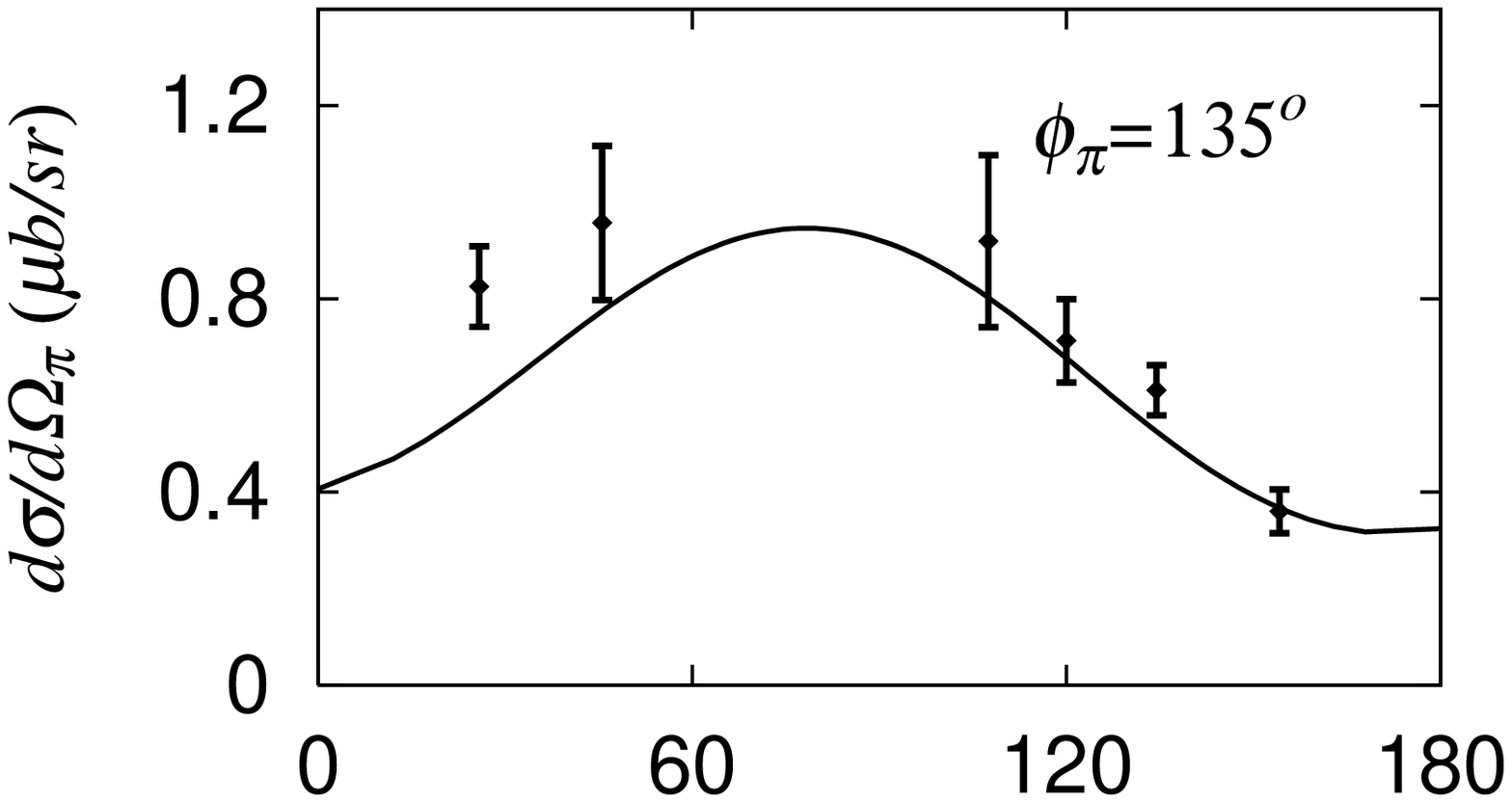,width=7cm}
\epsfig{file=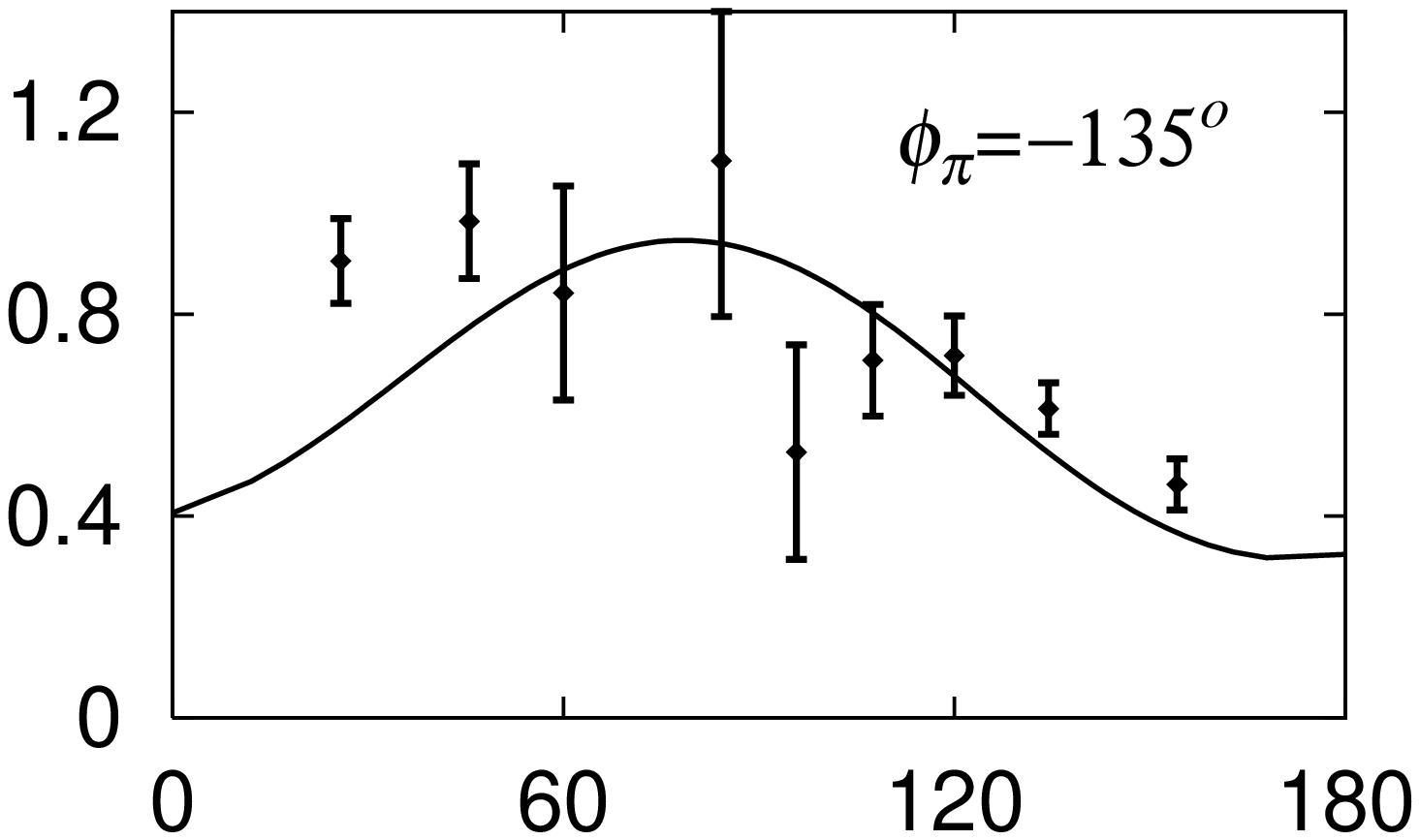,width=7cm}
           }
\vspace*{-0.5cm}
\centerline{
\epsfig{file=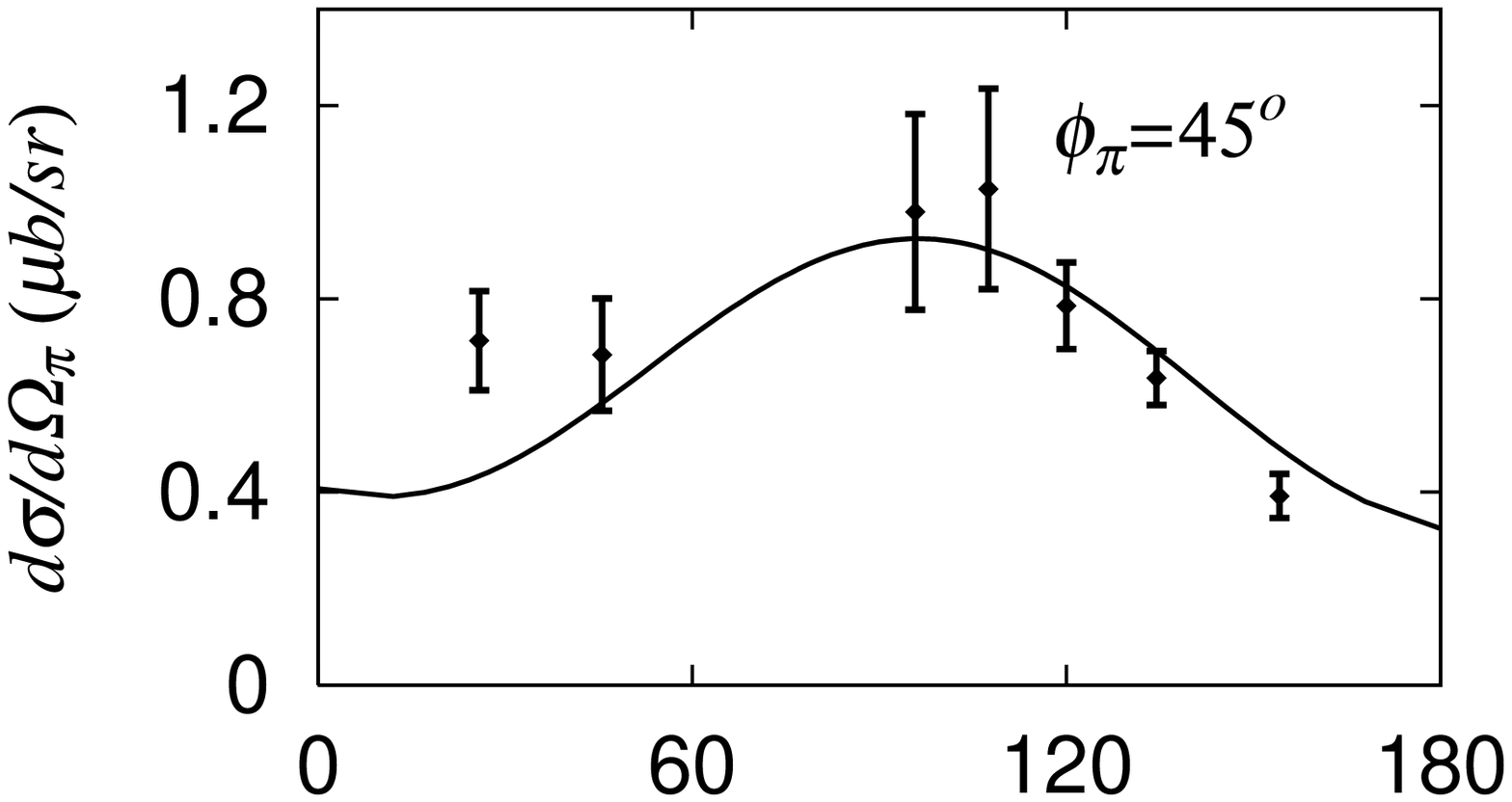,width=7cm}
\epsfig{file=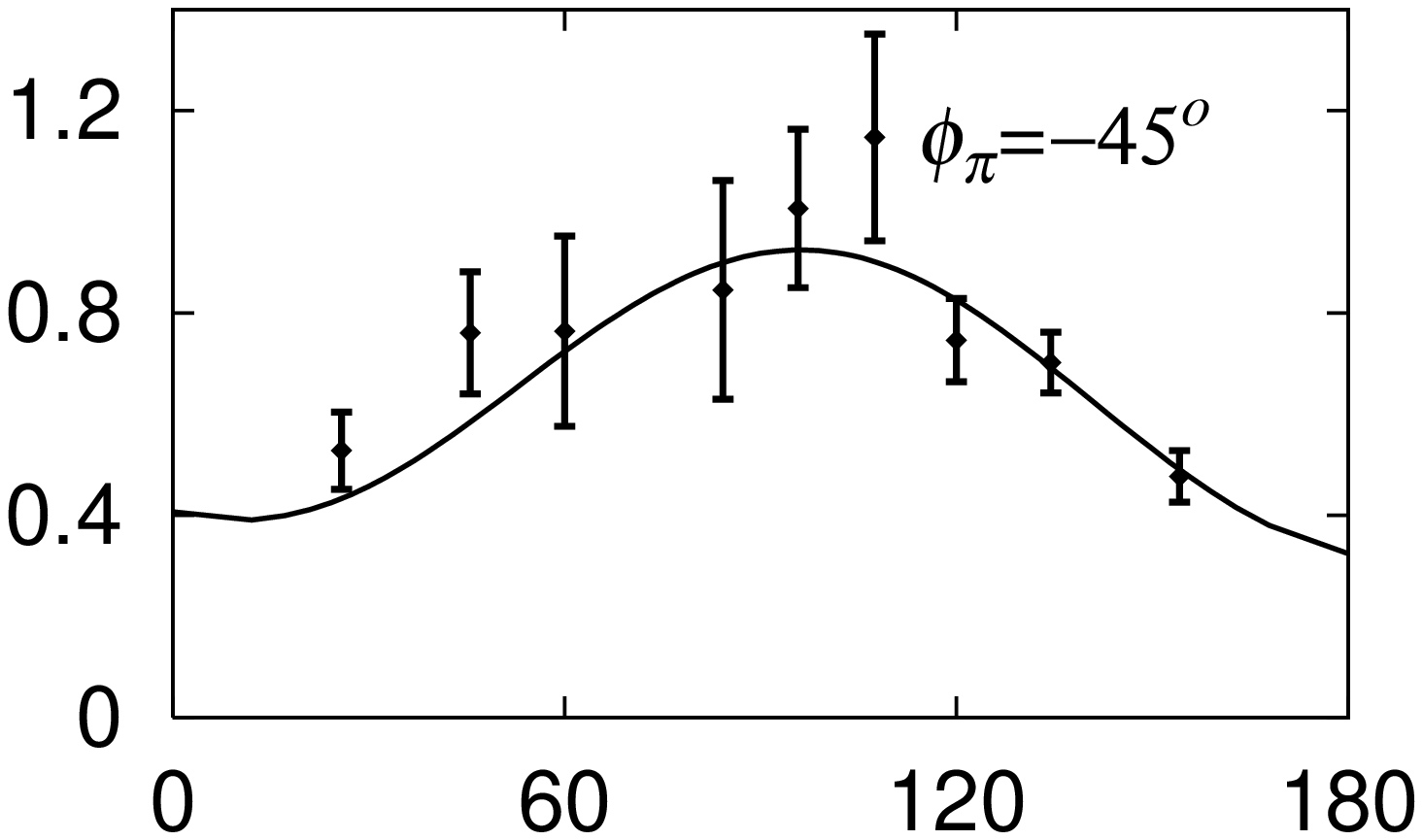,width=7cm}
           }
\vspace*{-0.5cm}
\centerline{
\epsfig{file=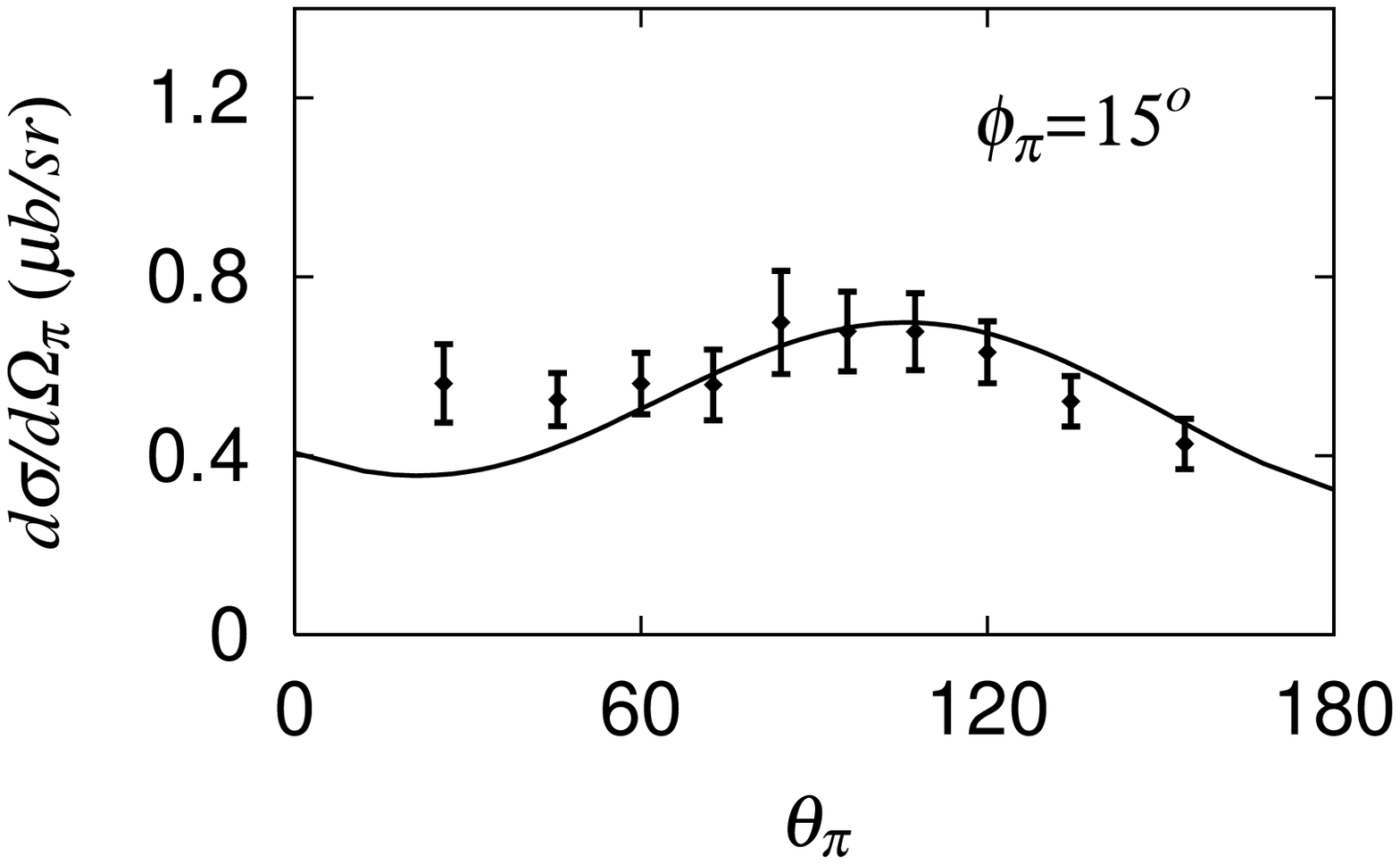,width=7cm}
\epsfig{file=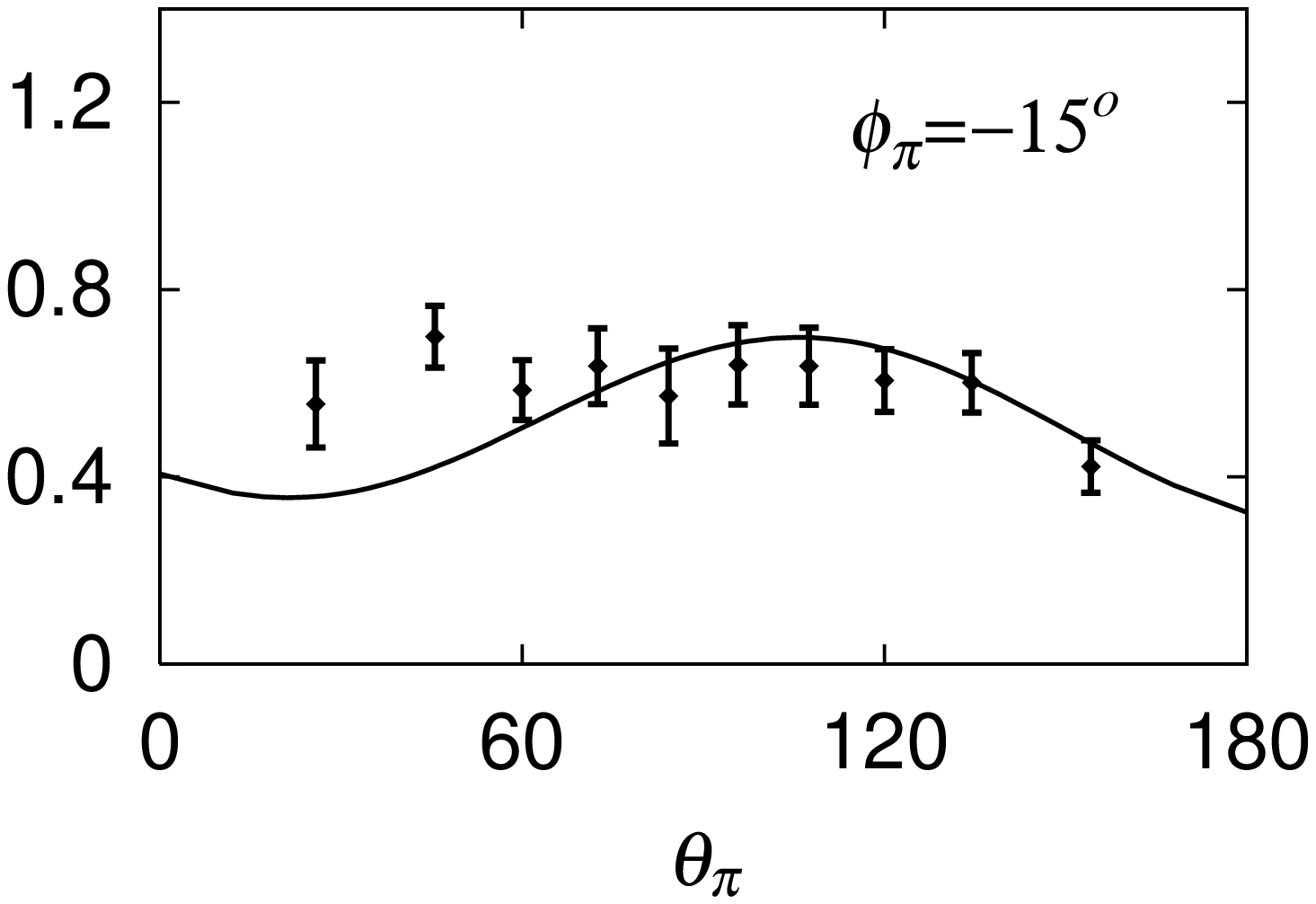,width=7cm}
           }

\caption[]{The predicted $p(e,e'\pi^0)$ differential cross sections 
at $W=1235$ MeV, $Q^2=2.8$ (GeV/c)$^2$
 are compared with the  JLab data \cite{JLAB}.}
\end{figure}

\newpage
\begin{figure}
\centerline{
\epsfig{file=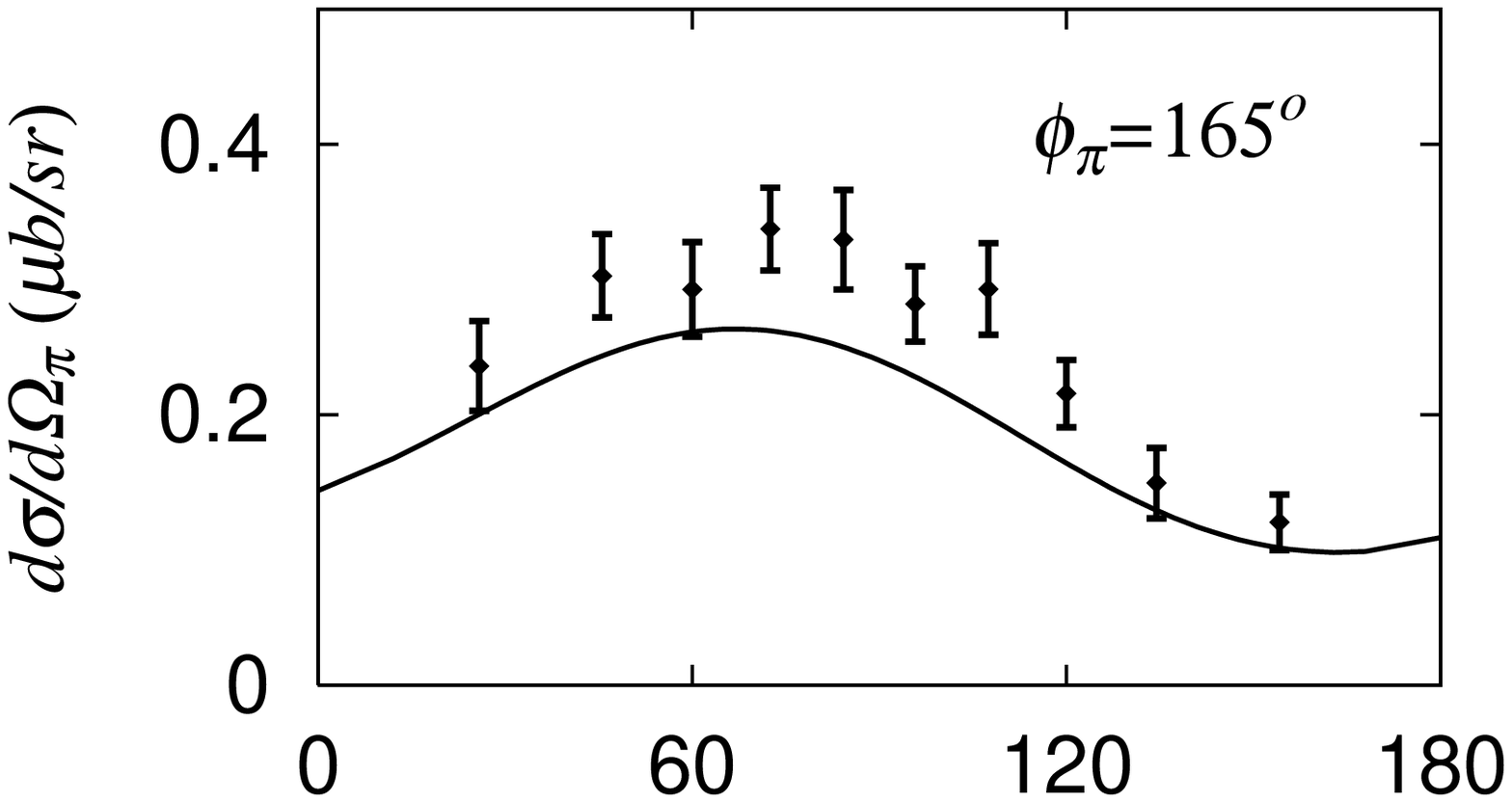,width=7cm}
\epsfig{file=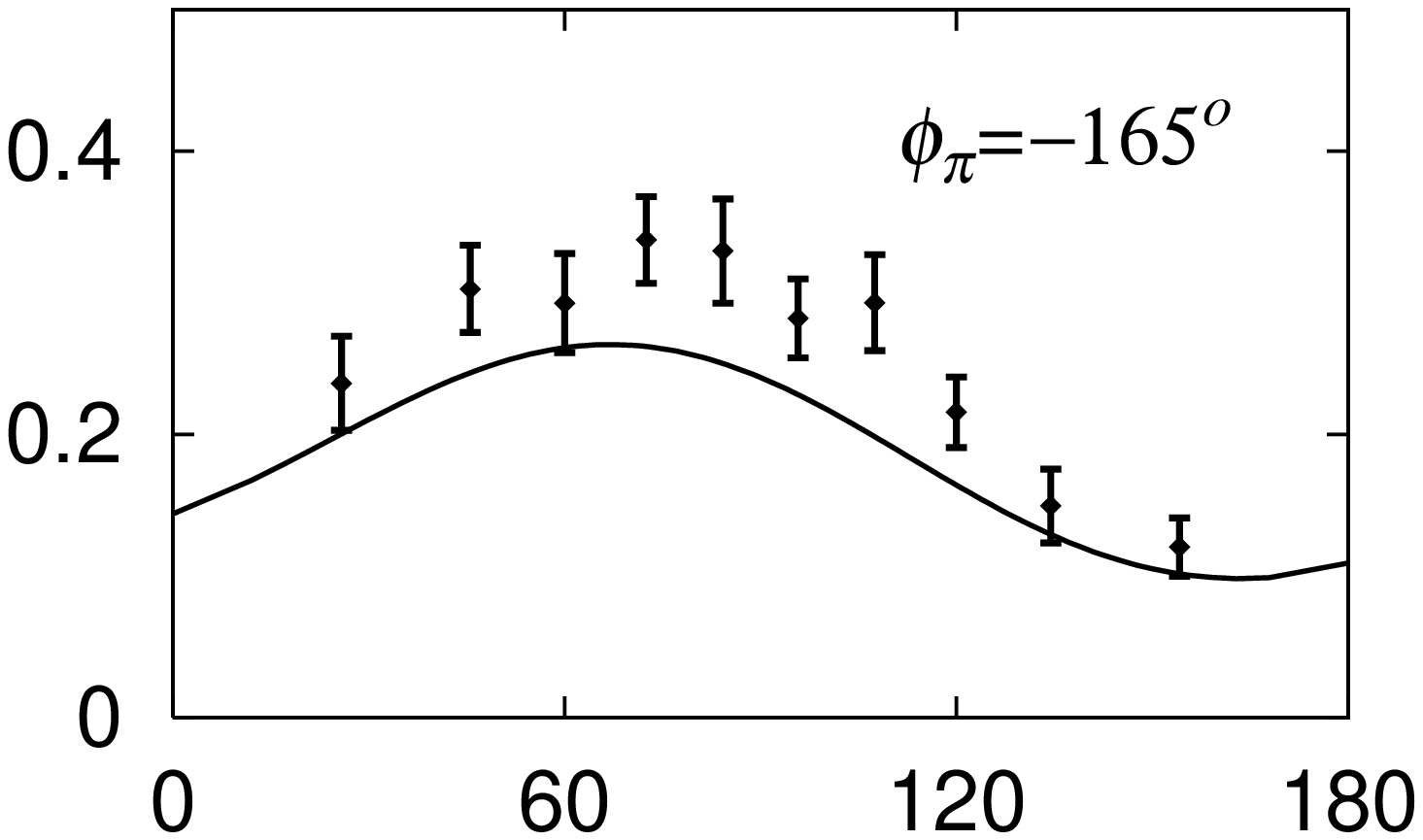,width=7cm}
           }
\vspace*{-0.5cm}
\centerline{
\epsfig{file=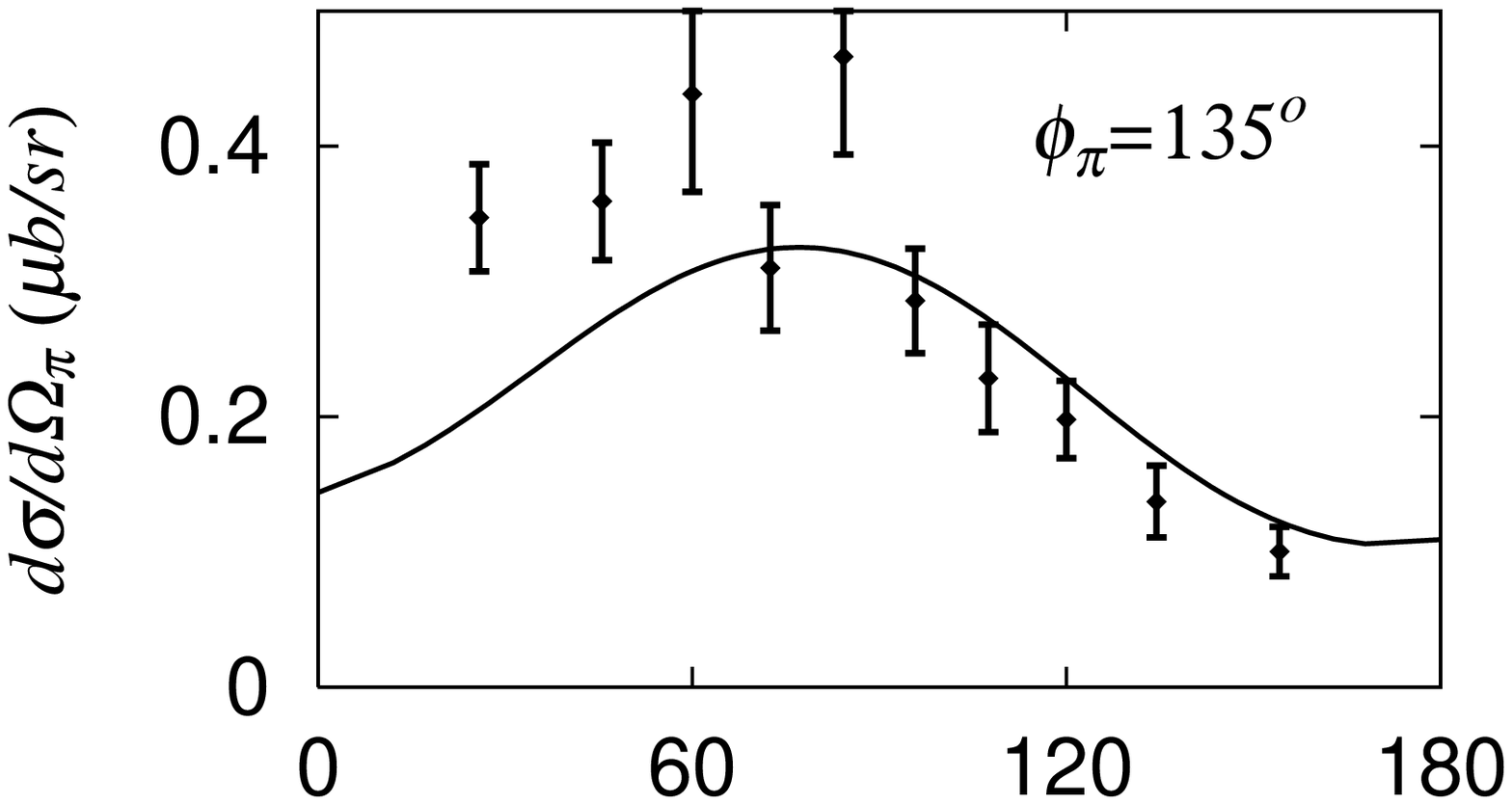,width=7cm}
\epsfig{file=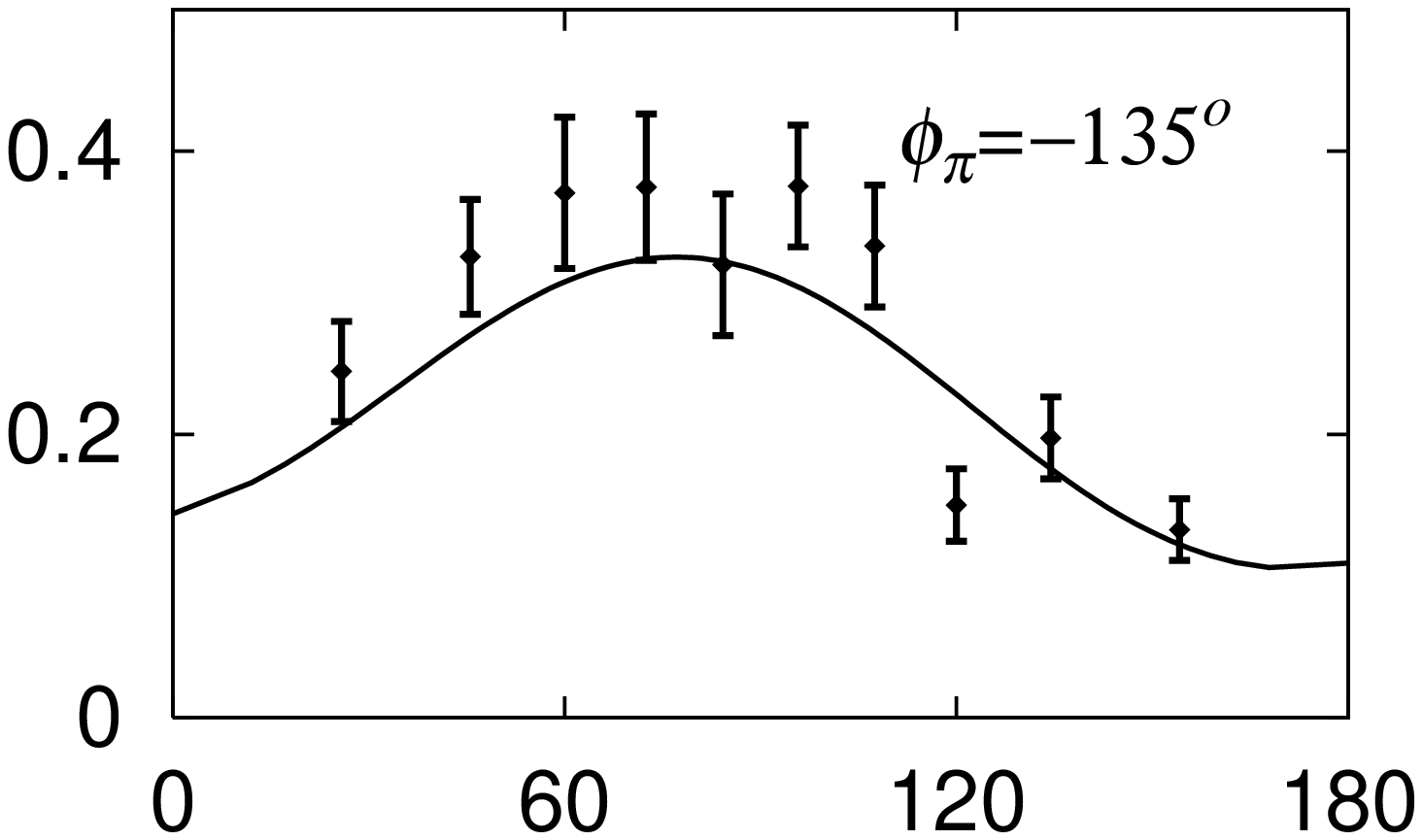,width=7cm}
           }
\vspace*{-0.5cm}
\centerline{
\epsfig{file=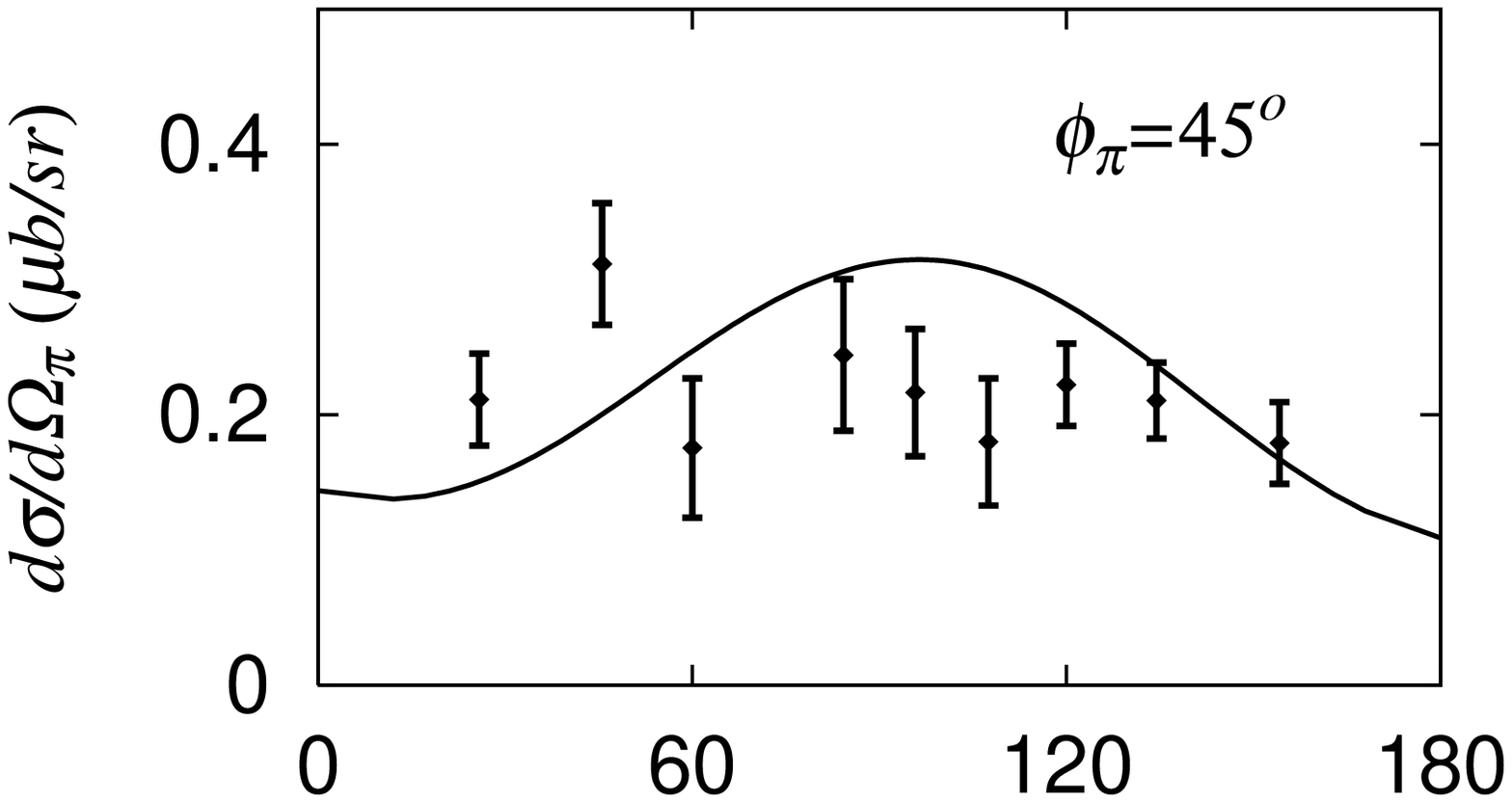,width=7cm}
\epsfig{file=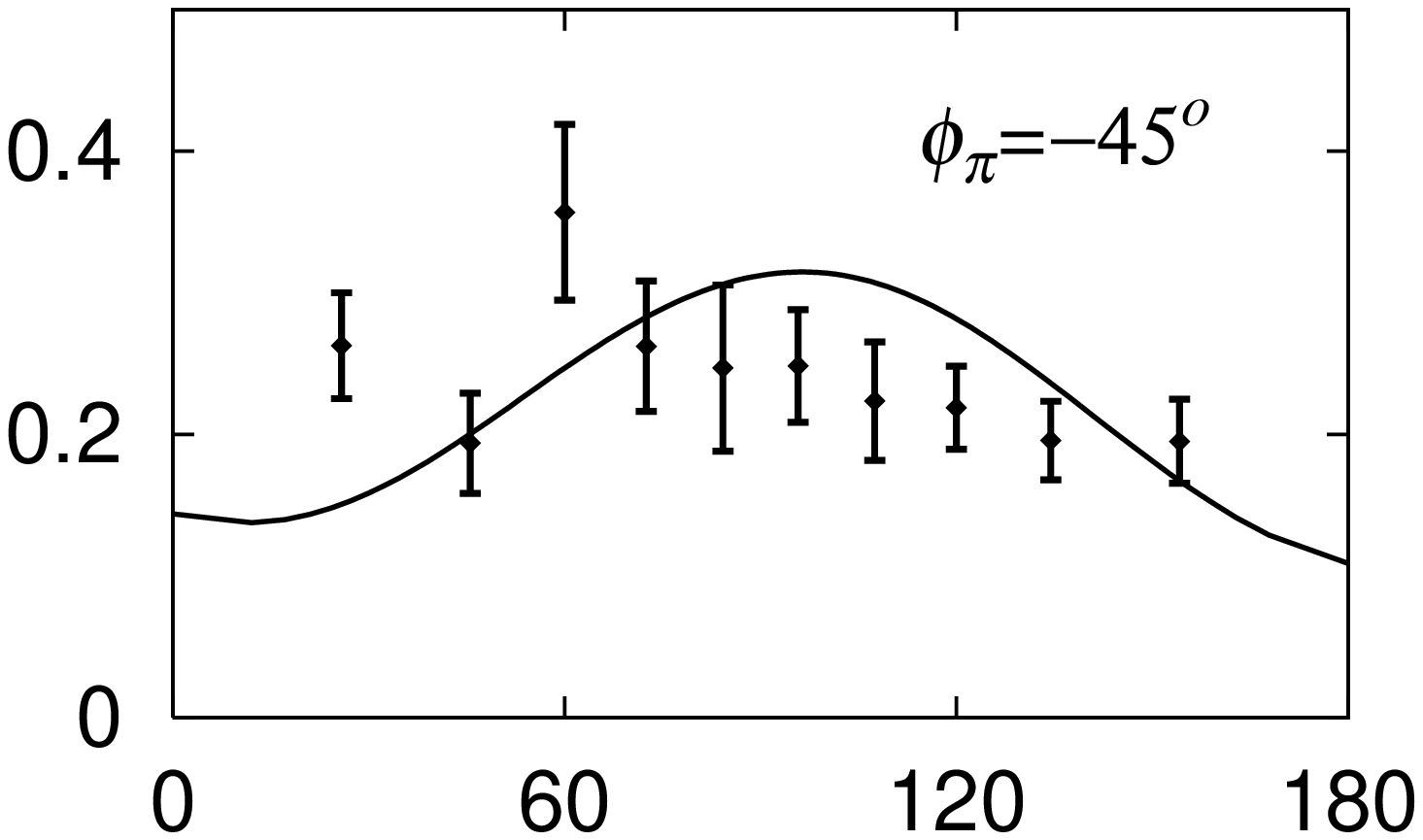,width=7cm}
           }
\vspace*{-0.5cm}
\centerline{
\epsfig{file=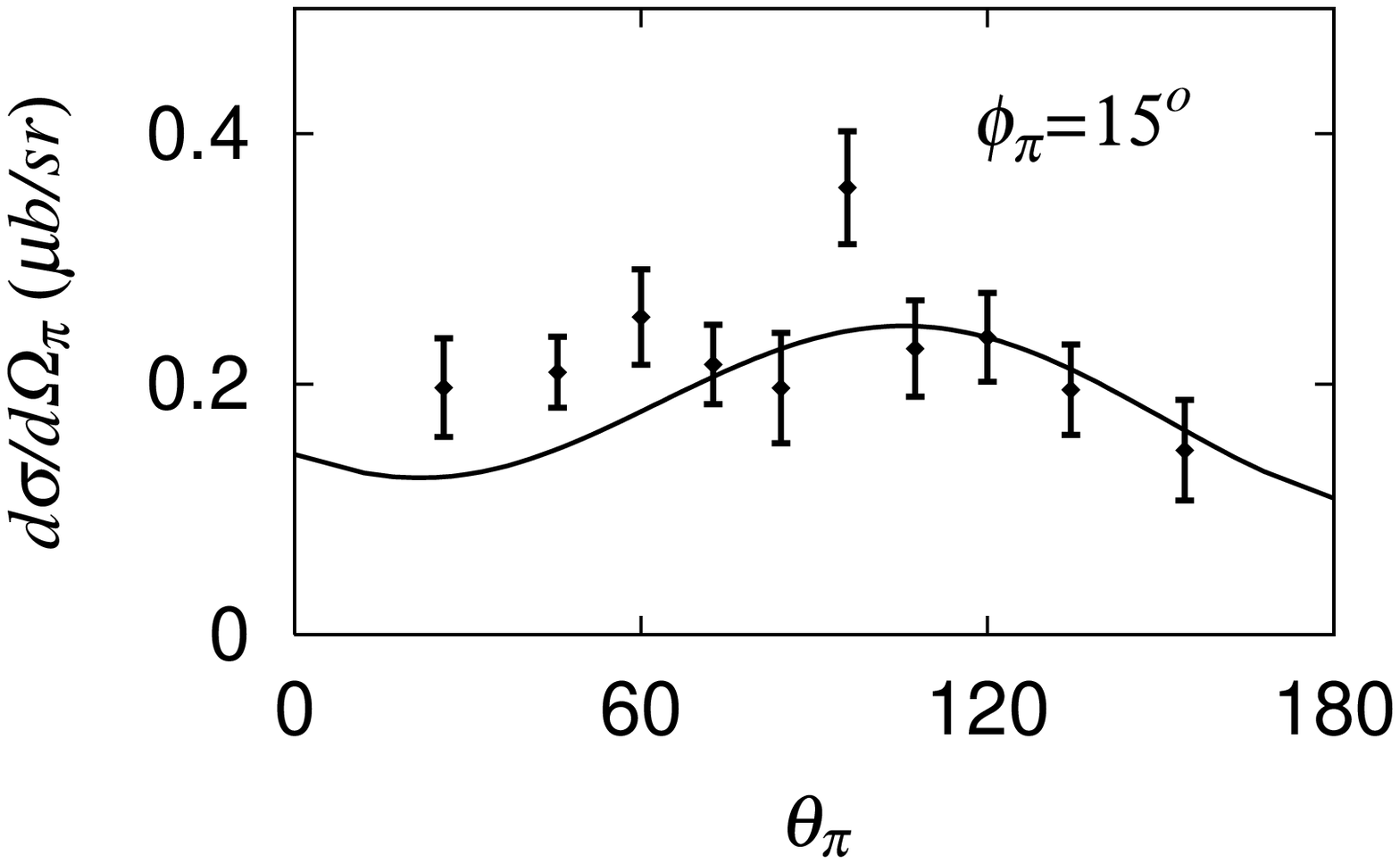,width=7cm}
\epsfig{file=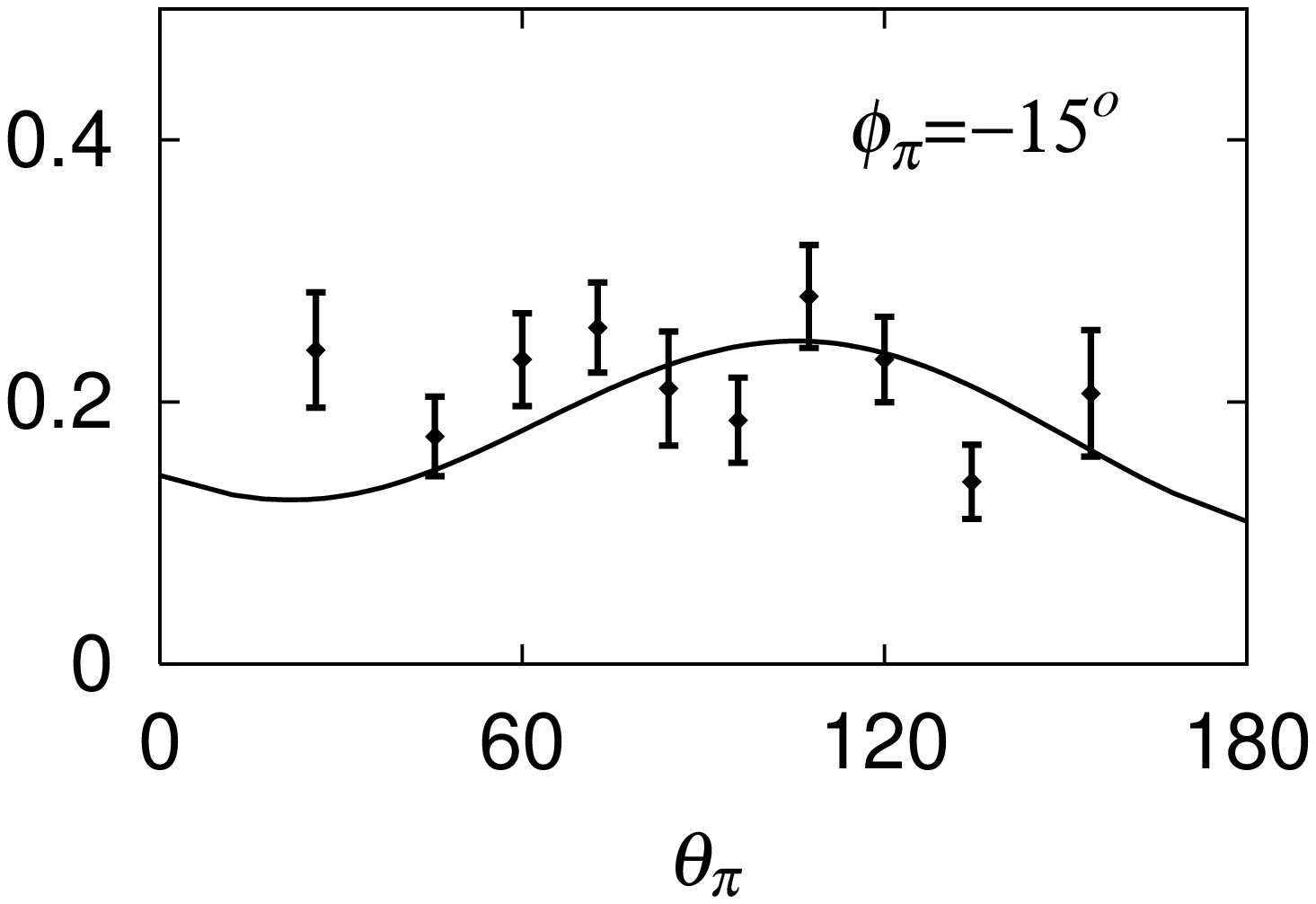,width=7cm}
           }
\caption[]{Same as Fig.7 except for $Q^2=$ 4 (GeV/c)$^2$.}
\end{figure}

\newpage
\begin{figure}
\centerline{
\epsfig{file=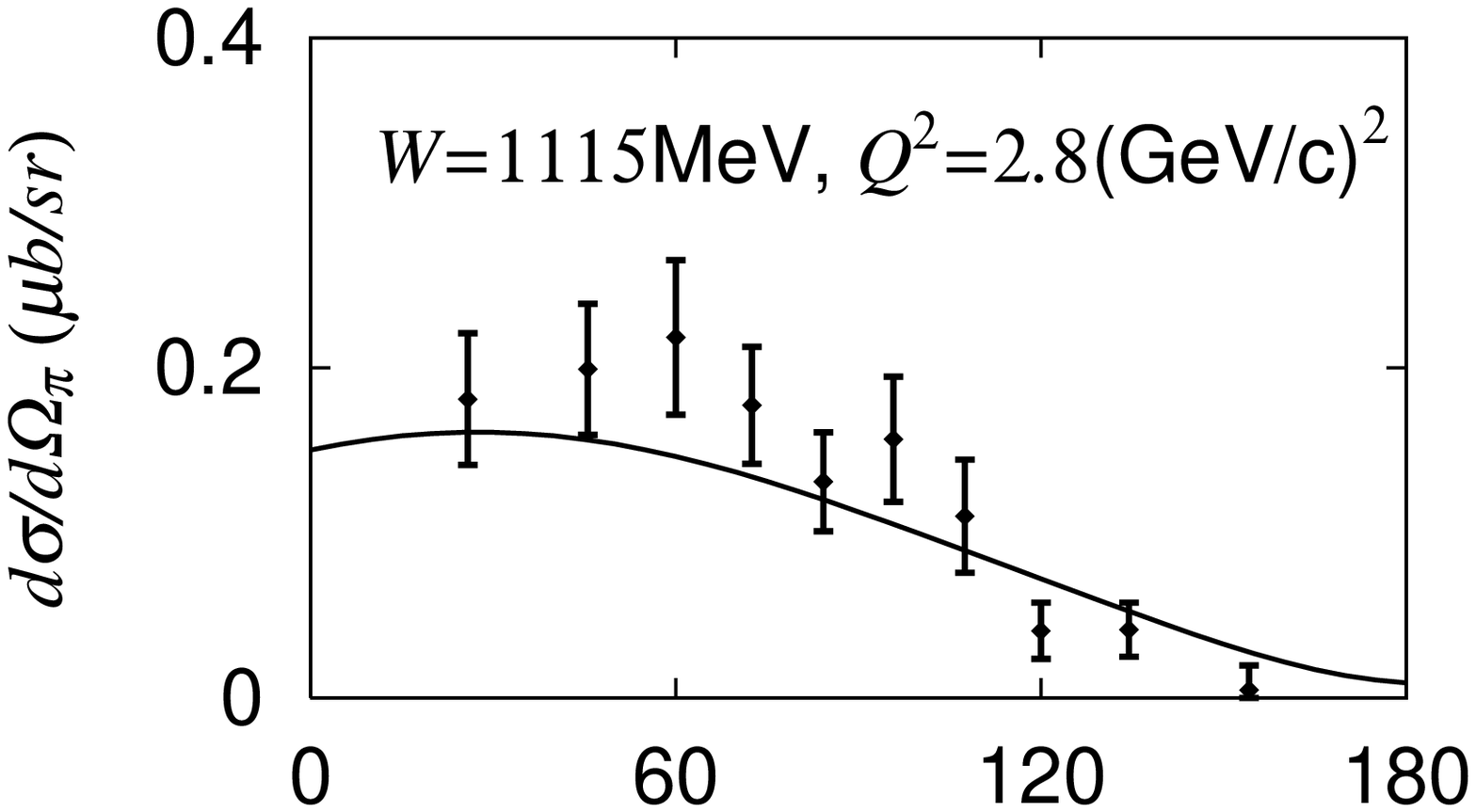,width=7cm}
\epsfig{file=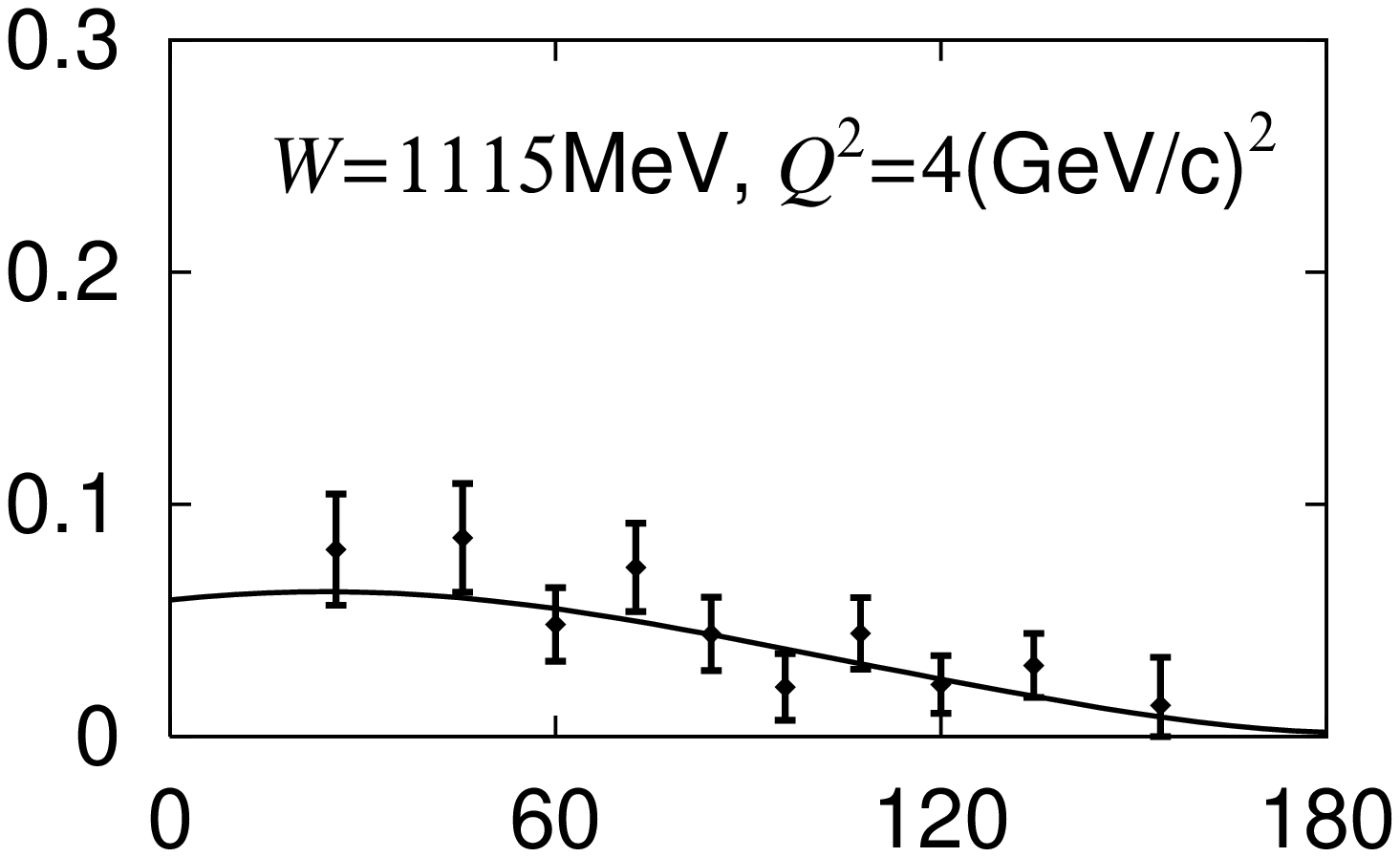,width=7cm}
           }
\vspace*{-0.5cm}
\centerline{
\epsfig{file=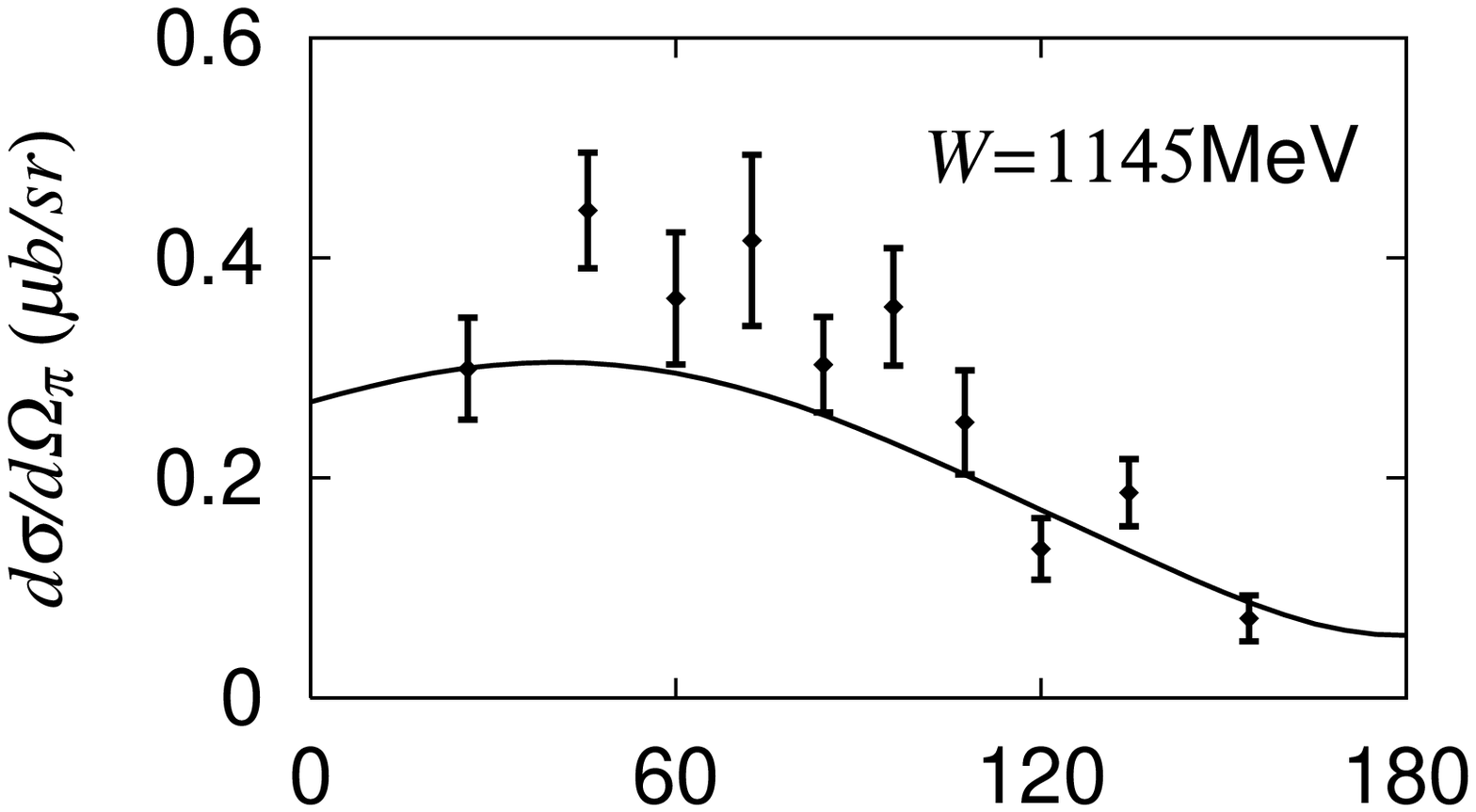,width=7cm}
\epsfig{file=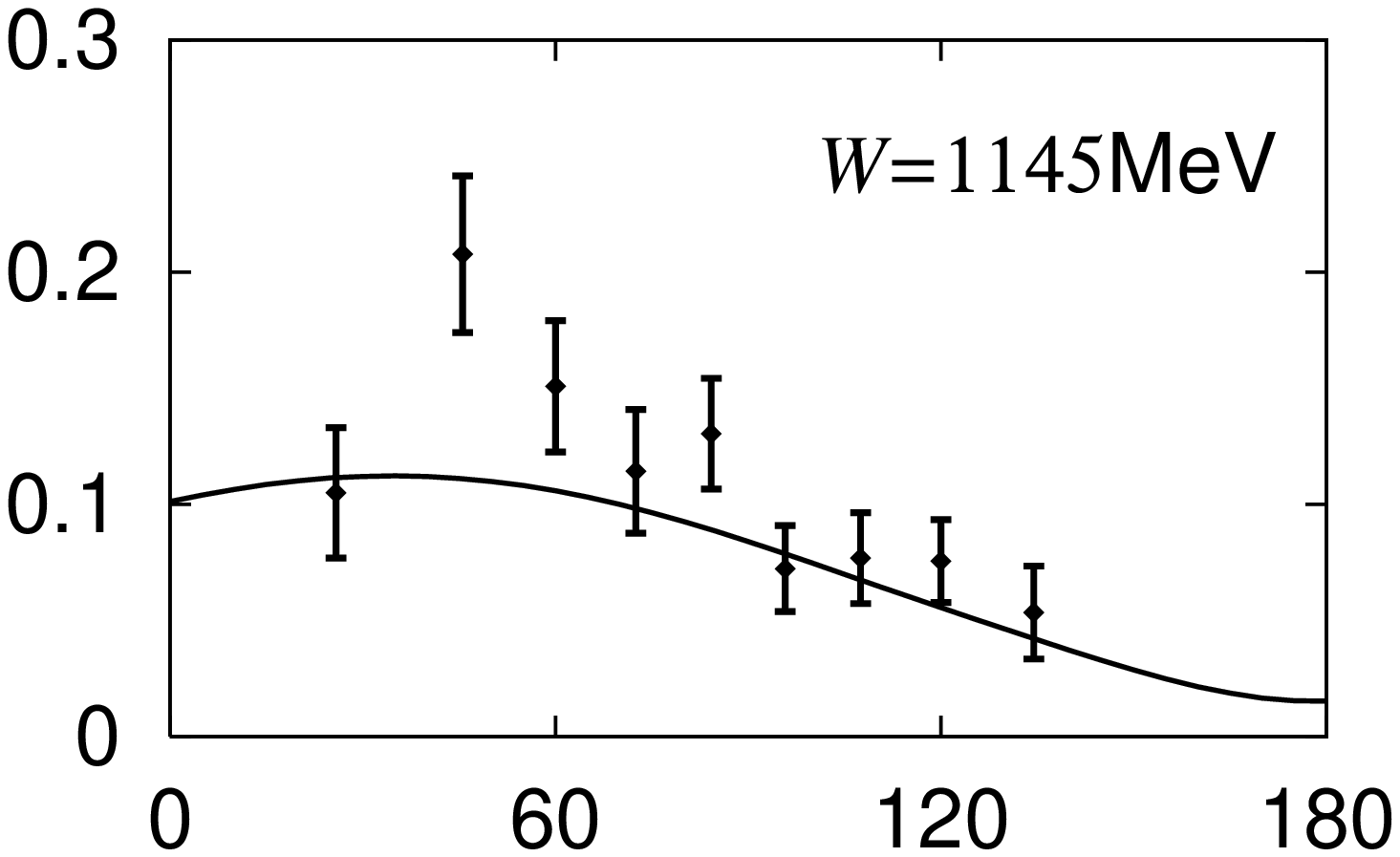,width=7cm}
           }
\vspace*{-0.5cm}
\centerline{
\epsfig{file=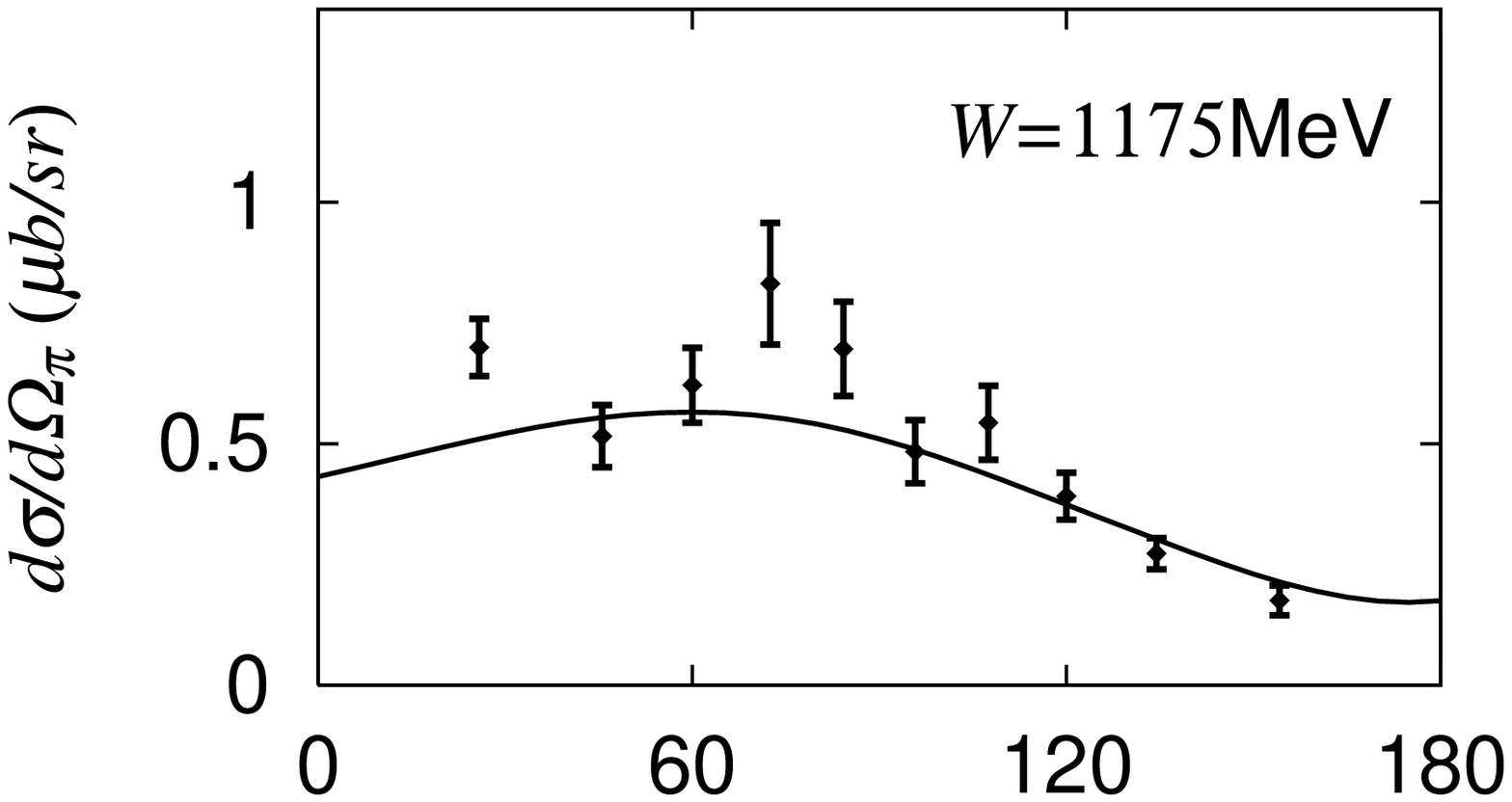,width=7cm}
\epsfig{file=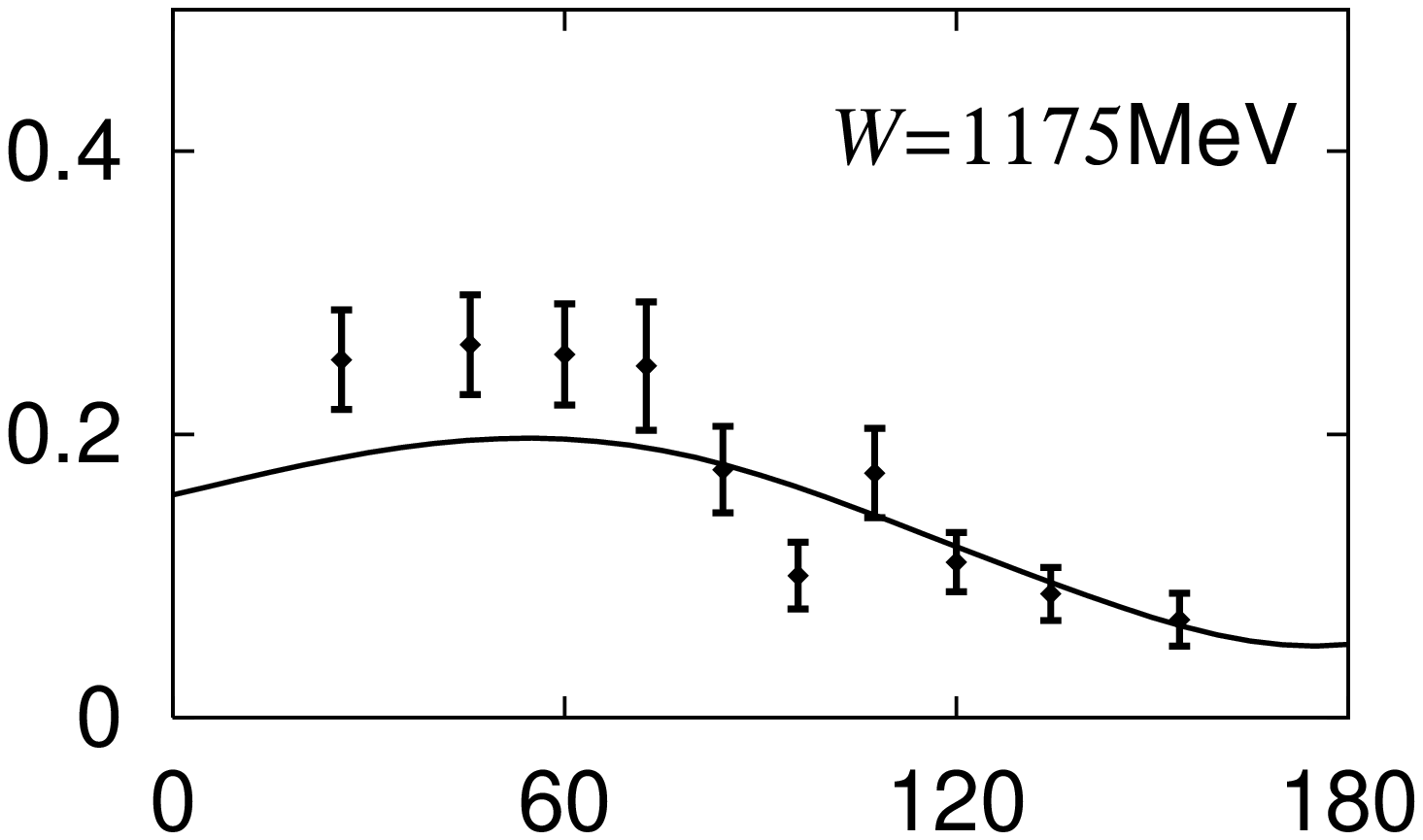,width=7cm}
           }
\vspace*{-0.5cm}
\centerline{
\epsfig{file=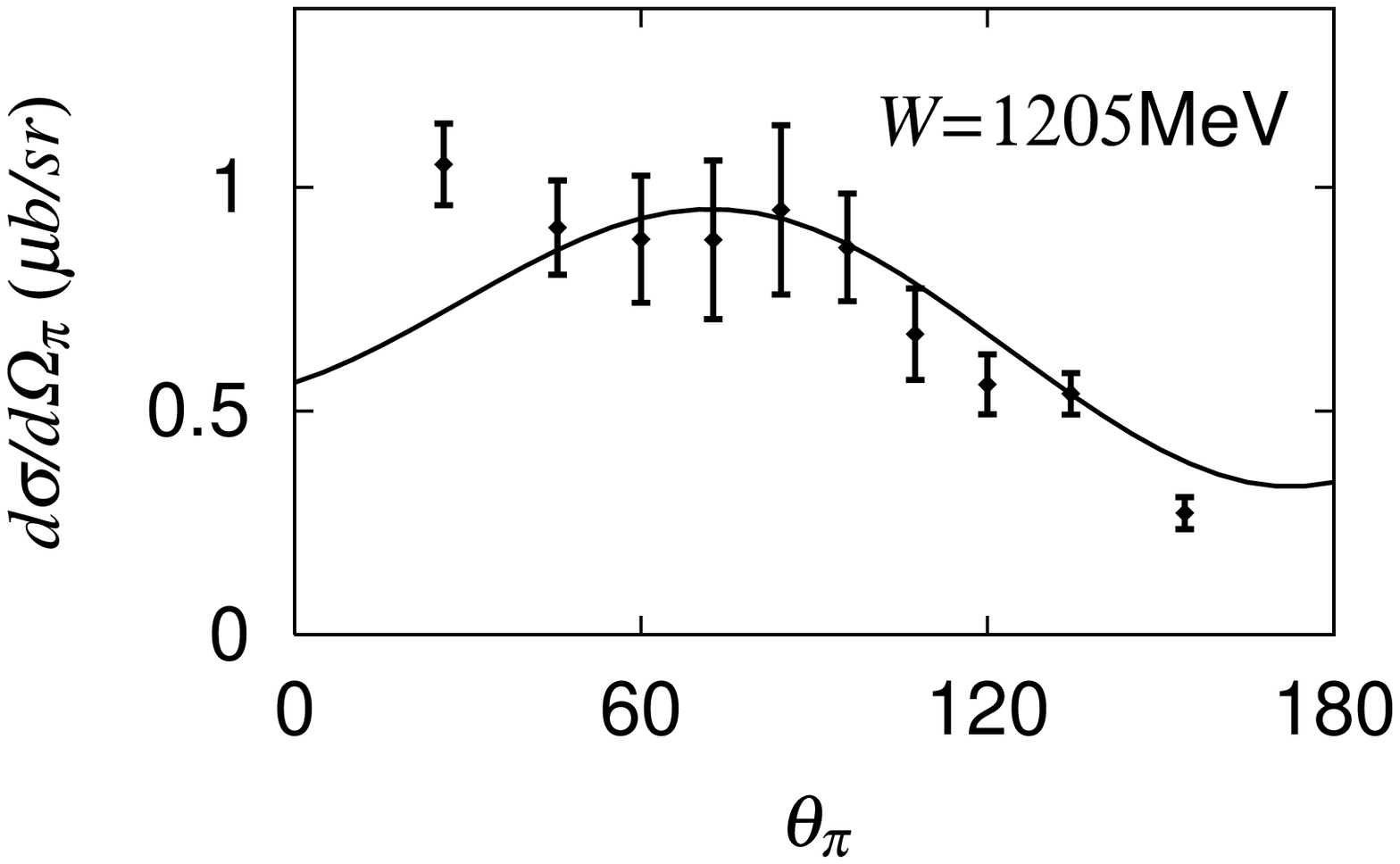,width=7cm}
\epsfig{file=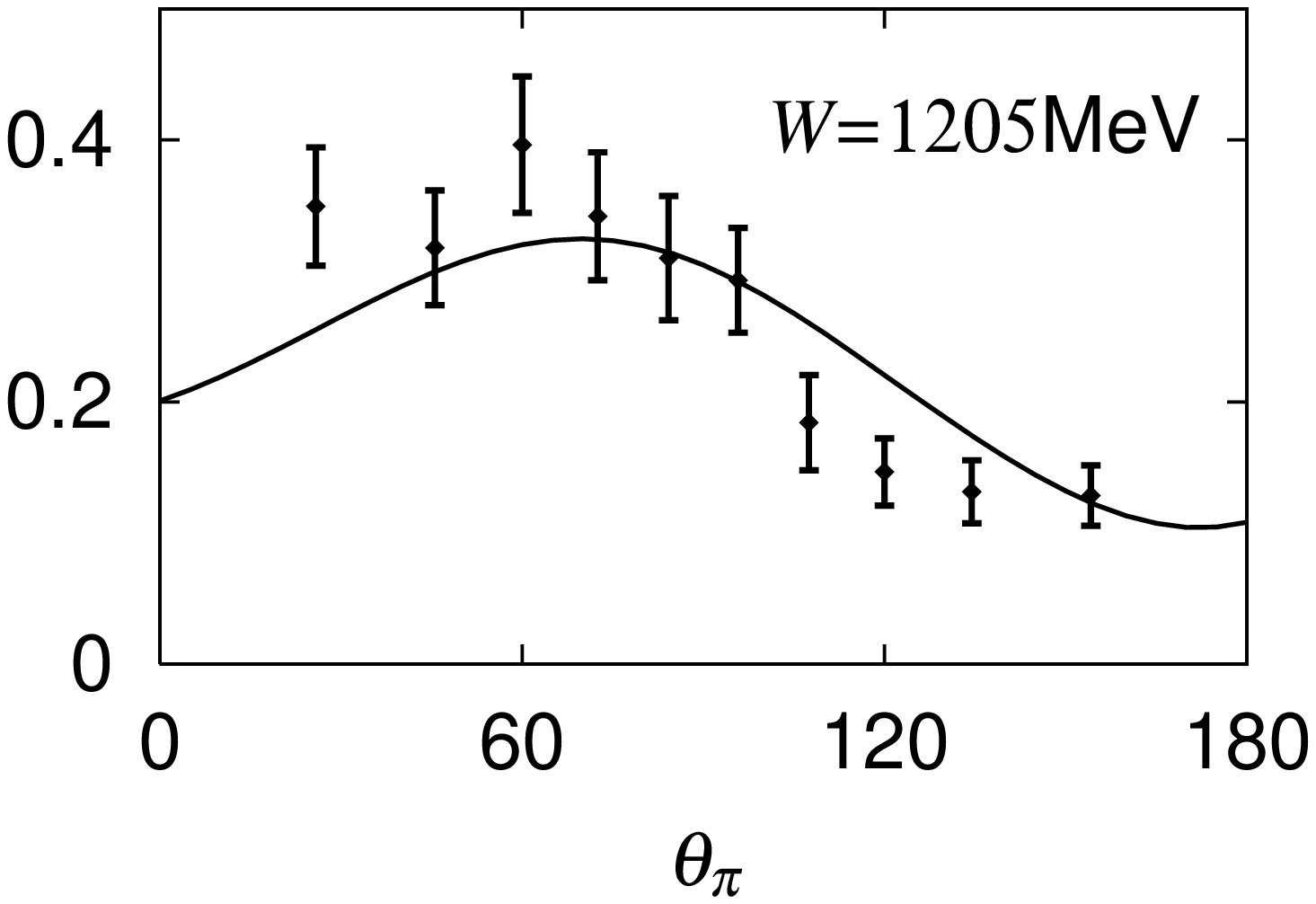,width=7cm}
           }
\vspace*{0.1cm}
\caption[]{The predicted $p(e,e'\pi^0)$ differential cross sections 
at $\phi_\pi=135^0$  with $Q^2=2.8$(left) and $4$(right) (GeV/c)$^2$
 are compared with the  JLab data \cite{JLAB}.}
\end{figure}

\newpage
\begin{figure}[h]
\centerline{\epsfig{file=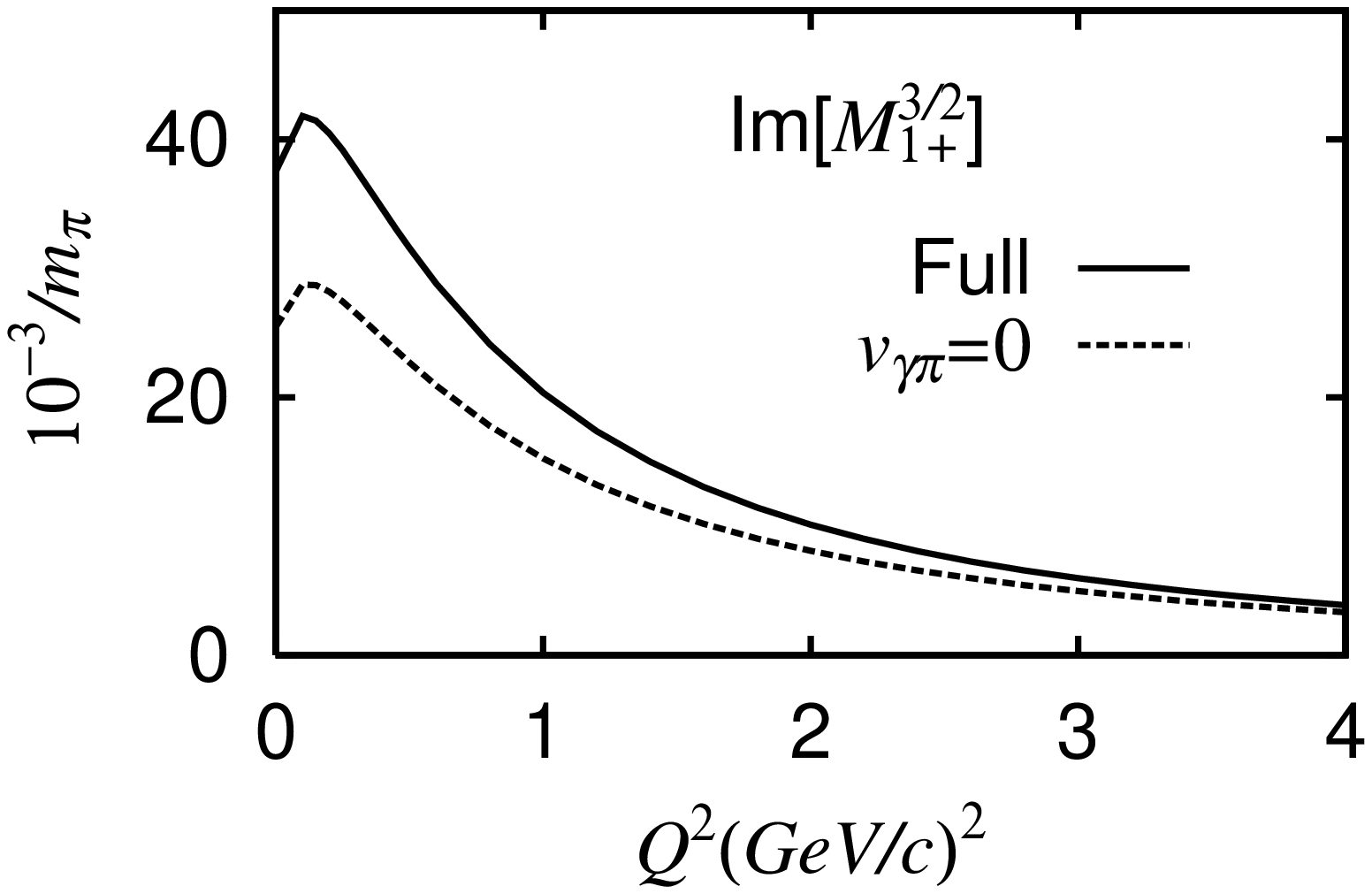,width=7cm}}
\centerline{\epsfig{file=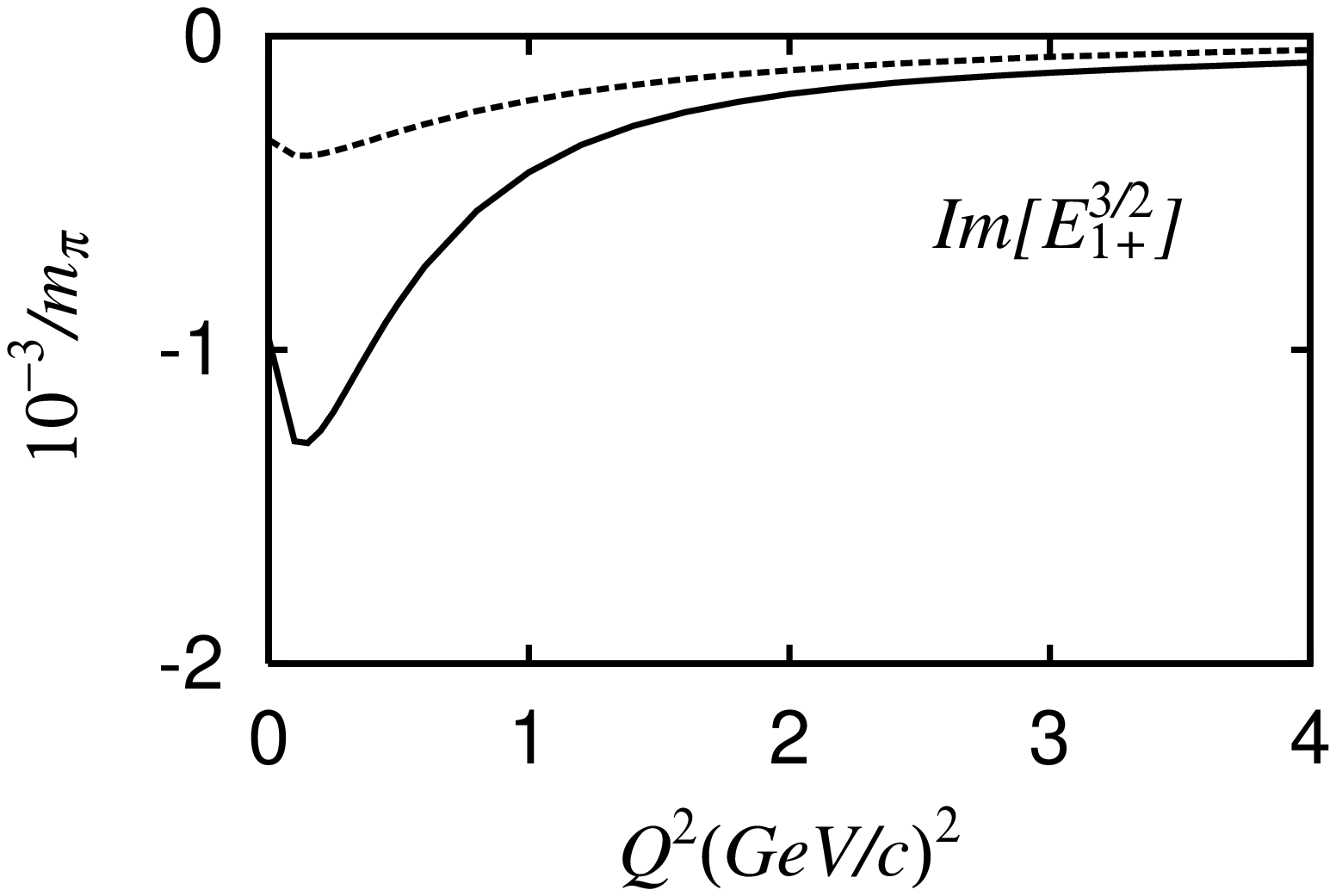,width=7cm}}
\centerline{\epsfig{file=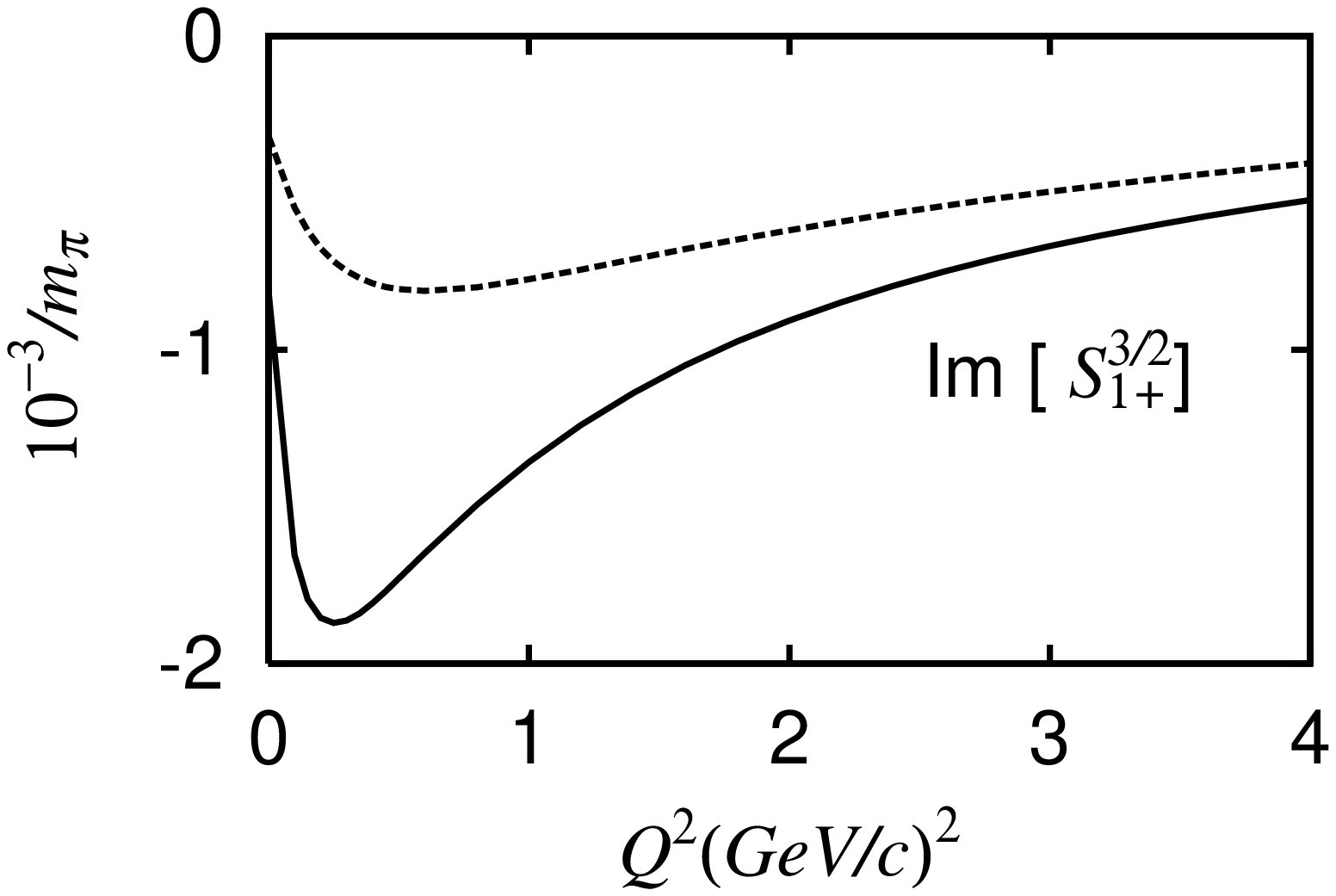,width=7cm}}
\vspace*{0.5cm}
\caption[]{$Q^2$ dependence of the imaginary($\mbox{Im}$) parts of the
$\gamma^* N \rightarrow \pi N$  multipole 
amplitudes $M_{1+}^{3/2}$
   $E^{3/2}_{1^+}$, and $S_{1+}^{3/2}$ 
    at $W=1236$ MeV. The solid curves
    are from our full calculations, and the dotted curves are from calculations
    with the non-resonant interaction $v_{\gamma\pi}$ set to zero. The real
     parts at W=1236 MeV are negligibly small and are omitted.}
 \end{figure}

\newpage
\begin{figure}
\centerline{\epsfig{file=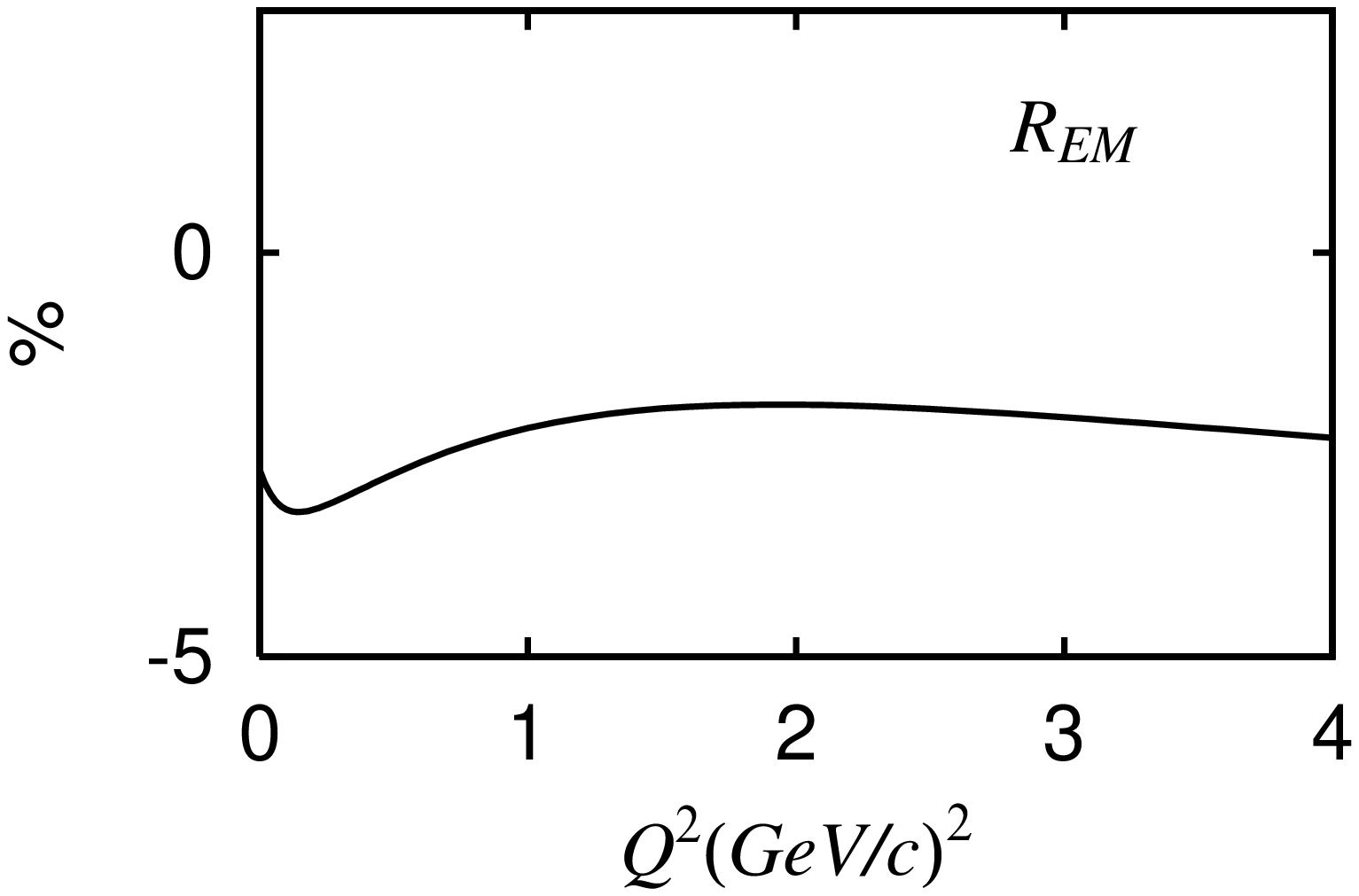,width=7cm}
            \epsfig{file=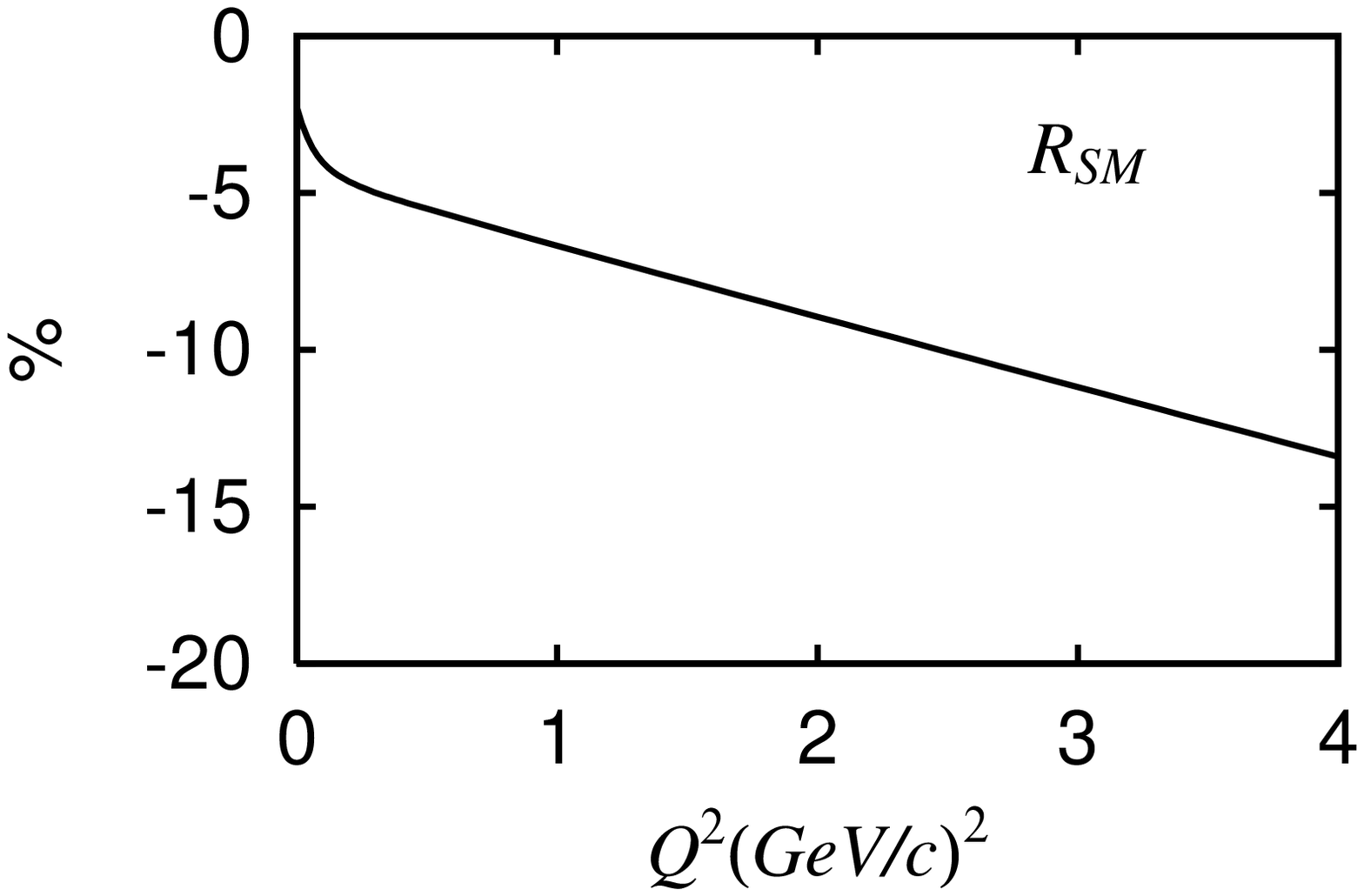,width=7cm}}
\vspace*{0.5cm}
\caption[]{The predicted $Q^2$-dependence of the
E2/M1 ratio \rem and C2/M1 ratio \rsm of 
the dressed $\gamma N\rightarrow \Delta$ form factors.}
\end{figure}

\newpage 
\begin{figure}
\begin{center}
\epsfig{file=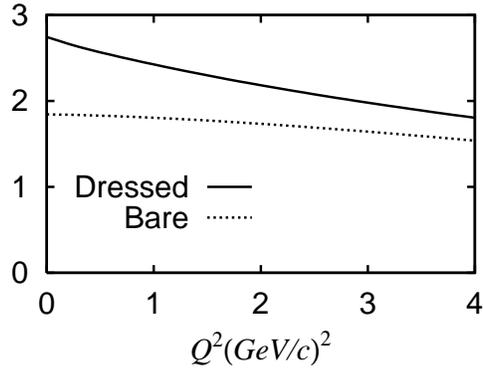,width=7.4cm}
\vspace*{0.5cm}
\caption[]{The ratio between the M1 form factor of the 
$\gamma N \rightarrow \Delta$ transition and the proton
dipole form factor $G_D$  defined by Eq. (\ref{frmdip}).
The solid curve is $G^*_{M}(Q^2)/G_D(Q^2)$ for the dressed M1 form factor, and
the dotted curve is $G_M(Q^2)/G_D(Q^2)$ for the bare M1 form factor }
\end{center}
\end{figure}

\newpage
\begin{figure}[h]
\centerline{\epsfig{file=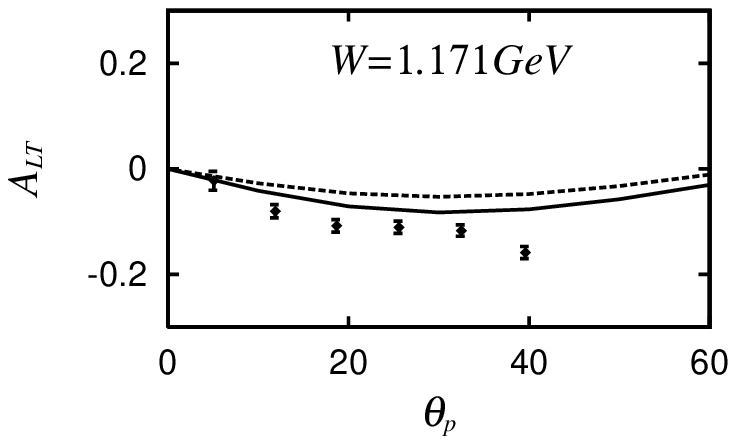,width=7cm}}
\centerline{\epsfig{file=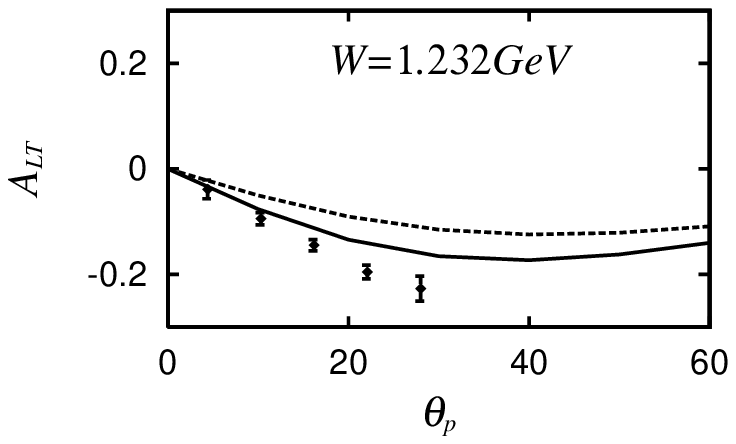,width=7cm}}
\centerline{\epsfig{file=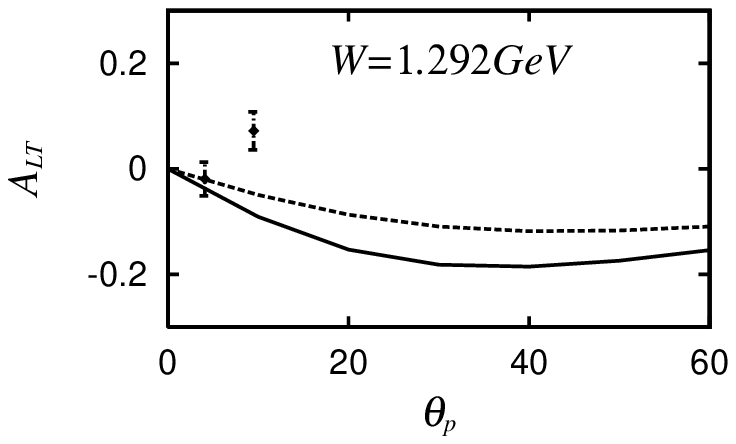,width=7cm}}
\vspace*{0.5cm}
\caption[]{The predicted asymmetry $A_{LT}$ (Eq.~(\ref{alt})) of the 
  $p(e,e'\pi^0)p$ reaction at $Q^2=0.126(GeV/c)^2$ are compared with the
  data from MIT-Bates \cite{MIT}. The dotted curves are obtained from
setting $G_M(Q^2)=G_C(Q^2)=0.$ }
\end{figure}

\newpage

\begin{figure}[h]
\centerline{\epsfig{file=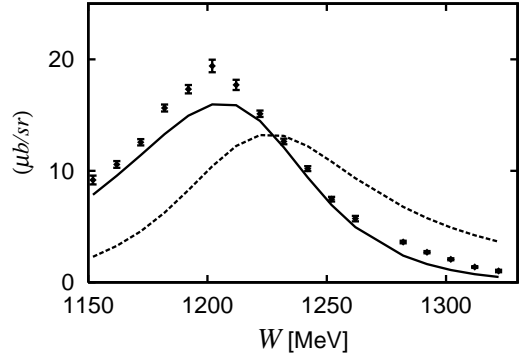,width=7cm}}
\vspace*{0.5cm}
\caption[]{ The predicted $R_{TT}$(Eq. (\ref{rtt}) at $\theta_\pi=180^0$ are compared
with the data from MIT-Bates \cite{MIT}. The dotted curve is obtained from
setting the non-resonant interaction $v_{\gamma\pi}$ to zero.}
\end{figure}

\begin{figure}[h]
\centerline{\epsfig{file=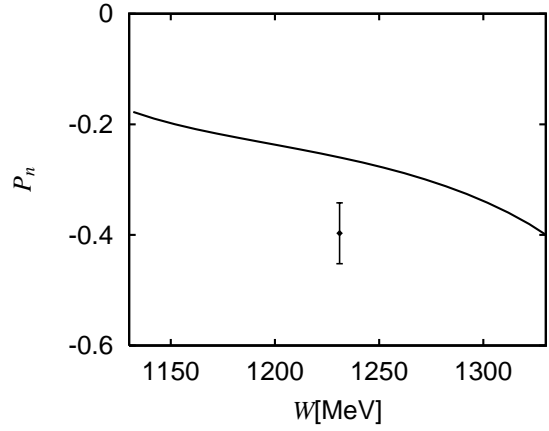,width=7cm}}
\vspace*{0.5cm}
\caption[]{The predicted induced proton polarization $P_n$ at
$\theta_\pi=180^0$ and the polarization vector $\vec{n}$ perpendicular to
the recoiled proton momentum are compared with the data from MIT-Bates \cite{MIT}.}
\end{figure}

\newpage
\begin{figure}[h]
\centerline{
\epsfig{file=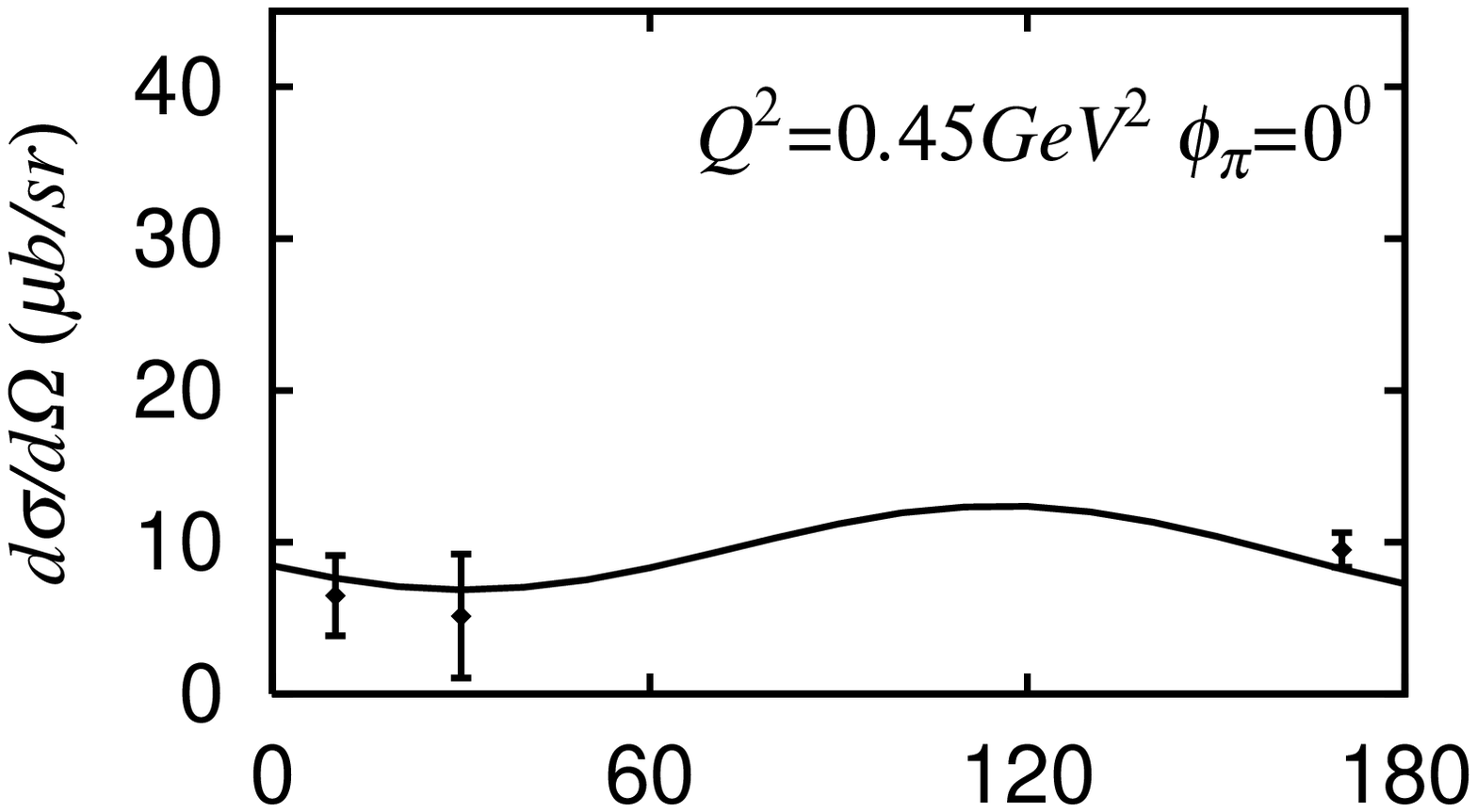,width=6cm}
\epsfig{file=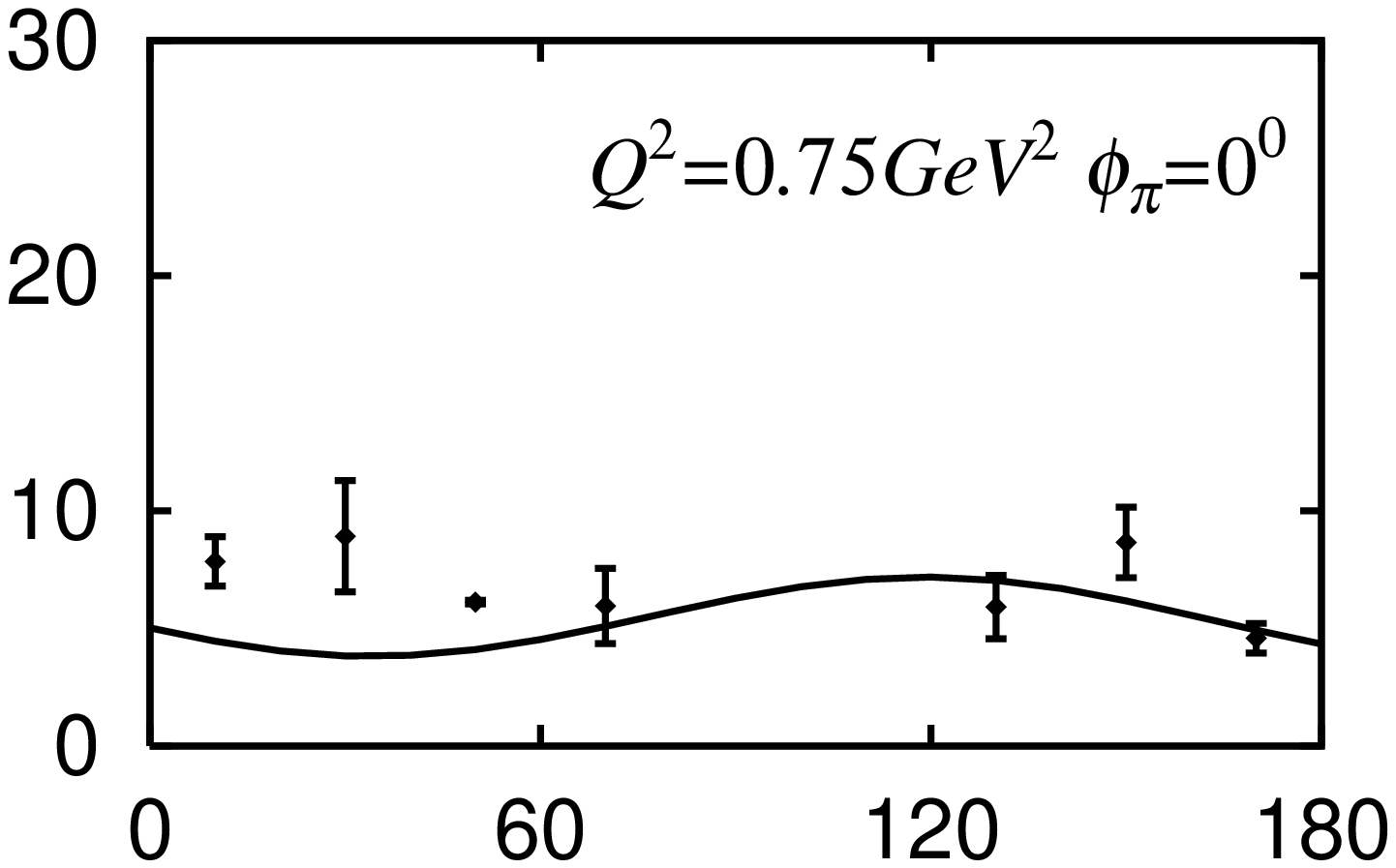,width=6cm}
}
\vspace*{-0.5cm}
\centerline{
\epsfig{file=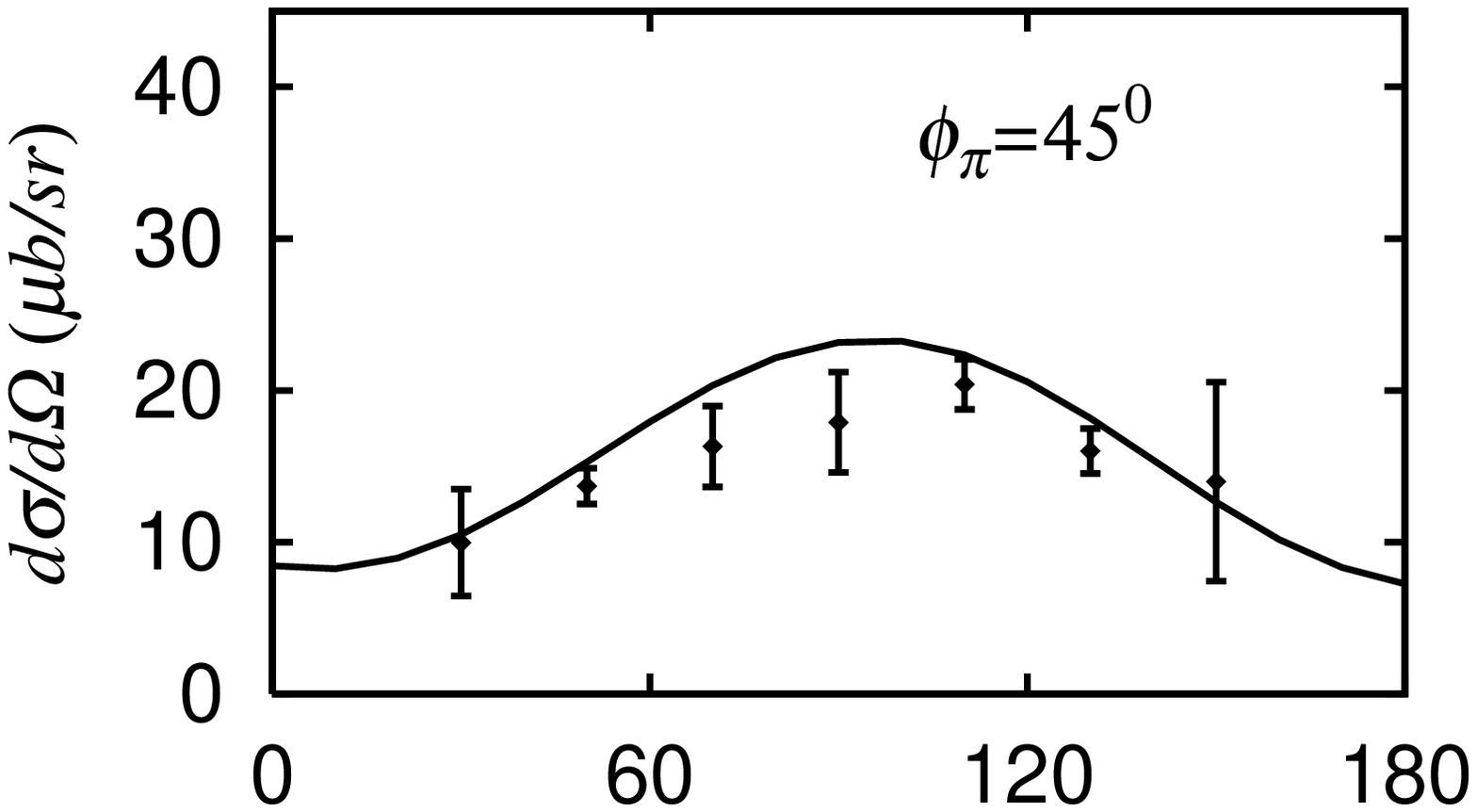,width=6cm}
\epsfig{file=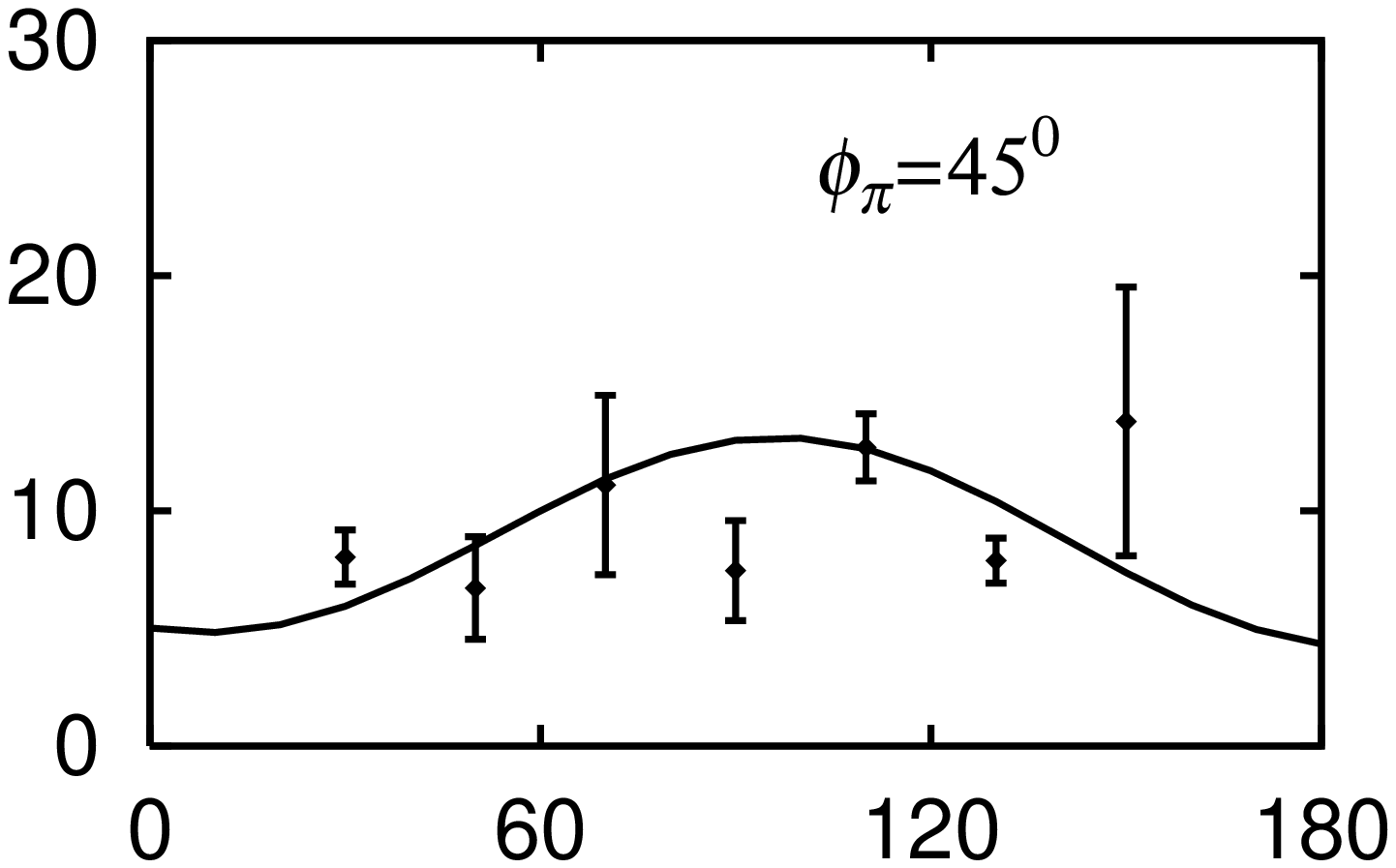,width=6cm}
}
\vspace*{-0.5cm}
\centerline{
\epsfig{file=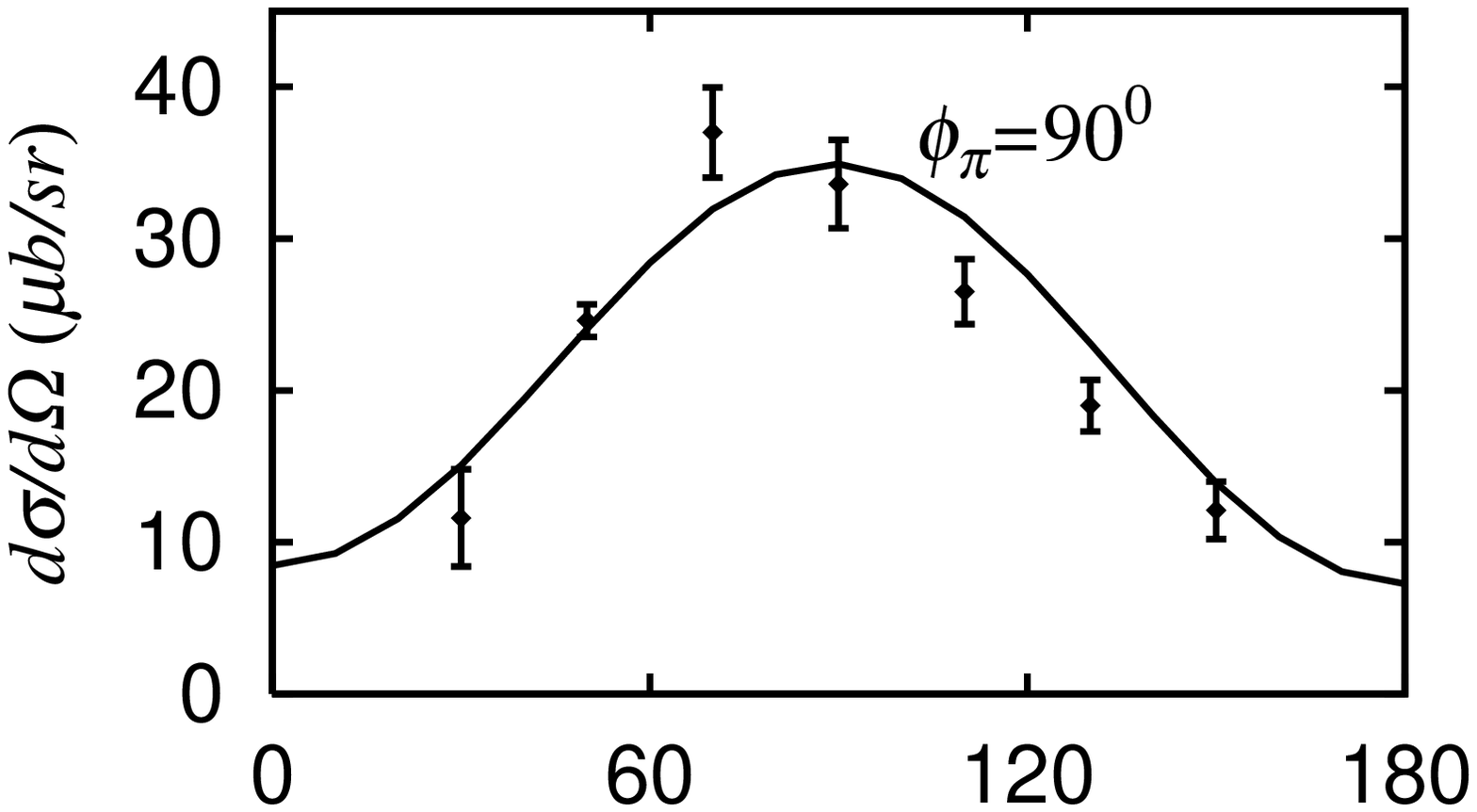,width=6cm}
\epsfig{file=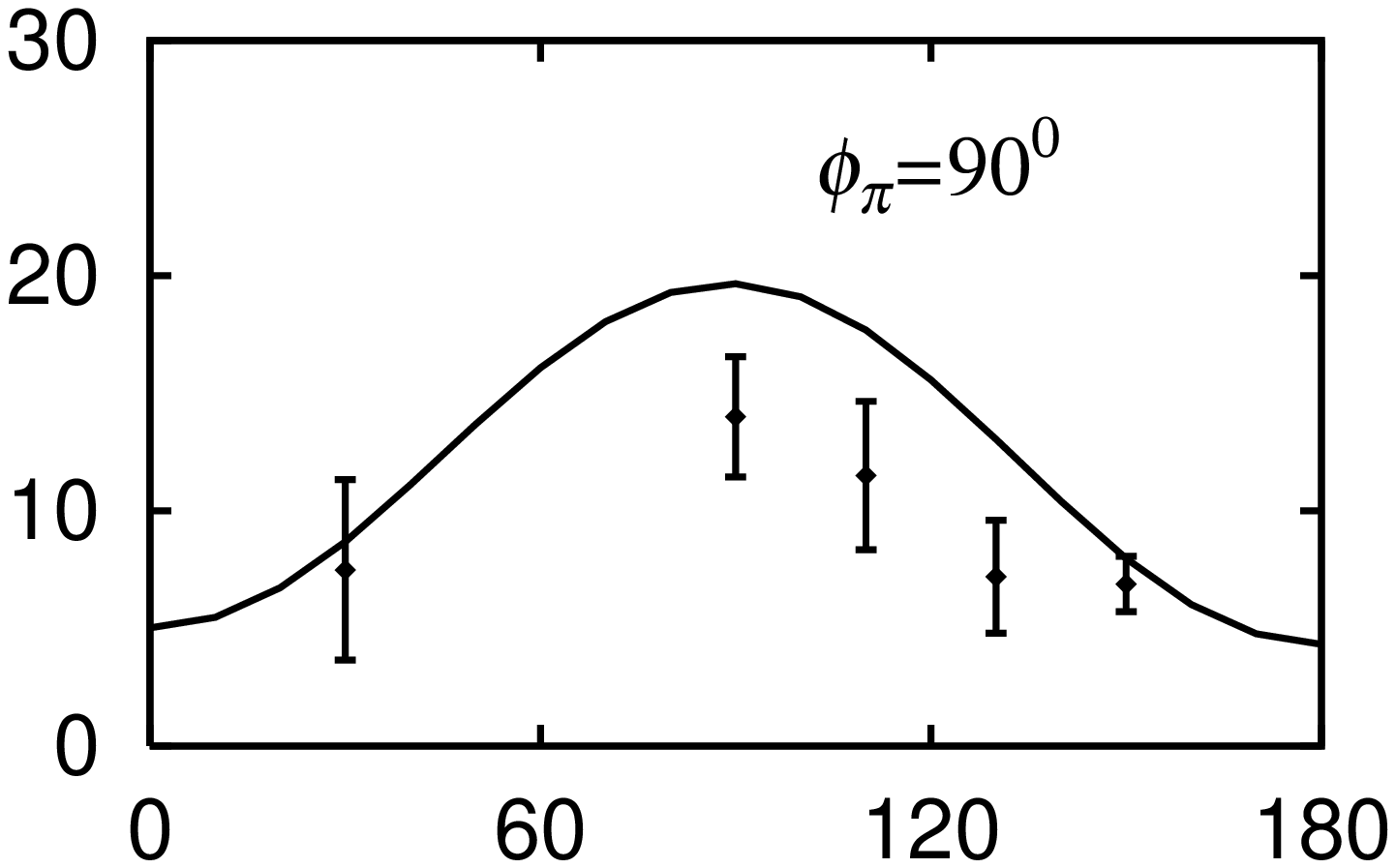,width=6cm}
}
\vspace*{-0.5cm}
\centerline{
\epsfig{file=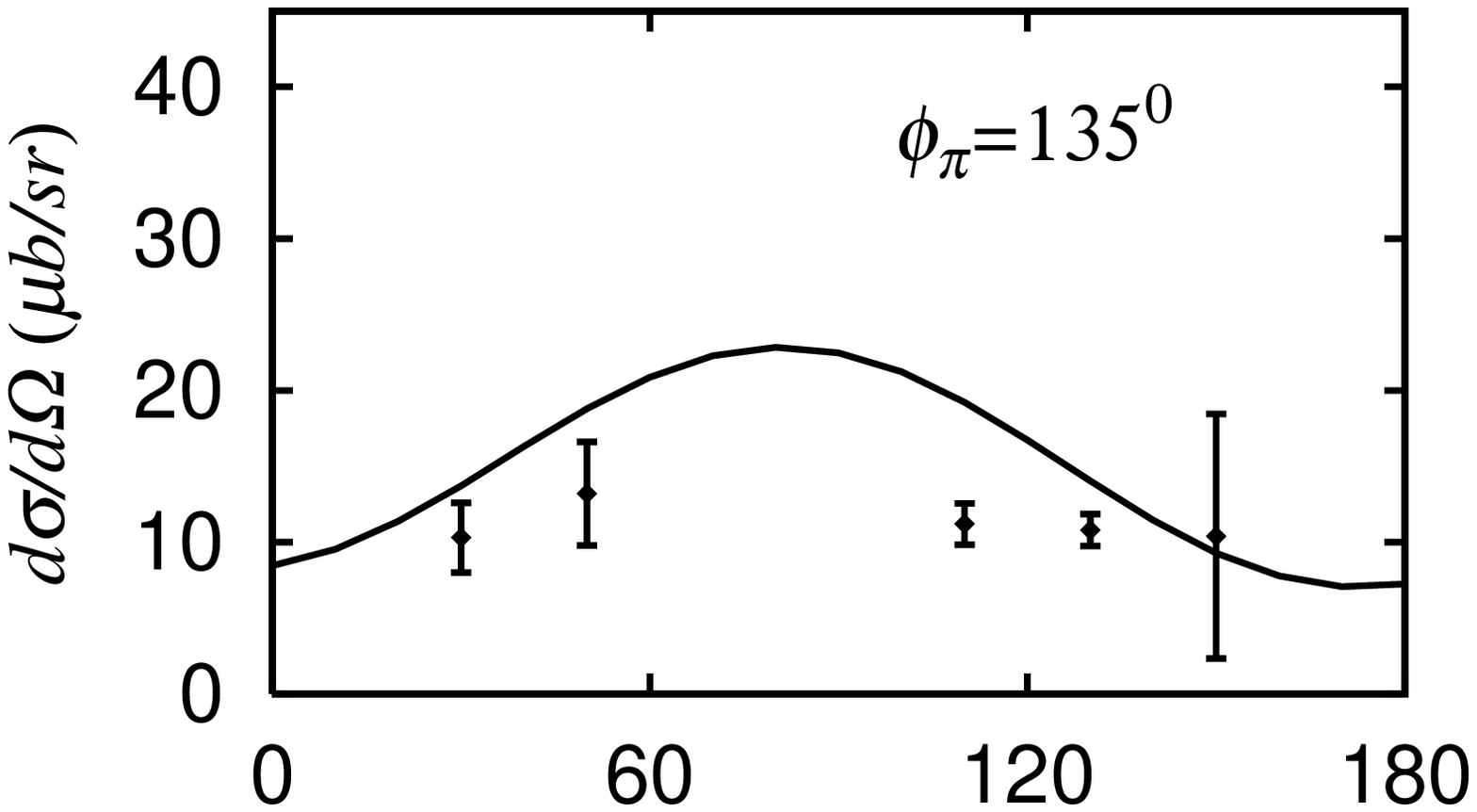,width=6cm}
\epsfig{file=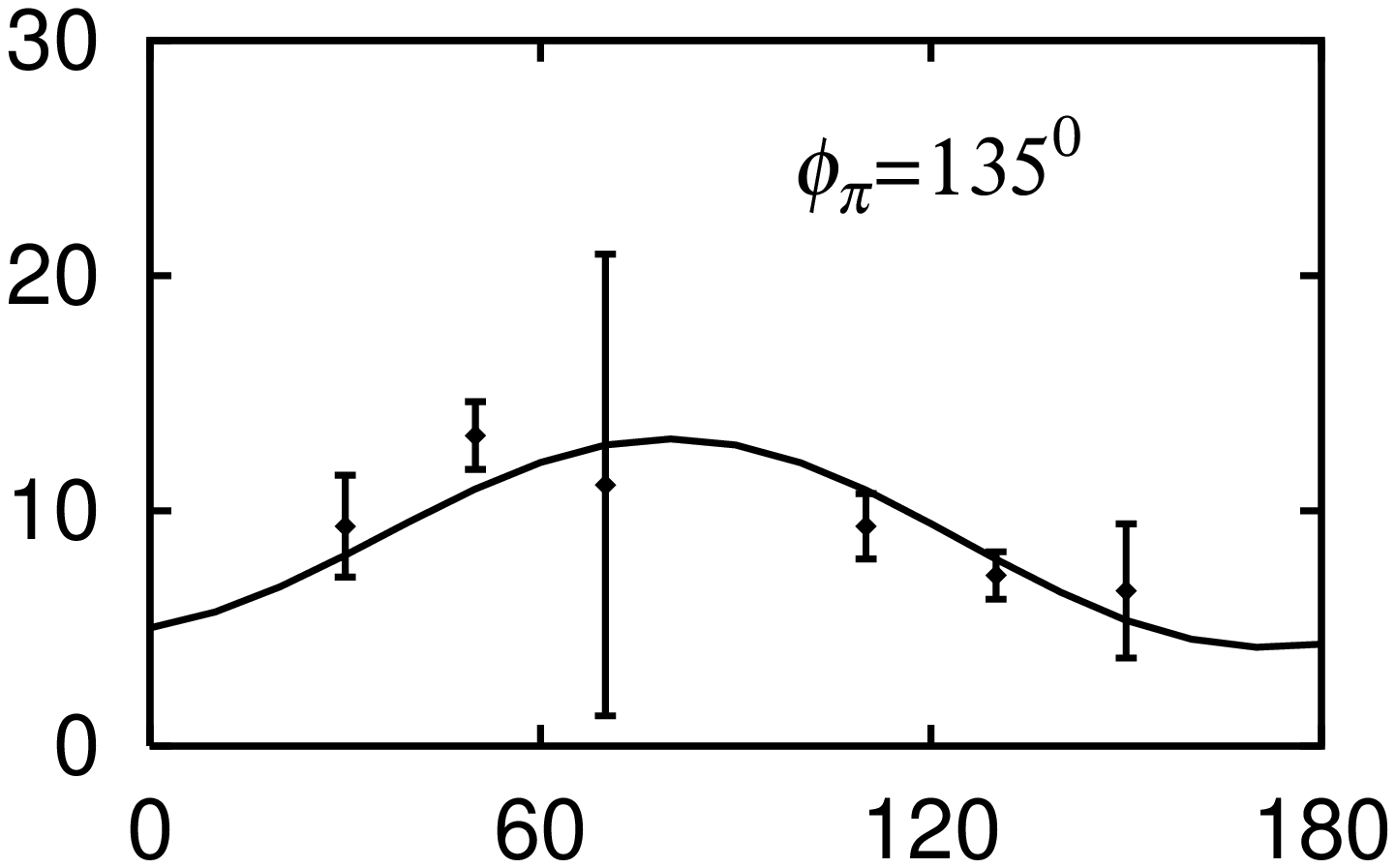,width=6cm}
}
\vspace*{-0.5cm}
\centerline{
\epsfig{file=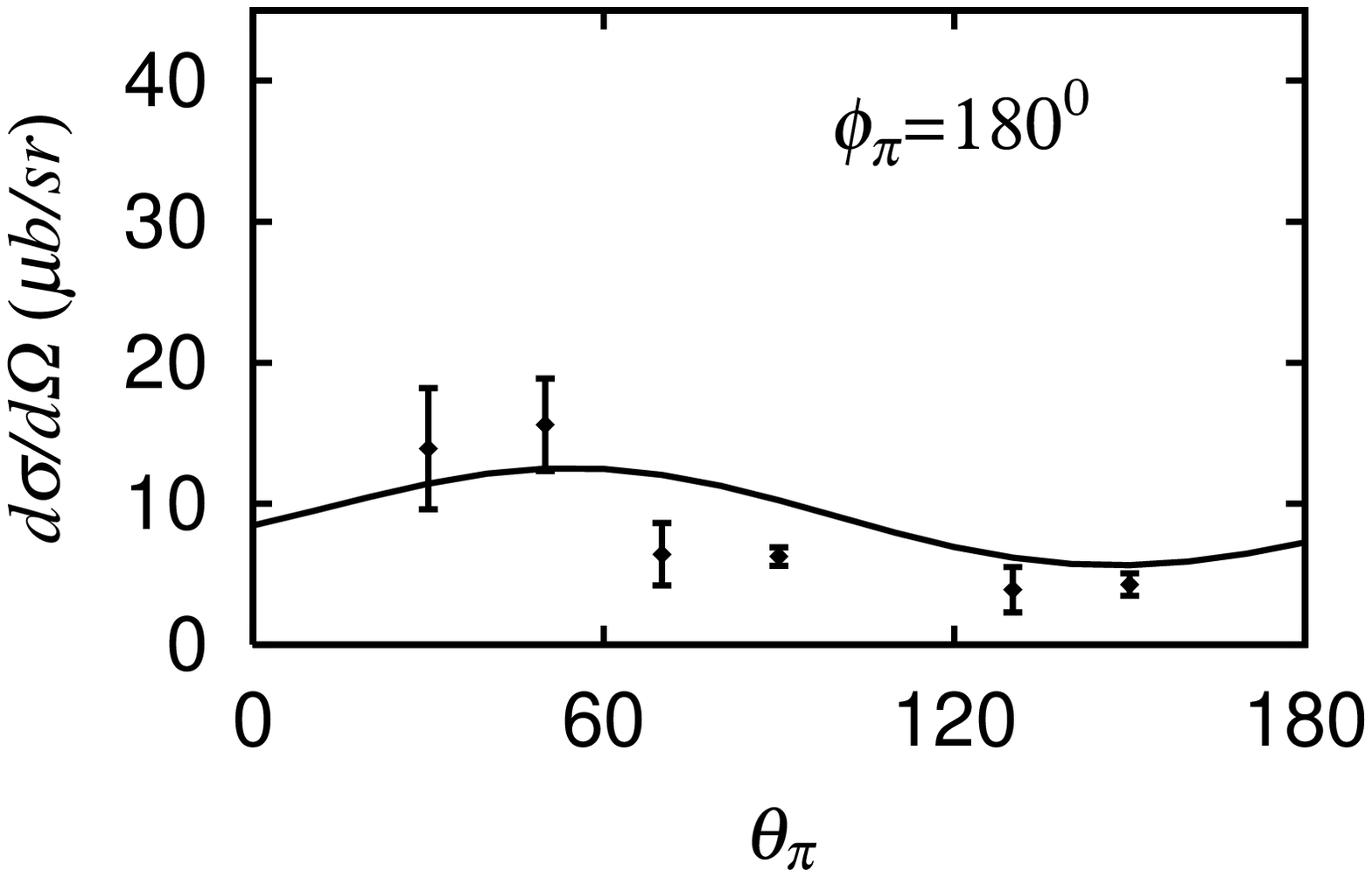,width=6cm}
\epsfig{file=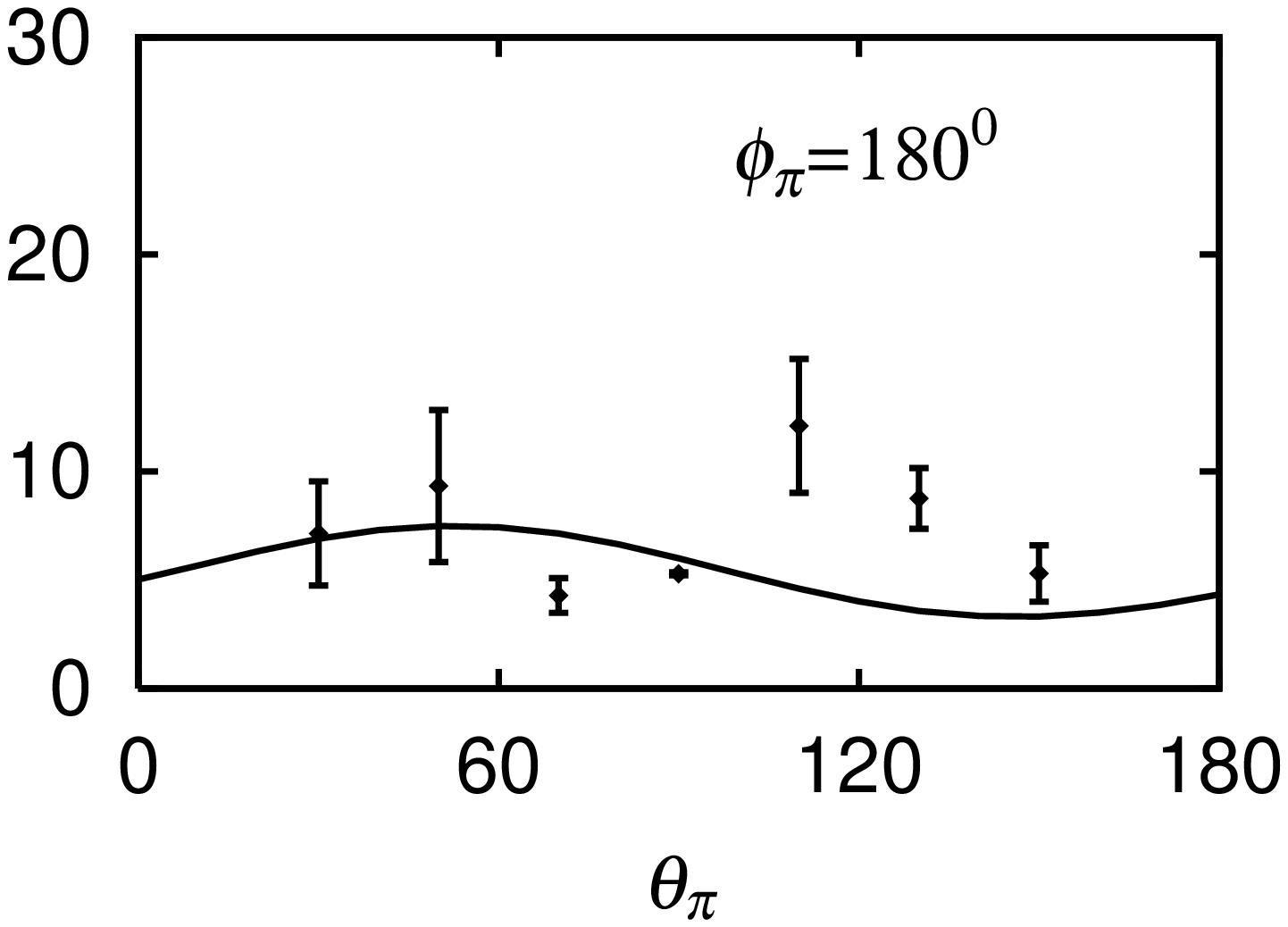,width=6cm}
}
\caption[]{The predicted differential cross sections of $p(e,e'\pi^0)$
reaction at $Q^2=$0.45(left), 0.75(right) (GeV/c)$^2$ 
and $W=$1232 MeV are compared with
the data \cite{bonn}.}
\end{figure}

\newpage

\begin{figure}
\centerline{
\epsfig{file=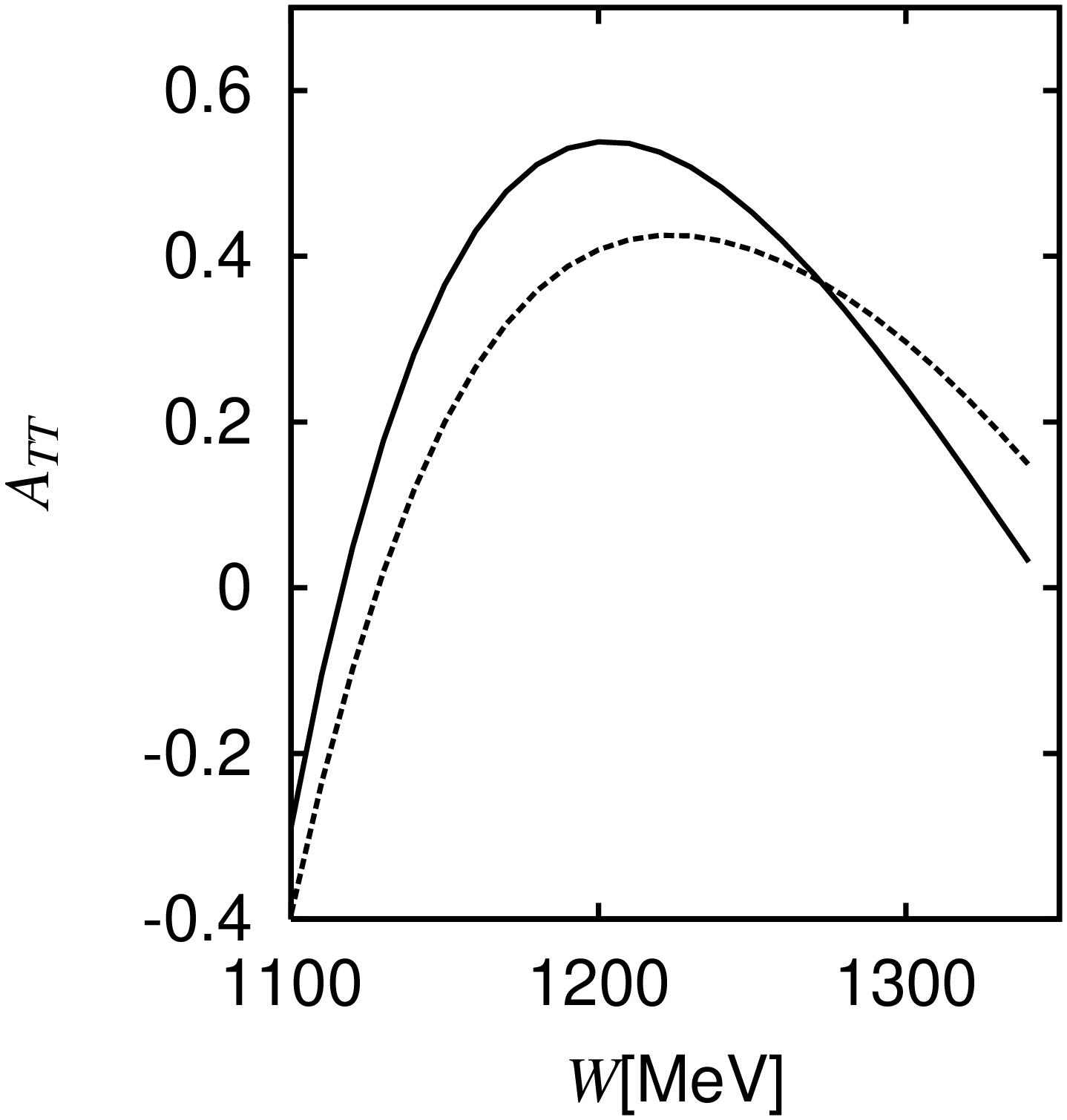,width=7cm}
\epsfig{file=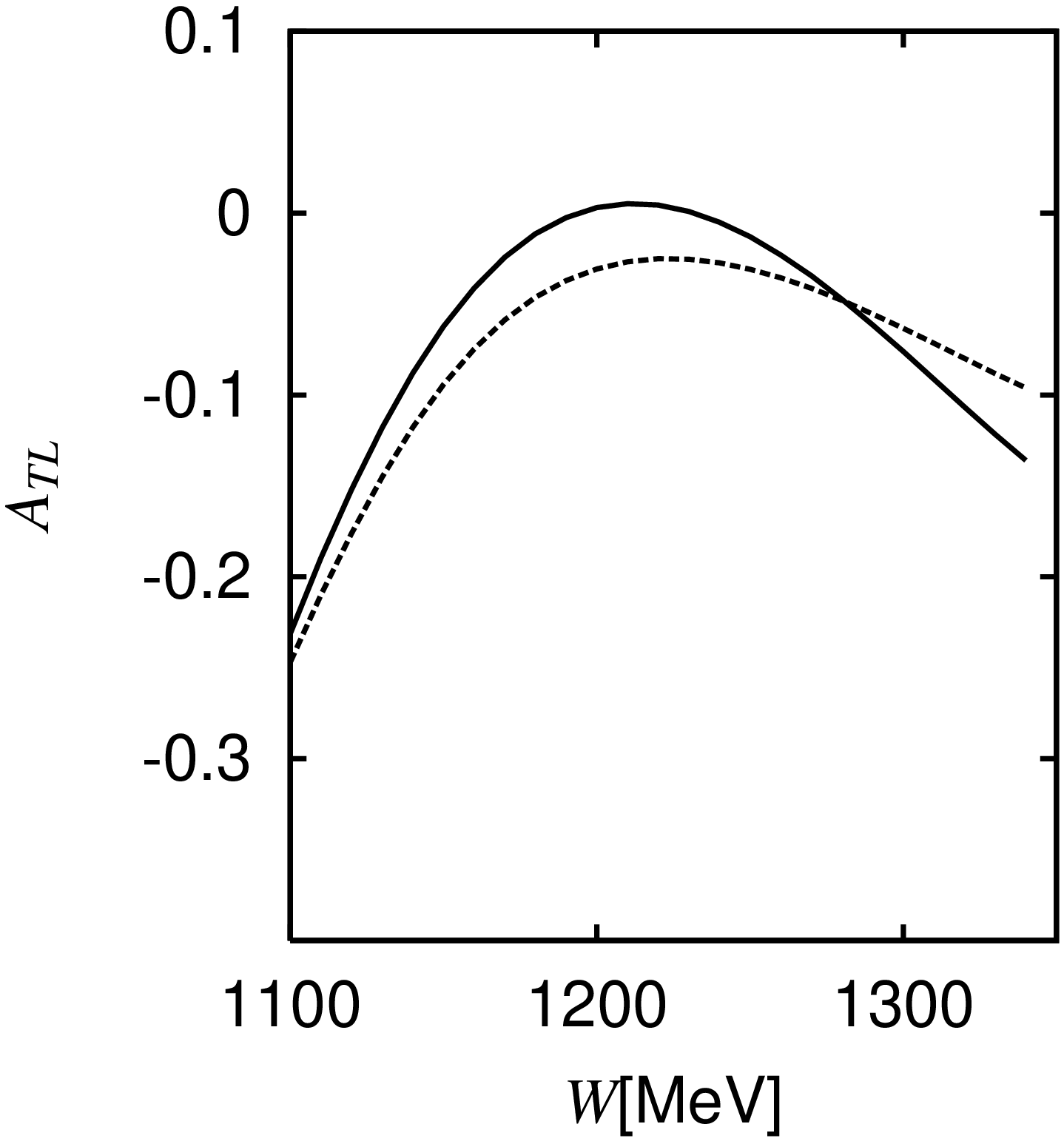,width=7cm}
}
\vspace*{0.5cm}
\caption[]{The predicted $A_{TT}$ and $A_{TL}$, as defined by Eqs. (2.25b) and
(2.25c) of Ref. \cite{nl1}, for the inclusive $\vec{p}(\vec{e},e')$ reaction
 at $Q^2=0.11$ (GeV/c)$^2$. 
The dotted curves are obtained when the $\gamma^* N \rightarrow \pi N$
multipole amplitudes $S_{1^+}^{3/2}$ and
 $E_{1^+}^{3/2}$ s
are not included in the calculation.}
\end{figure}

\end{document}